%% file: main.tex
\documentclass[
    twocolumn,
	prd,
        amssymb,
	preprintnumbers,
	secnumarabic,
	nofootinbib,
	superscriptaddress]{revtex4-1}

\pdfoutput=1

\usepackage{graphicx}
\usepackage{enumitem}
\usepackage{latexsym}
\usepackage{amsfonts}
\usepackage{amssymb}
\usepackage{pifont}
\usepackage{color}
\usepackage{xcolor}
\usepackage{amsmath}
\usepackage{slashed}
\usepackage{dcolumn}
\usepackage{verbatim}
\usepackage{float}
\usepackage{multirow}
\usepackage{xspace}
\usepackage[normalem]{ulem}
\usepackage{hyperref}

\definecolor{hypershade}{rgb}{0.3,0.3,0.8}
\hypersetup{
  pdfauthor={Nirmal Raj},
  pdftitle={Clumpy nuggety boom},
  pdfsubject={Clumpy nuggety boom},
  colorlinks=true,
  citecolor=red,
  urlcolor=blue,
  linkcolor=hypershade
}

\input{universalnewcommands.tex}

\allowdisplaybreaks

\begin{document}

\title{Supernovae and superbursts by dark matter clumps}

\author{Nirmal Raj}
\email{nraj@iisc.ac.in}
\affiliation{Centre for High Energy Physics, Indian Institute of Science, C. V. Raman Avenue, Bengaluru 560012, India}

\date{\today}

\begin{abstract}

Cosmologies in which dark matter clumps strongly on small scales are unfavorable to terrestrial detectors that are as yet unexposed to the clumps.
I show that sub-hectometer clumps could trigger thermonuclear runaways by scattering on nuclei in white dwarf cores (carbon and oxygen) and neutron star oceans (carbon), setting off Type Ia-like supernovae and x-ray superbursts respectively.
I consider two scenarios: ``dark clusters" that are essentially microhalos, and ``long-range dark nuggets", essentially macroscopic composites, with long-range Yukawa baryonic interactions that source the energy for igniting explosions.
I constrain dark clusters weighing between the Planck mass and asteroid masses, and long-range dark nuggets over a wider mass range spanning forty orders of magnitude.
These limits greatly complement searches I had co-proposed in Ref.~\cite{NSvIR:clumps2021} for scattering interactions of dark clumps in neutron stars, cosmic rays, and pre-historic minerals.

\end{abstract}

\maketitle

\section{Introduction}

Dark matter continues to elude non-gravitational detection.
One reason for why extensive terrestrial searches for the scattering interactions of particle dark matter (DM) have been futile could be that it is not distributed smoothly in the halo, but clumped in substructure, with detectors never having been exposed to DM clumps.
Cold DM, expected to clump down to milli-solar mass~\cite{Zybin:1999ic,Hofmann:2001bi,Berezinsky:2007qu}, 
may also clump below kiloparsec-scales if hierarchical clustering leaves small halos intact~\cite{BergstromGondolo:1998jj,vandenBosch:2017ynq,vandenBosch:2018tyt}.
Such clumps (``subhalos", ``microhalos", ``miniclusters", and so forth) could dominate the DM population in non-standard cosmologies that enhance small-scale power~\cite{ErickcekSigurdson,Barenboim:2013gya,FanWatson,drorcodecay,inflatflucs,Buckley:2017ttd,nussinovcluster,Barenboim:2021swl,Domenech:2023afs}. 
Whatever be the origin, the plausible clumping of DM compels us to rethink detection of its scattering on Standard Model (SM) states. 
In Ref.~\cite{NSvIR:clumps2021}, with co-authors I had proposed three strategies:
heating of neutron stars (NSs) by the transfer of kinetic energy during clump transits, 
exploiting the effects of strongly-interacting DM on cosmic rays and vice-versa,
and recasting searches for DM tracks in geological materials deep underground.
Here I will investigate one more: the triggering of thermonuclear explosions in compact stars by DM clumps depositing kinetic energy.

Type-Ia like supernovae are thought to originate from carbon-oxygen white dwarfs (WDs) accreting material from a binary companion and exceeding the Chandrasekhar mass, igniting runaway carbon fusion that unbinds the WD.
A suitable deposition of energy sourced by DM could trigger a similar explosion. 
This idea was first investigated for primordial black hole (PBH) DM undergoing dynamical friction in the WD material~\cite{GrahamRajendVarelaPBHWD}, with asteroid-mass PBHs ruled out from the observed survival of WDs, and though these limits could lift due to hydrodynamical instabilities in the wake of the transiting PBH quenching nuclear energy generation~\cite{MonteroCamachoHirata2019}, this caveat itself may lift for ignition via detonation (as opposed to deflagration)~\cite{SteigerwaldDetonationPBH:2021vgi}. 
Other DM mechanisms of explosive energy deposition in WDs have since struck the literature: 
gravitational potential energy shed via nucleon scattering as particle DM captured by the WD collapses in it ~\cite{Leung:2013pra,GrahamRajendVarelaPBHWD,Bramante:2015cua,FedderkeWDCHAMPs:2019jur,SteigerwaldProfumo:2022pjo},
super-heavy DM annihilating, decaying or scattering on nuclei~\cite{Graham:2018efk,MACROSidhu2020},
transiting composite DM accelerating WD nuclei in them and emitting bremmstrahlung radiation~\cite{Acevedo:2020avd}, 
and
Hawking radiation from black holes formed inside WDs via DM collapse~\cite{Acevedo:2019gre,Janish:2019nkk,FedderkeWDCHAMPs:2019jur}.
Ref.~\cite{StellarShocksDas2022} briefly mentions the possibility of tightly bound DM substructure triggering supernovae via impact shock waves. 
Some of these studies set limits based on the observed rates of supernovae.
I will use the existence of WDs to set limits on clump properties.

I will also set limits using ``superbursts", hours-long $\mathcal{O}(10^{42})$ erg x-ray bursts observed in NSs either isolated or hosted in low-mass or ultra-compact x-ray binaries~\cite{supburst:MINBARcatalog:2020,supburst:catalog2023}.
Type-I bursts in NSs are typically produced by surface accretion of mass from a binary companion that then ignites nuclear fuel into explosive burning. 
Superbursts are a special kind that last $10^3$ times longer and emit $10^3$ times more energy, with an accretion rate $> 10\%$ of the Eddington limit,  
suggesting that their ignition occurs deeper in the NS, in fact, in a layer of carbon in the NS ocean $\mathcal{O}$(10)~m below the surface~\cite{supburst:Cflashes:Cumming:2001wg,supburst:underZanding:2017ugu}.
DM clumps that reach this layer with sufficient energy to trigger superbursts more often than carbon accumulation can, are then ruled out.
Ref.~\cite{MACROSidhu2020} set limits on composite ``macro" DM using a single superburst source, while I avail of catalogues of recurring superbursts.

I consider two scenarios of clumpy DM.
The first, ``dark clusters", is that of subhalos comprised of DM particles that scatter on SM species, as treated in Ref.~\cite{NSvIR:clumps2021}; for simplicity I analyze only elastic nuclear recoils.
It will be seen that dark clusters of sub-cm size are required to trigger thermonuclear explosions. 
The second, ``long-range dark nuggets", is that of macroscopic composites that interact with baryons through a long-range Yukawa ``fifth force" as considered in Ref.~\cite{GreshamLeeZurek:2022biw} (although in that study the DM state could also be a point particle). 
These states could be thought of as DM ``nuclei", ``nuggets", or ``blobs"~\cite{nuggetsWitten:1984rs,nucleiKrnjaic:2014xza,nucleiDetmoldMcCullough:2014qqa,nucleiHardy:2014mqa,boundstateWise:2014jva,boundstateWise:2014ola,nucleiHardy:2015boa,nucleiGresham:2017zqi,nuggetsGresham:2017cvl,nucleiMcDermott:2017vyk,nuggetsGresham:2018anj,BaiLongLunuggets:2018dxf,BlobsGrabowska:2018lnd}, and would reduce to dark clusters in the limit of vanishing coupling and force range.
The Yukawa potential energy between compact stars and long-range dark nuggets can now act as the primary energy source for igniting thermonuclear explosions; the long-range fifth-force also serves to increase the flux of infalling nuggets compared to that due to gravity alone.
In both scenarios it will be seen that many (tens of) orders of magnitude in the space of DM clump mass, size and effective coupling are constrained.
My limits highly complement other probes of DM clumps and composites:
gravitational microlensing~\cite{ECOlocation1,ECOlocation2}, 
pulsar timing arrays~\cite{PTA:Dror:2019twh,PTALee:2021zqw}, 
accretion glows in molecular clouds~\cite{dMACHOS:Bai:2020jfm},
heating of cold inter-stellar material~\cite{Wadekar:2022ymq},
stellar impacts~\cite{StellarShocksDas2022},
gravitational waves~\cite{Croon:2022tmr},
and even underground detection~\cite{nussinovcluster,DDclumps:WidrowStiff:2001dq,DDclumps:Kamionkowski:2008vw,bradleyclumpSun,xmaslights:Bramante:2018qbc,xmaslights:Bramante:2018tos,BlobsGrabowska:2018lnd,xmaslights:Bramante:2019yss,Acevedo:2020avd,xmaslights:DEAPCollaboration:2021raj,Acevedo:2021kly,snowmass:Carney:2022gse,Jacobs:2014yca,Ebadi:2021cte,Acevedo:2021tbl}.
Long-range dark nuggets somewhat larger than those considered here could also trigger runaway helium fusion in the cores of red giant branch stars (RGBs), observable in the RGB luminosity function of globular clusters~\cite{Dessert:2021wjx}. A careful treatment of this effect would make for an interesting study in the future.
Thermal emission signatures of DM heating compact stars kinetically is an active area of study~\cite{NSvIR:Baryakhtar:DKHNS,NSvIR:Raj:DKHNSOps,NSvIR:SelfIntDM,NSvIR:Bell2018:Inelastic,NSvIR:GaraniGenoliniHambye,NSvIR:Queiroz:Spectroscopy,NSvIR:Bell2019:Leptophilic,NSvIR:Hamaguchi:RotochemicalvDM2019,NSvIR:GaraniHeeck:Muophilic,NSvIR:Pasta,NSvIR:Riverside:LeptophilicShort,NSvIR:Marfatia:DarkBaryon,NSvIR:Riverside:LeptophilicLong,NSvIR:Bell:Improved,NSvIR:DasguptaGuptaRay:LightMed,NSvIR:GaraniGuptaRaj:Thermalizn,NSvIR:Bell:ImprovedLepton,NSvIR:Bell2020improved,NSvIR:Queiroz:BosonDM,NSvIR:anzuiniBell2021improved,NSvIR:Lin2021:spin1med,NSvIR:zeng2021PNGBDM,NSvIR:HamaguchiEWmultiplet:2022uiq,NsvIR:HamaguchiMug-2:2022wpz,NSvIR:IISc2022,NSvIR:PseudoscaTRIUMF:2022eav,NSvIR:InelasticJoglekarYu:2023fjj,globularGaraniRajReynosoC:2023esk,snowmass:Berti:2022rwn}. 
More generally, dark matter interactions with compact stars are extensively reviewed in Ref.~\cite{reviewdarkincompact}.

This paper is laid out as follows.
In Sec.~\ref{sec:boomphysics} I review the conditions for runaway fusion in WD cores and NS crusts.
In Sec.~\ref{sec:clumpsnuggetsbooms} I describe the set-up of dark clusters and long-range dark nuggets, and the physics of their encountering compact stars and triggering thermonuclear explosions.
In Sec.~\ref{sec:results} I show and explain the limits thus obtained.
In Sec.~\ref{sec:disc} I provide further discussion and conclude.

\section{Thermonuclear Runaway}
\label{sec:boomphysics}

I now present the general requirements for triggering runaway fusion in a WD core or NS ocean, 
following the discussion in Ref.~\cite{FedderkeWDCHAMPs:2019jur}.
Two conditions must be met.
First, a mimimum energy $Q_{\rm dep}$ must be deposited to raise the temperature of a critical mass $M_{\rm crit}$ of density $\rho$ to a critical $T_{\rm crit}$ that sustains nuclear fusion:
\bea
\nn && \textsc{Condition 1}   \\
&& Q_{\rm dep} \geq M_{\rm crit} (\rho, T_{\rm crit}) \bar c_p (\rho, T_{\rm crit}) T_{\rm crit}~.
\label{eq:cond1}
\eea
Here I have assumed that the temperature prior to heating $\ll T_{\rm crit}$, and $\bar c_p \simeq c^{\rm e}_p/2 + c^\gamma_p/4 + c^{\rm ion}_p$ is the average isobaric specific heat capacity with the various contributions given by
\beq
 c^\ell_p (\rho, T_{\rm crit}) = \frac{a_\ell b_\ell}{u} \bigg(\frac{T_{\rm crit}}{E_{\rm F}}\bigg)^{\alpha_\ell} \bigg[1 - \bigg(\frac{m_e}{E_{\rm F}}\bigg)^2 \bigg]^{\beta_\ell}~, 
\label{eq:cp}
\eeq
where $u$ is the atomic mass unit, 
$m_e$ the electron mass,  
and $a_\ell = \{\pi^2,4\pi^4/5,5/2\}$, 
$b_\ell = \{\sum X_i Z_i/A_i, \sum X_i Z_i/A_i,\sum X_i/A_i, \}$ (with $X_i$, $Z_i$, $A_i$ respectively the mass fraction, charge and atomic number of the ion species $i$),
$\alpha_\ell = \{1, 3, 0\}$,
$\beta_\ell = \{-1, -3/2, 0\}$ for
the \{electronic, radiative, ionic\} contributions.
In practice $c^{\rm ion}_p$ is negligible.
The Fermi energy $E_{\rm F} = [m^2_e+(3\pi^2 n_e)^{2/3}]^{1/2}$ with $n_e = \rho b_{\rm e}/u$.
The trigger energy in Eq.~\eqref{eq:cond1} ranges between $\mathcal{O}(10^{17}-10^{24})$~GeV for WD central densities corresponding to WD masses ranging from 1.4 $M_\odot$ down to 0.8 $M_\odot$. 
Eq.~\eqref{eq:cond1} is necessary but not sufficient for runaway fusion.

The critical mass $M_{\rm crit} = 4\pi \rho \lambda_{\rm trig}^3/3$ is set by the second condition. 
Namely, the rate of energy gain via nuclear fusion must exceed the rate of energy loss via diffusion over the volume set by the ``trigger length" $\lambda_{\rm trig}$:\\
\bea
\nn && \textsc{Condition 2}  \\
&& \dot Q_{\rm nuc} > \dot Q_{\rm diff}~.
\label{eq:cond2}
\eea
Writing $\dot Q_{\rm nuc} = M_{\rm crit} \dot S_{\rm nuc}$ and $\dot Q_{\rm diff} \simeq 4\pi k \lambda_{\rm trig} T_{\rm crit}$ for per-mass nuclear energy deposition rate $\dot S_{\rm nuc}$ (the estimation of which involves numerical simulations of flame propagation with a nuclear reaction network~\cite{TimmesWoosley1992}) and thermal conductivity $k$, we obtain 
\bea
\nn \lambda_{\rm trig} &=& \sqrt{\frac{3 k T_{\rm crit}}{\rho \dot S_{\rm nuc}(\rho, T_{\rm crit})}}~ \\
&=& \begin{cases} 
\lambda_1~(\frac{\rho}{\rho_1})^{-2} \ \ \ \ \ \ \ \ \ \ \ \ \ \ \ \ \ \ \  \ , \rho \leq \rho_1 \\
\lambda_1~(\frac{\rho}{\rho_1})^{\ln(\lambda_2/\lambda_1)/\ln(\rho_2/\rho_1)}~, \rho_1 < \rho \leq \rho_2
\end{cases}
\label{eq:lambdatrig}
\eea
where for WDs  
$\{\lambda_1^{\rm WD},\lambda_2^{\rm WD}\} = \{1.3 \times 10^{-4}~{\rm cm}, 2.5 \times 10^{-5}~{\rm cm} \}$ 
and $\{\rho_1,\rho_2\} = \{2 \times 10^{8}~{\rm g/cm}^3, 10^{10}~{\rm g/cm}^3 \}$.
I have adopted this analytic form from Ref.~\cite{FedderkeWDCHAMPs:2019jur},
obtained by fitting to Figure 6 of Ref.~\cite{TimmesWoosley1992} that is restricted to $\rho_1 \leq \rho \leq \rho_2$ and then extrapolating to lower densities assuming reasonable density-scalings of $k$ and $\dot S_{\rm nuc}$.
This fit is for $T_{\rm crit}$ = 0.5 MeV, which I will take as my critical temperature, although it could be lower~\cite{TimmesWoosley1992}.
This fit also assumes, as I will, equal C and O masses in WDs.
In the NS ocean, the mass fraction of carbon is 10\%~\cite{supburst:Cflashes:Cumming:2001wg}, implying $\rho \to 0.1 \rho$ in Eq.~\eqref{eq:lambdatrig}.
 Though superbursts occur via pure C burning, I expect the scalings of Eq.~\eqref{eq:lambdatrig} to hold as seen from Ref.~\cite{TimmesWoosley1992}, Table 3, for conductive burning.
 
To translate between $\rho$ and WD masses $M_{\rm WD}$, I use the analytical fit~\cite{FedderkeWDCHAMPs:2019jur}
\bea
 \label{eq:rhoWDvMWD}
 && \bigg(\frac{\rho_{\rm WD}}{1.95\times 10^6 \ {\rm g/cm}^3}\bigg)^{2/3} +1 \approx \bigg[ \sum_{i=0}^6 c_i \bigg(\frac{M_{\rm WD}}{M_\odot}\bigg)^i\bigg]^{-2}~, \\
\nn && \{c_i\} = \{1.003, -0.309, -1.165, 2.021, -2.060, 1.169, -0.281\}~.
\eea
As my benchmark NS (with $v_{\rm esc}^{\rm NS} = \sqrt{2GM_{\rm NS}/R_{\rm NS}}$ the surface escape speed) I take
\beq
M_{\rm NS} = 1.4~M_\odot, ~ R_{\rm NS} = 10.99~{\rm km} \Rightarrow v_{\rm esc}^{\rm NS} = 0.61~,
\label{eq:NSbenchmark}
\eeq
as predicted by the Prakash-Ainsworth-Lattimer (PAL) equation of state of high-density matter~\cite{EoSPrakash:1988md}. 
To obtain my results I will use the NS crust density profile predicted by this model~\cite{crustdensityprofileRRIIIA}.

\begin{figure*}[htb!]
    \centering
    \includegraphics[width=0.75\textwidth]{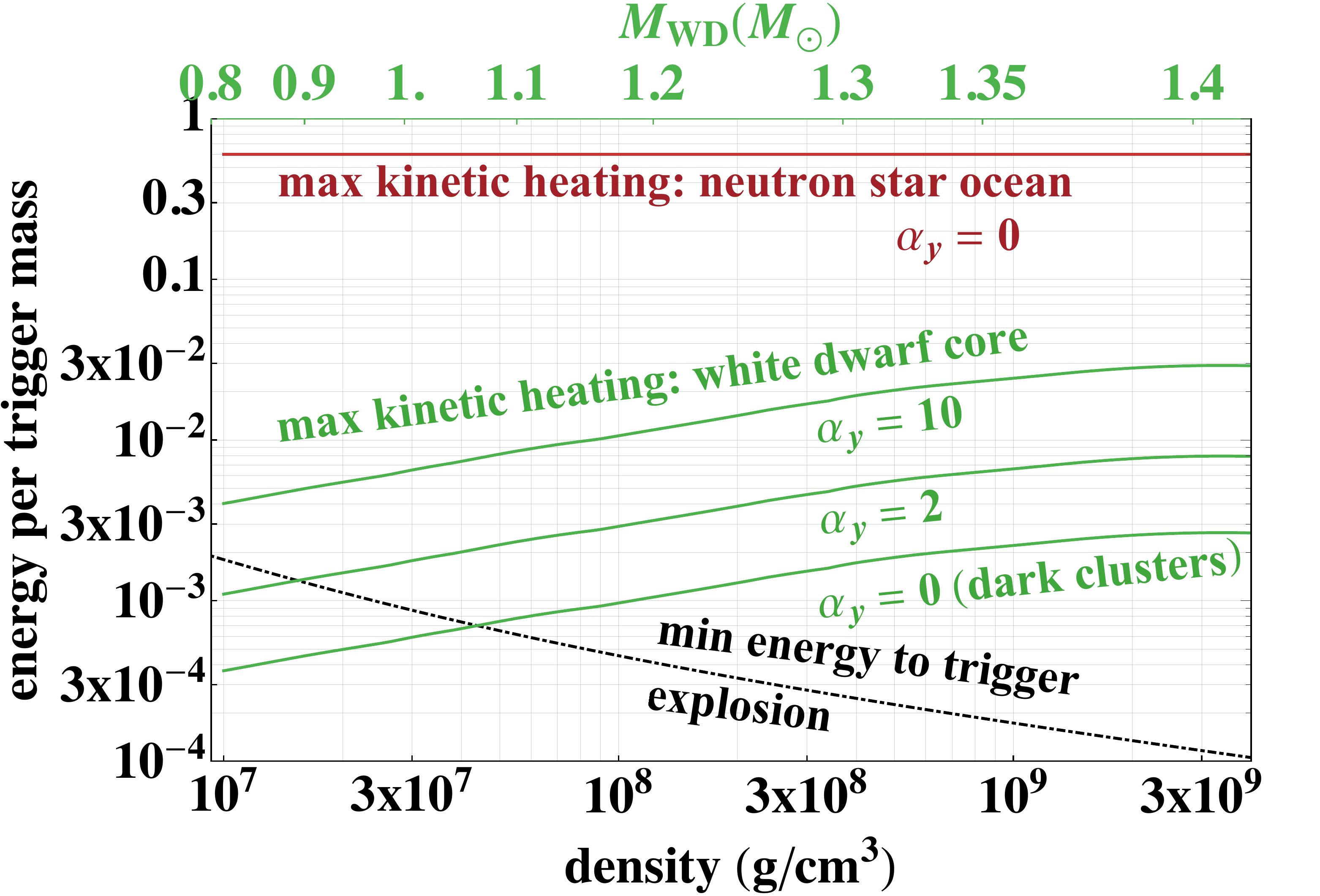}
    \caption{As a function of white dwarf central density are shown the trigger energy per trigger mass, $\bar c_p T_{\rm crit}$ (black dot-dashed), the maximum recoil energy of nuclei per trigger mass ($\approx v^2_{\rm esc}$) in white dwarf cores (green) for recoils induced by dark clusters ($\alpha_{\rm y} = 0$) and long-range dark nuggets ($\alpha_{\rm y} \neq 0$), and in neutron star oceans by dark clusters.
    The WD masses in the top x-axis ticks correspond to WD central densities in the bottom x-axis.
    Kinetic heating by clumpy dark matter is always capable of triggering superbursts in NSs, but can trigger supernovae only for WDs above some mass. 
    See Sec.~\ref{sec:clumpsnuggetsbooms} for further details.
  }
    \label{fig:EtrigvEkinvrho}
\end{figure*}

\section{Dark matter-induced explosions}
\label{sec:clumpsnuggetsbooms}

\subsection{Dark clusters}
\label{subsec:clumps}

Assuming dark clusters of uniform mass $M$ and radius $R$, the impact parameter of their encounters with compact stars of mass $M_\star \gg M$ and radius $R_\star$, accounting for gravitational focusing, is 
\beq
 b^{\rm cl}_{\rm max} = R_\star \sqrt{1+\bigg(\frac{v_{\rm esc}}{v_\chi}\bigg)^2} + R~,
\label{eq:bmaxclump}
\eeq
which at Galactic position $r$ gives an encounter rate{\footnote{Thanks to gravitational slingshot, this rate could be slightly higher in binary systems~\cite{Brayeur:2011yw} such as the sites of some superbursts. 
Moreover as the dark clusters are taken to be non-rigid, there may in principle be tidal effects in their accretion, but these are negligible for the small dark cluster sizes that are of interest here~\cite{NSvIR:clumps2021}.}} of
\bea
\label{eq:encounterrateclump}
\Gamma^{\rm cl}_{\rm meet} (r) &=&  f_\chi \frac{\rho_\chi(r)}{M} v_\chi(r) \pi (b^{\rm cl}_{\rm max})^2 \\
\nn  &=& \frac{f_\chi}{2 \ {\rm Gyr}}\bigg(\frac{10^{-15}~M_\odot}{M}\bigg)~,~R \ll R_\star,~ r = r_\odot~.
\eea

Here $f_\chi$, $\rho_\chi$ and $v_\chi$ are the fraction of dark clusters making up the DM, the DM density and DM speed, and 
$v_{\rm esc} = \sqrt{2 G M_\star/R_\star}$ is the escape speed at the stellar surface.
The second line of Eq.~\eqref{eq:encounterrateclump} is in the limit of small clumps at the solar position $r_\odot = 8.3$~kpc, where $\rho_{\chi}^\odot = 0.4$~GeV/cm$^3$, $v_{\chi}^\odot = 300$~km/s.
I will be primarily interested in the case of $f_\chi = 1$.

Through elastic center-of-momentum-frame isotropic scattering with DM particles of mass $m_\chi$ in the clusters, the typical rest-frame recoil energy of C or O ions of mass $m_{\rm T}$ is
\bea
\nn \bar E^{\rm cl}_{\rm rec} &\simeq& (2 z + z^2) \mu^2_{\rm T\chi}/m_{\rm T}\\
  &\simeq& m_{\rm T} v_{\rm esc}^2, \ \ m_{\rm T} \ll m_\chi , v_{\rm esc} \ll 1~,
  \label{eq:Edepclump}
\eea
where $\mu_{\rm T\chi}$ is the ion-$\chi$ reduced mass and $1+z = (1-v^2_{\rm esc})^{-1/2}$ is the blueshift of infalling DM.
As a function of $m_\chi$, $\bar E^{\rm cl}_{\rm rec}$ is greatest in the limit $m_\chi \gg m_{\rm T}$, where the maximum kinetic energy deposited into a trigger mass is simply $\simeq (2 z + z^2) M_{\rm trig} \simeq M_{\rm trig} v^2_{\rm esc}$.
This maximum is achieved when every ion in the trigger volume encounters a clump particle, i.e. when the clump is optically thick to an ion: 
\beq
\frac{\sigma_{\rm T\chi}}{m_\chi} \geq \frac{\pi R^2}{M}~.
\label{eq:optickthick}
\eeq
This condition is valid so long as the reduced cross section above is smaller than the ``ceiling" set by the outer layers essentially stopping the clump before it reaches the trigger volume. 
I elaborate on this in Sec.~\ref{subsec:clumps}. 

In Figure~\ref{fig:EtrigvEkinvrho} I comparatively plot two quantities as a function of white dwarf density: the maximum energy deposited per trigger mass, $2 z + z^2$, for both WD cores and NS oceans, and the minimum energy per mass to trigger runaway, $Q_{\rm dep}/M_{\rm trig} = \bar c_p T_{\rm crit}$ from Eq.~\eqref{eq:cond1}. 
(The former quantity for NS oceans of course does not depend on the WD density, and is hence depicted by a horizontal line.)
On the top x-axis the ticks are WD masses corresponding to the core densities in the bottom x-axis (Eq.~\eqref{eq:rhoWDvMWD}).
The curves labelled $\alpha_{\rm y} = 0$ correspond to dark clusters, as will become clear in Sec.~\ref{subsec:nuggets}.
One sees that only WDs with masses $\gsim 1.05 M_\odot$ can be triggered into thermonuclear runaway by elastic scattering of dark clusters.
Superbursts, on the other hand, can be comfortably triggered by elastic recoils due to the deeper gravitational potentials of NSs.

I have used the WD central density to estimate the trigger mass, which is justified as the WD density reduces by an $\Oc(1)$ factor in the outermost regions of the WD~\cite{BellImprovedWD:2021fye}. 
This reduces the effective radius where explosions occur by a small amount, potentially affecting constraints at the largest cluster masses (where the encounter rate is low) and smallest cluster sizes (just above the trigger volume). 
As this effect will not be visible over the many orders of magnitude of parameters constrained, I will simply adopt the central WD density for my estimates.

 \subsection{Long-range dark nuggets}
 \label{subsec:nuggets}

Now consider macroscopic DM composites $\mathcal{X}$ that interact with nucleons by exchanging a real spin-0 state $\phi$ of mass $\mu$, 
\beq
\mathcal{L} \supset g_n \phi \bar n n + g_\chi \phi \bar{\mathcal{X}} \mathcal{X}~,
\label{eq:Lag}
\eeq
inducing an attractive potential
\beq
V_{\rm Yuk}(r) =  - \alpha_{\rm y} \frac{G M_\star M}{r} e^{-\mu r}, \ \  \alpha_{\rm y}  = \frac{g_n g_\chi}{4\pi G m_n M}~.
\label{eq:VYuk}
\eeq

As mentioned in the Introduction, this potential sources energy that may be used to trigger supernovae and superbursts.
As seen shortly, this potential also serves to increase the incident flux of nuggets by augmenting gravitational focusing.
I particularly consider the region $\mu^{-1} \gg R_\star$, where this scenario is interesting. 
For $\mu^{-1} \ll R_\star$ and $\alpha_{\rm y} \ll 1$ my results are comparable to the gravity-only scenario just seen in Sec.~\ref{subsec:clumps}. 

For nuggets of size $\ll R_\star$, the maximum impact parameter for stellar intersections is~\cite{GreshamLeeZurek:2022biw}
\bea
\label{eq:bmaxnug}
 b^{\rm nug}_{\rm max} &=& {\rm min}(b_{\rm max, in}, b_{\rm max, out})~, \\
\nn b_{\rm max, in} &\approx& R_\star (1+z) \frac{v_{\rm esc}}{v_\chi}  \\
 \nn &\times& \bigg[1 + \alpha_{\rm y}e^{-\mu R_\star}\bigg( 1 + \alpha_{\rm y}e^{-\mu R_\star} \frac{v^2_{\rm esc}}{4} \bigg) \bigg]^{1/2}~,\\
\nn b_{\rm max, out} &\approx& \frac{w}{\mu}\sqrt{\bigg(1 + \frac{\mu R_\star}{w} \frac{v^2_{\rm esc}}{v^2_\chi} \bigg)}~, \\
\nn {\rm with} \ && w = {\rm max}\bigg(\log\bigg[\frac{\alpha_{\rm y}}{ \frac{2}{\mu R_\star}\frac{v^2_\chi}{v^2_{\rm esc}} + \frac{1}{\log \alpha_{\rm y}}} \bigg], 2 + \log 2\bigg)~,
\eea
where the analytic approximations were estimated in the limit of $v^2_\chi \ll v^2_{\rm esc}/2 \ll 1$ and $\mu^{-1} \gg R_\star$.
Eq.~\eqref{eq:bmaxnug} is obtained from (i) taking the orbit equation in Schwarzschild co-ordinates~\cite{GreshamLeeZurek:2022biw}, 
\bea
  M^2 \dot r^2 &=& (E-V_{\rm eff,+})(E-V_{\rm eff,-})~, \\
\nn \frac{V_{\rm eff, \pm}}{M} &=& -\frac{GM_\star\alpha_{\rm y}e^{-\mu r}}{r} \\
\nn && \pm \sqrt{\bigg(1-\frac{2GM_\star}{r} \bigg)\bigg(\frac{(L/M)^2}{r^2}+1\bigg)}~,
 \label{eq:orbiteqn}
\eea
(ii) locating two ``centrifugal barriers" $r_b$ in the effective potential from the conditions 
\bea
\nn V_{\rm eff,+}(r_b)|_{L=\gamma_\chi M b_{\rm max}v_\chi} &=& \gamma_\chi M\ \ [dr/d\tau = 0]~, \\
V'_{\rm eff,+}(r_b)|_{L=\gamma_\chi M b_{\rm max}v_\chi} &=& 0 \ \ \ \ \ \  \  [{\rm local \ max}]~,
\label{eq:barrierconditions}
\eea
where $\gamma_\chi = (1-v^2_\chi)^{-1/2}$, 
(iii) setting the solved $r_b = R_\star$.
The ``inner" barrier is gravity-driven and the ``outer" is fifth-force-driven.

The nugget-star encounter rate is then obtained by simply making the replacement in Eq.~\eqref{eq:encounterrateclump}
\beq
\Gamma^{\rm nug}_{\rm meet} = \Gamma^{\rm cl}_{\rm meet}|_{b^{\rm cl}_{\rm max} \ra b^{\rm nug}_{\rm max}}~.
\label{eq:encounterratenugget}
\eeq
In the set-up above, DM nugget transits deposit energy in the star through tidal excitations of seismic oscillations~\cite{GreshamLeeZurek:2022biw}, however I will assume additional short-range scattering interactions that dominate energy deposition, i.e. the ``maximal heating" scenario considered in Ref.~\cite{GreshamLeeZurek:2022biw}.
As in the case of dark clusters, the individual constituents $\chi$ of the nuggets scatter on the star-bound targets\footnote{I expect coherence in scattering over neither the target nucleus nor the projectile nugget. My parameter space of interest is $\alpha_{\rm y} > 1$, where the effects of the fifth force are apparent, and in this region the momentum transfer $q \gg R^{-1}_{\rm nuc}, R^{-1}_{\rm nug}$, where $R_{\rm nuc} \simeq \mathcal{O}(1)$~fm is the nuclear size and $R_{\rm nug}$ is the nugget size required for successful thermonuclear explosion. See, e.g., Ref.~\cite{BlobsGrabowska:2018lnd}.}, but now at higher speeds due to acceleration by both gravity and the fifth force in Eq.~\eqref{eq:VYuk}. 
For a target state of mass $\bar m_{\rm T}$, the typical rest-frame recoil energy is 
\bea
\label{eq:Edepnug}
\bar E^{\rm nug}_{\rm rec} &\simeq& \frac{\bar m_{\rm T}m^2_\chi}{\bar m_{\rm T}^2+m^2_{\chi}+2m_\chi \bar m_{\rm T}(1+\tilde z)}(2 \tilde z + \tilde z^2)~\\
\nn &\simeq& \begin{cases}
m_\chi \tilde z \frac{1+\tilde z/2}{1+\tilde z}  \ \ \ ,~~ 1 \ll m_\chi/\bar m_{\rm T} \ll (1+\tilde z) \\
\bar m_{\rm T} (2\tilde z + \tilde z^2)    ,~~   (1+\tilde z) \ll m_\chi/\bar m_{\rm T}~,
\end{cases}
\eea
where 
\bea
\nn 1 + \tilde z &=& (1 + z)\bigg(1+\alpha_{\rm y}e^{-\mu R_\star}\frac{v_{\rm esc}^2}{2}\bigg) \\
 &\approx& 1 + \frac{v_{\rm esc}^2}{2}(1+\alpha_{\rm y}e^{-\mu R_\star})~.
\eea

It can be seen that for $v_\chi \ll v_{\rm esc} \lsim 1$, Eq.~\eqref{eq:bmaxnug} $\ra$ Eq.~\eqref{eq:bmaxclump} and Eq.~\eqref{eq:Edepnug}$\ra$ Eq.~\eqref{eq:Edepclump} in the limit $\alpha_{\rm y} \ra 0$, $\mu \ra \infty$.
Thus for a force range $\mu^{-1} \gg R_\star$, the kinetic energy of the incoming nugget at the stellar surface is $\simeq (1+\alpha_{\rm y})$ times that from a purely gravitational pull.
This implies WDs of mass $< 1.05 M_\odot$ can be triggered, unlike for dark clusters as discussed in Sec.~\ref{subsec:clumps}.
I have depicted this in Fig.~\ref{fig:EtrigvEkinvrho}, where increasing $\alpha_{\rm y}$ is seen to help explode lighter WDs.
Of course, non-zero $\alpha_{\rm y}$ trivially ignites superbursts in NS oceans.
For simplicity I have assumed here elastic nuclear scattering, arising for instance from nugget constituents scattering with geometric cross sections (as discussed in Sec.~\ref{subsec:limitsclumps}), however it is also possible that nuggets transferring high momenta resolve nucleons or partons.
In such inelastic scattering, the fraction of nugget kinetic energy transferred could be as high as $y_{\rm max} \gsim 0.1$~\cite{Agashe:2015xkj}.

\begin{figure*}[htb!]
    \centering
    \includegraphics[width=0.75\textwidth]{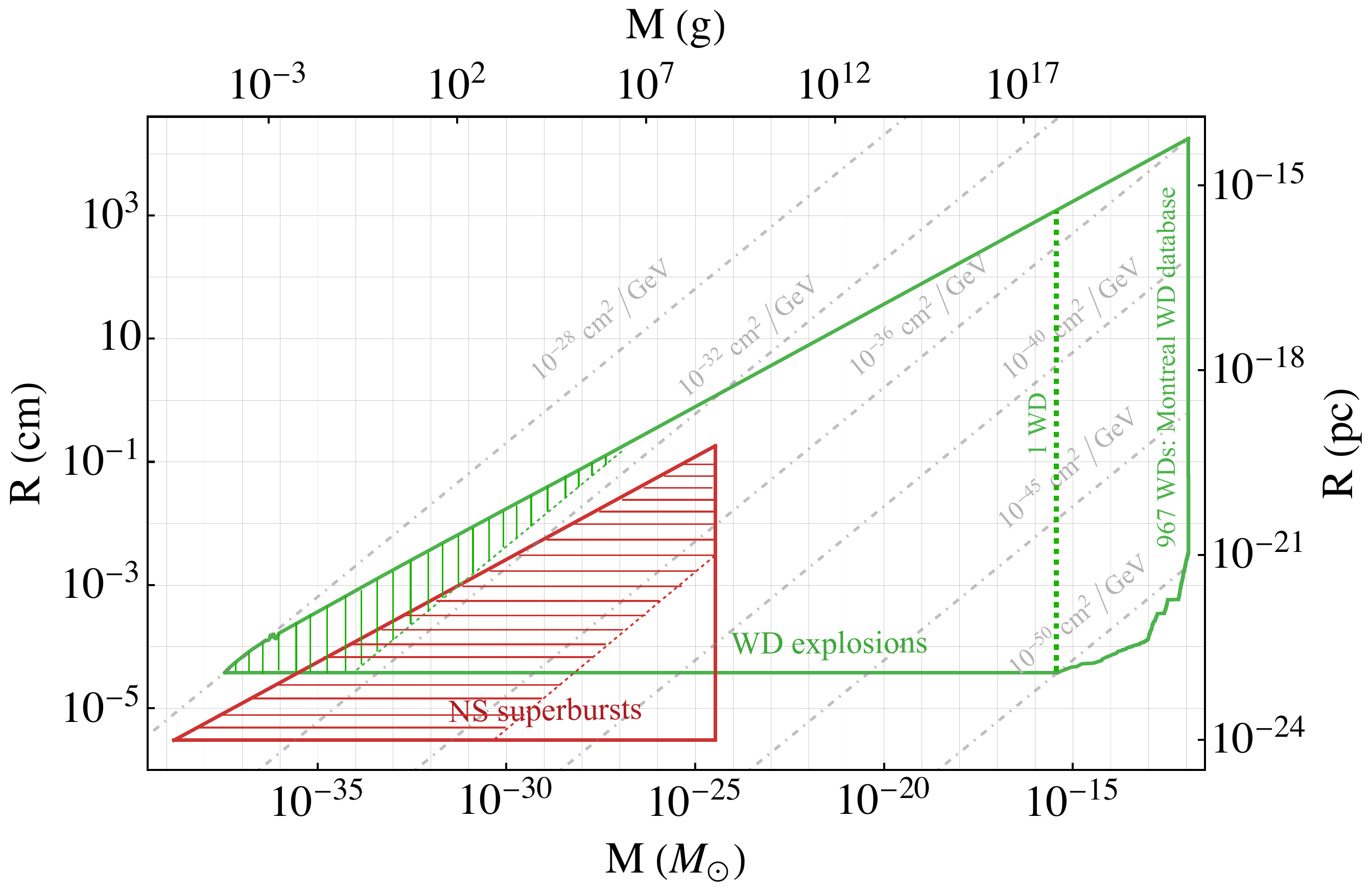}
    \caption{Dark cluster masses and radii excluded by their triggering of Type Ia-like supernovae in carbon-oxygen WDs and x-ray superbursts in NS carbon oceans. 
    Overlaid are contours of the minimum reduced cross sections for nuclear scattering on cluster constituents in order for all nuclei in a trigger mass to recoil, so that the total energy transferred is guaranteed to cause explosion (Fig.~\ref{fig:EtrigvEkinvrho}).
    The regions shaded with green vertical lines and red horizontal lines rely on a ``saturated overburden effect" by which the point-like scattering cross sections are high enough to prevent the deceleration of the dark cluster constituents by the outer layers of the compact star.
    These regions likely require cross sections saturated by unitarity with contributions from higher partial waves.
    The maximum dark cluster mass limited by WD explosions, depending on the null hypothesis, could vary between the dotted green vertical line that considers a single WD of mass 1.41~$M_\odot$ in the Montreal WD Database, and the solid green vertical line to its right accounting for the 967 WDs for which lifetimes are known.  
    Molecular gas cloud heating and CMB distortions, not shown here, limit the reduced cross section to about $\sigma_{\rm T\chi}/m_\chi < 10^{-27}$~cm$^2$/GeV.
    For the tiny region that is constrained here for $M < 10^{-38} M_\odot$, limits from Ohya and DEAP-3600 detectors may potentially apply for $m_\chi \gg 10^{10}$~GeV. 
    See Secs.~\ref{subsec:clumps} and \ref{subsec:limitsclumps} for further details.
    }
    \label{fig:RvM}
\end{figure*}

\begin{figure*}[htb!]
    \centering
\makebox[\textwidth][c]{ \includegraphics[width=0.6\textwidth]{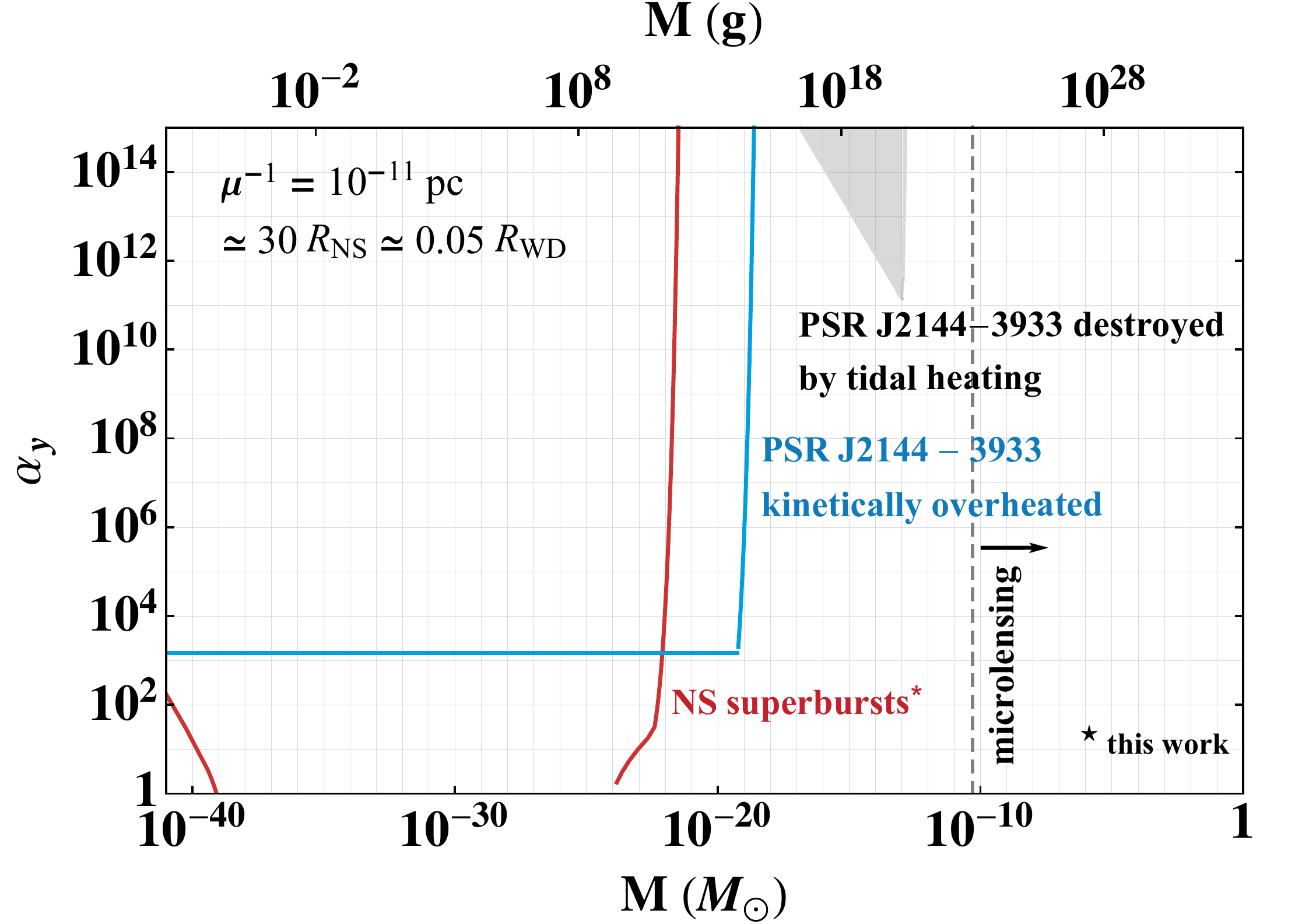} \includegraphics[width=0.6\textwidth]{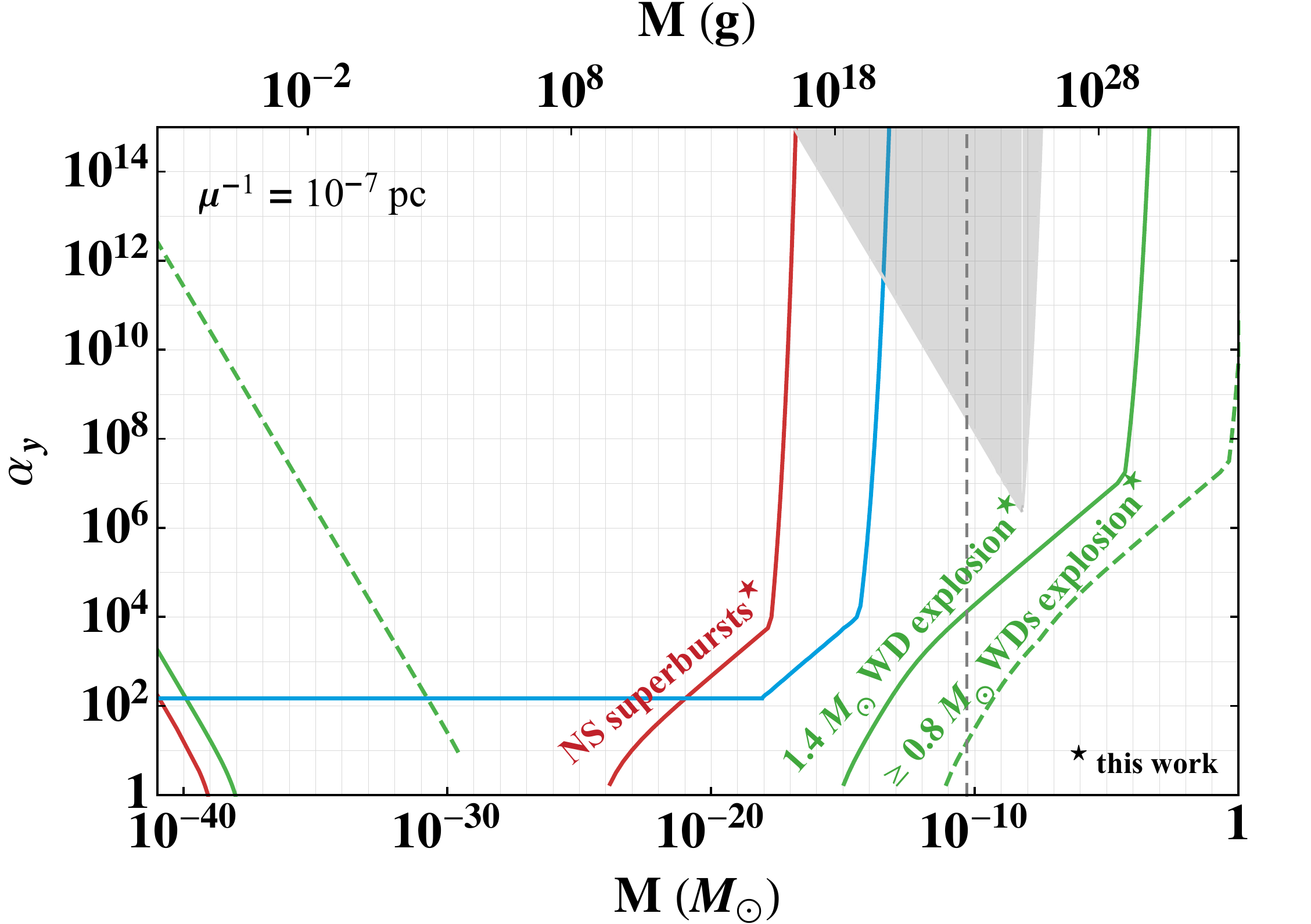}} \\
\vspace{.2 in}
 \makebox[\textwidth][c]{ \includegraphics[width=0.6\textwidth]{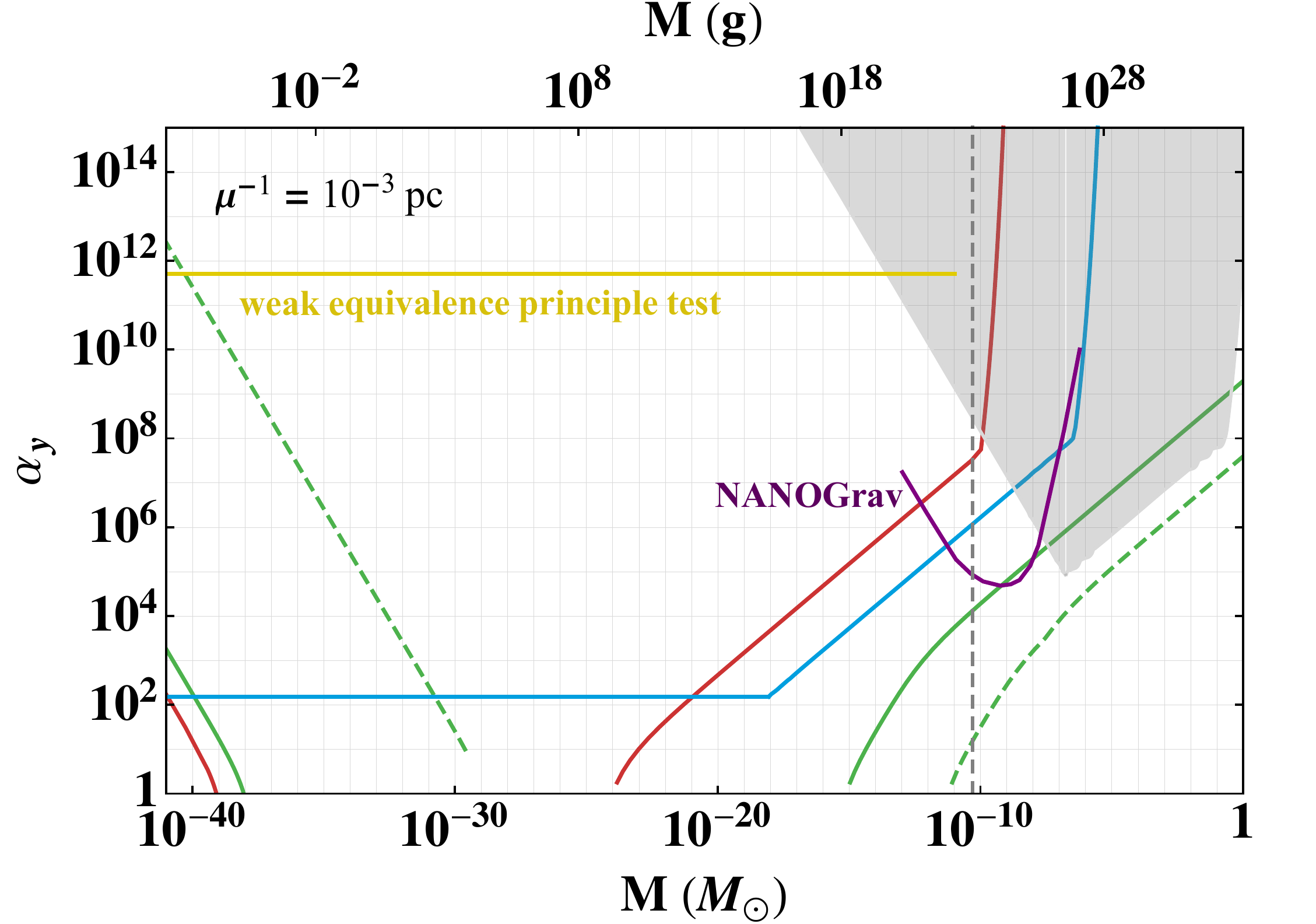}  \includegraphics[width=0.6\textwidth]{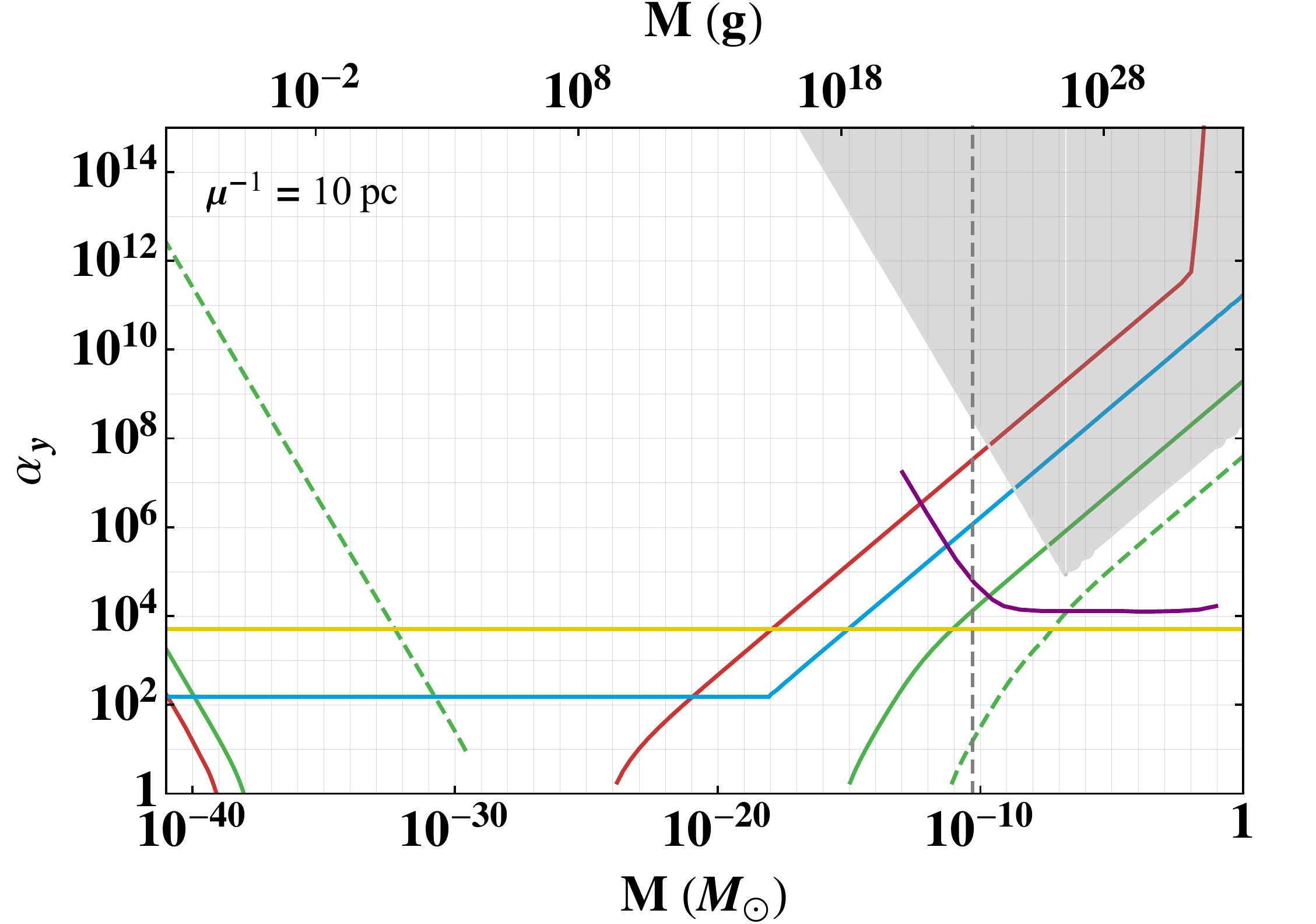} }
    \caption{Long-range dark nugget effective couplings and masses excluded by their triggering of supernovae and superbursts for various values of the fifth-force range, assuming nugget sizes just big enough to accomplish ignition. 
    Limits from other studies are also shown; the microlensing limits assume that nuggets comprise all the DM, but would lift rapidly for a DM sub-component.
     See Sec.~\ref{subsec:limitsnuggets} for further details.}
    \label{fig:alphavM}
\end{figure*}

\begin{table}
    \centering
    \begin{tabular}{|c|c|c|}
    \hline
        superburst & r (kpc) & $t_{\rm recur}$ (yr)  \\
        \hline
     4U 1820+30  & 1.2 & 2.5~\cite{MACROSidhu2020} \\
      4U 0614+091   & 11.5  & 4.8~\cite{supburst:catalog2023} \\
         GX 17+2  & 2.4 & 1~~~\cite{supburst:catalog2023} \\
       4U 1636$-$536  & 5.3 & 1~~~\cite{supburst:trecur:CummingZandPage2005} \\
      Ser X$-$1  & 5 & 1~~~\cite{supburst:trecur:CummingZandPage2005} \\
       Aql X$-$1 & 5 & 1~~~\cite{supburst:trecur:CummingZandPage2005} \\
        \hline
    \end{tabular}
    \caption{Recurring superbursts~\cite{supburst:catalog2023} used to set limits in this work. 
    Their Galactic position $r = \sqrt{r^2_{\odot} + d^2_\oplus - 2 r_\odot d_\oplus \cos \ell \cos b}$ was obtained using information on the distance to the source $d_\oplus$, its longitude $\ell$ and latitude $b$ from SIMBAD~\cite{SIMBAD}. 
    The recurrence times $t_{\rm recur}$ are taken from the references indicated.
    See Sec.~\ref{sec:results} for further details.}
    \label{tab:superbs}
\end{table}

\section{Results}
\label{sec:results}

\subsection{Limits on dark clusters}
\label{subsec:limitsclumps}

In Figure~\ref{fig:RvM} I show limits on the dark cluster size vs mass from the existence of WDs (region enclosed by green curves) and the observed frequency of NS superbursts (red).
For WDs I fix the upper bound on $R$ as $R < \lambda_{\rm trig}$ (Eq.~\eqref{eq:lambdatrig}) since the trigger condition Eq.~\eqref{eq:cond1} will be satisfied for $R \geq \lambda_{\rm trig}$.
Smaller clumps, too, would satisfy Eq.~\eqref{eq:cond1}: as seen in Fig.~\ref{fig:EtrigvEkinvrho}, for $M_{\rm WD} > 1.05 M_\odot$ dark clusters of size $\lambda_{\rm trig}$ would raise the temperature of the trigger volume to {\em above} $T_{\rm crit}$.
By comparing the trigger energy and maximum energy deposit in Fig.~\ref{fig:EtrigvEkinvrho}, one sees that for the heaviest WDs, the minimum triggering $R$ is about $\lambda_{\rm trig}/10^{1/3}$, thus my choice of the upper bound on $R$ is conservative by a factor of $\lsim 2$. 
For NSs, I take the upper bound to be $R \geq \lambda_{\rm trig}(\bar c_p T_{\rm crit}/(2 z + z^2))^{1/3}$ (Eqs.~\eqref{eq:cond1} and \eqref{eq:Edepclump}), ensuring sufficient energy deposition to ignite the ocean matter.
For WDs the smallest $R$ bounded in Fig.~\ref{fig:RvM} is the $\lambda_{\rm trig}$ that corresponds to a near-Chandrasekhar limit 1.41 $M_\odot$ WD for $M \in [10^{-38.5}, 10^{-16.5}] M_\odot$.
The upper end of this range comes from requiring 2.3 encounters (corresponding to the 90\% C.L. limit) with clumps of the solitary 1.41 $M_\odot$ WD in the Montreal White Dwarf Database (MWDD)~\cite{MontrealWDDatabase} for which the WD age (1.3 Gyr) is available.  
For higher $M$, I require that the total clump-WD encounters $N^{\rm cl-WD}_{\rm enc} = 2.3$~, with
\beq
N^{\rm cl-WD}_{\rm enc} = \sum_i \Gamma^{\rm cl}_{{\rm meet},i} \tau_{{\rm WD},i}~,
\label{eq:Nenc}
\eeq
where $\Gamma^{\rm cl}_{\rm meet}$ is from Eq.~\eqref{eq:encounterrateclump},
$\tau_{\rm WD}$ is the WD age, and $i$ runs over WDs in MWDD for which mass and cooling age information are available.
As $M$ is increased, $\Gamma^{\rm cl}_{\rm meet}$ decreases, and to compensate for this the limiting $R$ samples over larger WDs (in spite of the smaller gravitational focusing), which corresponds to lighter WDs, in turn corresponding to smaller WD central densities and hence larger $\lambda_{\rm trig}$ (Eq.~\eqref{eq:lambdatrig}), resulting in larger upper bounds on $R$.   
The right vertical line at $M = 10^{-12}~M_\odot$ is reached after sampling over 967 WDs; the upper bound on $R$ here corresponds to triggering WDs of mass 1.05 $M_\odot$, the minimum for which cluster-induced elastic nuclear recoils can result in WD explosion (Sec.~\ref{subsec:clumps}).
In obtaining these limits I set the DM density and velocity dispersion to their values in the solar neighborhood, an approximation that is justified since only 19 WDs in the sample lie outside the local kpc.
The null hypothesis inherent in setting these limits is that no Galactic WD has been destroyed by dark clusters in the past, or in other words the MWDD sample is assumed to be consistent with standard WD population models.
A more conservative limit may be set by considering a single WD, and this is depicted in Fig.~\ref{fig:RvM} with a vertical green dotted line along the maximum $M$ constrained by the sole 1.41 $M_\odot$ WD in the MWDD sample.

For superbursts the upper bound on $R$ is the $\lambda_{\rm trig}$ corresponding to the NS density at the bottom of the C ocean, $1.2\times 10^{10}$~g/cm$^3$, which I estimated by first obtaining crustal column depths by integrating over the NS crust density profile of PAL Model 1~\cite{crustdensityprofileRRIIIA}, and then by comparison with the maximum ignition column depth shown in Fig.~1 of Ref.~\cite{supburst:Cflashes:Cumming:2001wg}: $4 \times 10^{13}$ g/cm$^2$.
The right vertical line is from requiring the total clump-NS encounters $N^{\rm cl-NS}_{\rm enc} = 2.3$ using an equation analogous to Eq.~\eqref{eq:Nenc} with the replacement $\tau_{\rm WD} \ra t_{\rm recur}$, where $t_{\rm recur}$ is the timescale, typically $\mathcal{O}$(yr), of recurrence between repeating superbursts.
In Table~\ref{tab:superbs} I have listed estimates of $t_{\rm recur}$ of the 6 superburst sources that have recurred at least once, out of 16 observed in total~\cite{supburst:catalog2023}. 
In the encounter rate I set the DM densities at the superburst locations (in Table~\ref{tab:superbs}) to those obtained from an NFW profile with scale parameters taken from Ref.~\cite{PPPCookbook:2010xx};
I also set $v_\chi (r) = \sqrt{3 G M_{\rm NFW}(r)/2r}$, the Maxwell-Boltzmann dispersion speed for a mass $M_{\rm NFW}(r) = \int_0^r d^3r^\prime \rho(r^\prime)$ enclosed within $r$.
The source 4U 1820-30 resides in the globular cluster NGC 6624, where the DM densities could be potentially higher (and dispersion speeds lower), however due to the large uncertainties in the dynamics of such stellar systems~\cite{globularGaraniRajReynosoC:2023esk} I will conservatively take $\rho_\chi (r)$ in Eq.~\eqref{eq:encounterrateclump} to be the Galactic background one. 
Despite that, this superburst dominates the encounter rate due to its proximity to the Galactic Center, warranting its exclusive use by Ref.~\cite{MACROSidhu2020}.

The left diagonal lines in Fig.~\ref{fig:RvM} are from requiring that the fraction of the kinetic energy of the infalling cluster transferrable to the trigger volume (through either elastic or inelastic scattering) exceeds the trigger energy:
\beq
z M \bigg(\frac{\lambda_{\rm trig}}{R}\bigg)^3 \geq M_{\rm trig} \bar c_p T_{\rm crit}~.
\label{eq:RvMleftdiag}
\eeq
I overlay these plots with contours of constant $\sigma_{\rm T\chi}/m_\chi$ corresponding to the minimum required for the clumps to be optically thick to nuclei (Eq.~\eqref{eq:optickthick}).
There is a lower bound (ceiling) to $\sigma_{\rm T\chi}/m_\chi$ above which DM scattering with the outer stellar layers degrades its energy and thwart explosion in the relevant inner layer.
For WDs this arises from the non-degenerate envelope with thickness $\simeq 10^{-3} R_\star$ and maximum density $\simeq$ 100 g/cm$^3$~(\cite{text:paddytheoryastroII}.
For the corresponding column depth, this limit is $\sigma_{\rm T\chi}/m_\chi \gsim 4 \times 10^{-32}$~cm$^2$/GeV.
This value is larger than the ceiling reported in Ref.~\cite{Graham:2018efk} by  $\Oc(10^3)$ and Ref.~\cite{MACROSidhu2020} by $\Oc(10^{10})$, which I discuss in detail in Ref.~\ref{sec:disc}.
Similarly for NS crusts, using the column depth at the carbon ocean bed in Ref.~\cite{supburst:Cflashes:Cumming:2001wg}, the limit is  $\sigma_{\rm T\chi}/m_\chi \gsim 4 \times 10^{-38}$~cm$^2$/GeV.
For point-like scatters these ceilings break down for $\sigma_{\rm T\chi}$ above the ``saturated overburden" value, the point at which the mean-free path of DM particles equals the inter-nuclear distance, $({\rm overburden \ number \ density})^{-1/3}$~\cite{Bramante:2018qbc}.
That is, above some $m_\chi$ the energy degradation from scattering on a finite number of nuclei in the upper layers is insufficient to prevent an explosive trigger deeper down; this is $m_\chi \gsim 10^{15}$~GeV for both WDs and NSs.
This effect is akin to that seen for simple DM-nucleus scattering potentials that saturate the cross section at the geometric one, so that for high DM masses the overburden is ineffective~\cite{GoodmanWitten:1984dc,DigmanBarn:2019wdm}.
In Fig.~\ref{fig:RvM} I have shown the regions that rely on the saturated overburden effect using shading with green vertical (for WD explosions) and red horizontal lines (for NS superbursts). 
For this effect to be present, large cross sections are required but the scattering itself must be point-like, i.e., it cannot proceed through a finite-range mediator and the cross sections cannot be geometric, as these would imply scattering with all the targets within the cross sectional area along the path of the DM particle. 
Such a scenario, although not easily conceivable, is nevertheless achievable for cross sections that saturate unitarity but to which higher partial waves contribute, as discussed in Refs.~\cite{xmaslights:Bramante:2018tos} and \cite{DigmanBarn:2019wdm}.
Moreover, as shown in Ref.~\cite{resonantXuFarrar:2020qjk}, (non-Breit-Wigner) resonances that arise generically for attractive scattering potentials could make per-nuclear cross sections many orders of magnitude larger than the geometric limit even for point-like scattering of DM.
Thus the parameters in the horizontal and vertical shaded regions of Fig.~\ref{fig:RvM} are physically possible, but currently not quite amenable to model-building.
I remark here that Ref.~\cite{MACROSidhu2020} estimates the ceiling $\sigma_{\rm T\chi}/m_\chi$ of a tightly bound strongly-interacting DM candidate by treating the drag force it experiences, which is numerically equal to the ceiling obtained from the stopping power of point-like scattering.

Finally, the region $4 \times 10^{-19} < M/M_\odot < 10^{-13}$ (and for $R$ as high as $10^9$~cm) can be probed in the future by observing a single 10$^3$ Kelvin NS or $>$100 $10^4$ Kelvin NSs via the same kinetic heating mechanism considered here~\cite{NSvIR:clumps2021}.

Dark clusters can be potentially constrained using a number of complementary searches.
For dark clusters masses below about $10^{-38} M_\odot \simeq 10^{19}$~GeV, the integrated flux of dark clusters through terrestrial setups can become appreciable.
Searches for tracks in the Ohya etched plastic detector constrain $\sigma_{\rm T\chi}$ to be smaller than about $10^{-19}$~cm$^2$, for constituent DM particle masses $m_\chi$ between 10$^9-10^{19}$~GeV~\cite{Bhoonah:2020fys}.
Similarly, the DEAP-3600 experiment limits $\sigma_{\rm T\chi} \lsim 10^{-22}$~cm$^2$ for $m_\chi$ between $10^{18}-10^{19}$~GeV~\cite{xmaslights:DEAPCollaboration:2021raj}. 
As there is only a very small region satisfying $M < 10^{-38} M_\odot$ constrained in the $M$-$R$ space in Fig.~\ref{fig:RvM}, and as these terrestrial limits are either too high compared to the minimum $\sigma_{\rm T\chi}/m_\chi$ required for triggering thermonuclear explosions or may be avoided by choosing a suitable range of $m_\chi$, I do not display them in Fig.~\ref{fig:RvM}.
Searches for tracks in ancient mica limit $\sigma_{\rm T\chi} \lsim 10^{-17}$~cm$^2$ for $10^{12} < (m_\chi/{\rm GeV}) < 10^{26}$~\cite{Bhoonah:2020fys}, which is again either higher than my minimum required $\sigma_{\rm T\chi}/m_\chi$ or can be evaded by choosing $m_\chi$ sufficiently small. 
The limit from the heating of molecular gas clouds is $\sigma_{\rm T\chi}/m_\chi < 10^{-27}$~cm$^2$/GeV~\cite{gasclouds:Bhoonah:2018gjb} and from the distortion of the cosmic microwave background is $\sigma_{\rm T\chi}/m_\chi < 10^{-26}$~cm$^2$/GeV~\cite{snowmass:Carney:2022gse}, implying that my dark cluster scenario is well safe from them.
In the future, large volume neutrino detectors like SNO+ may gather enough integrated flux of dark clusters to set complementary limits~\cite{xmaslights:Bramante:2018tos,xmaslights:Bramante:2019yss}.
Investigations of this and other scenarios discussed in Sec.~\ref{sec:disc} are left to future work.

\subsection{Limits on long-range dark nuggets}
\label{subsec:limitsnuggets}

In Figure~\ref{fig:alphavM} I show limits on the effective fifth-force coupling vs dark nugget mass for four values of the force range $\mu^{-1}$.
For $\mu^{-1} = 10^{-11} \ {\rm pc} \ll R_{\rm WD}$, there are no WD limits; for $\mu^{-1} >$ 10 pc, the typical inter-WD~\cite{WDdistribs:Napiwotzki} or inter-NS distance~\cite{NSdistribs:BlaesMadau,NSdistribs:Sartore2010}, the use of Eq.~\eqref{eq:bmaxnug} may not be valid.   
I assume for minimality that the dark nuggets are of the size $\leq \lambda_{\rm trig}$ required to heat the nuclear material to $\geq T_{\rm crit}$ as described in Sec.~\ref{subsec:limitsclumps}; these sizes are well within the range of nugget models in the literature~\cite{BaiLongLunuggets:2018dxf,BlobsGrabowska:2018lnd,nuggetsGresham:2017cvl,GreshamLeeZurek:2022biw}.
{\em \`{A} la} dark clusters in Sec.~\ref{subsec:limitsclumps}, the mass lower bounds at 90\% C.L. come from requiring a total of 2.3 nugget-star encounters, where now in Eq.~\eqref{eq:Nenc} (and its NS equivalent with $\tau_{\rm WD} \to t_{\rm recur}$) I use the encounter rate in Eq.~\eqref{eq:encounterratenugget}.
For WDs I show two cases corresponding to different nugget sizes: destroyal of all WDs of masses $\geq 0.8 M_\odot$ and that of a single 1.41 $M_\odot$ WD.
The slopes of these curves steepen above some $\alpha_{\rm y}$ because the impact parameter in Eq.~\eqref{eq:bmaxnug} transitions from being set by the ``inner" centrifugal barrier to the ``outer" (Sec.~\ref{subsec:nuggets}).
Across the four panels, more parameter space is constrained as $\mu^{-1}$ is increased because $b_{\rm max, out}$ correspondingly increases (Eq.~\eqref{eq:bmaxnug}).

Again as in Sec.~\ref{subsec:clumps}, the mass upper bounds come from requiring that the incoming kinetic energy of the nugget, $\tilde z M$, exceeds the trigger energy in Eq.~\eqref{eq:cond1}.  
For 0.8~$M_\odot$ WDs this occurs for $\alpha_{\rm y} \geq 9.7$.
Note that the denser the trigger region, the smaller is the minimum nugget mass required to trigger it.
Note also that as $\alpha_{\rm y} \ra 0$ the minimum and maximum masses in Fig.~\ref{fig:alphavM} correspond to those in Fig.~\ref{fig:RvM}.

I also show in Fig.~\ref{fig:alphavM} the mass limit from gravitational microlensing surveys, to which the dark nuggets appear as point-like lenses as they are far smaller than the Einstein radius~\cite{ECOlocation1,ECOlocation2}. 
This only applies to $f_\chi = 1$; for $f_\chi \lsim 10^{-2}$ these limits go away, while the lower bound on $M$ weakens $\propto f_\chi$ due to fewer nugget-star encounters (Eq.~\eqref{eq:encounterratenugget}).
Also shown are limits derived in Ref.~\cite{GreshamLeeZurek:2022biw}:
(1) From kinetically overheating the coldest observed ($\lsim 4 \times 10^4$ K) NS, PSR J2144-3933~\cite{coldestNSHST}, i.e., by requiring $\tilde z M \Gamma^{\rm nug}_{\rm meet} \leq$ the luminosity of this NS. 
The maximum $M$ limitable is obtained by requiring $1/\Gamma^{\rm nug}_{\rm meet} < $500 yr, the cooling timescale of an NS of this temperature.
More parameter space can be potentially probed by surveying sky regions for excess populations of NSs that are at even higher temperatures~\cite{NSvIR:clumps2021}. 
(2) From destroying PSR J2144-3933 by depositing more energy via tidal heating than about 10\% of its gravitational binding energy, which occurs for $\tilde z M > 3\times10^{-3} M_\odot$.
The maximum $M$ limitable now corresponds to $1/\Gamma^{\rm nug}_{\rm meet}$ = 300 Myr, the age of the NS. 
(3) From recasting 11-yr NANOGrav pulsar timing data to look for phase shifts induced by long-range dark nuggets transiting observational lines of sight.
These limits are only in the bottom panels because smaller $\mu^{-1}$ exponentially suppress the new phase shifts in pulse arrival times.
They weaken at small $M$ since the effects of gravity and the fifth-force lessen, and also weaken at large $M$ for $\mu^{-1} = 10^{-3}$~pc as the typical nugget-pulsar distance increases, which decreases the new phase shift; for $\mu^{-1} = 10$~pc the limit on $\alpha_{\rm y}$ plateaus because here the fifth force effectively rescales the gravitational constant (Eq.~\eqref{eq:VYuk}).
N. B. For pulsar timing limits to apply the nuggets need only be smaller than their impact parameter ($> 10^9$~km), thus they apply for nuggets larger than those required to trigger thermonuclear explosions, and larger than the Einstein radius where microlensing limits go away. 
(4) From weak equivalence principle tests measuring the difference in acceleration of two celestial bodies approaching the Galactic Center, giving the limit $\alpha_{\rm y} (\mu^{-1}/{\rm pc})^2 \lsim 5 \times 10^5$ so long as $(\rho_\chi/M)^{-1/3} \ll \mu^{-1} \ll r_\odot$.

My limits are seen to complement all of the above in regions spanning several decades of parameters.
In particular, although my limits have qualitatively similar shapes to the one from heating PSR J2144$-$3933 at high masses, I am able to constrain much higher $M$ with WD explosions.
This is because my mass upper bound is set by the Gyr timescale of the ages of $\sim$ 10$^3$ WDs, as opposed to the 500 yr timescale of cooling of a single NS.

\section{Discussion}
\label{sec:disc}

I have set limits on macroscopic dark matter clumps/composites by their triggering of Type Ia-like supernovae and superbursts, requiring that the maximum recoil energy they can induce via scattering on all target nuclei in a triggerable volume exceeds the minimum energy needed to ignite that volume.
This is possible when the clump is optically thick to a stellar nucleus, given that the clump constituents are not too softened by scattering with the stellar outer layers, which constitute the overburden here.

The overburden reduced cross section I find for white dwarfs, $4 \times 10^{-32}~$cm$^2$/GeV, is 4000 times higher than that reported in Ref.~\cite{Graham:2018efk} and about 10 orders higher than in Ref.~\cite{MACROSidhu2020}.
I attempt to explain these discrepancies here.
The ceiling from the overburden may be obtained as $1/(\rho_{\rm env} L_{\rm env})$, where $\rho_{\rm env}$ and $L_{\rm env}$ are the density and thickness of the WD's non-degenerate envelope.
In Ref.~\cite{Graham:2018efk} $L_{\rm env}$ is taken as 50~km and the relation $\rho_{\rm env} \simeq 10^{-3} \rho_{\rm WDc}$ is used, with the WD central density taken as $3\times 10^8$~g/cm$^3$ as appropriate for the 1.25~$M_\odot$ benchmark WD considered for setting constraints.
With these values, one estimates the ceiling to be about $10^{-36}$~cm$^2$/GeV.
This roughly agrees with the ceiling in Ref.~\cite{Graham:2018efk}.
Despite the caption of Ref.~\cite{Graham:2018efk}'s Figure 6 mentioning  ``DM-carbon elastic scattering cross section", the cross sections displayed in this figure and Figure 11 are per-nucleon ones, as stated elsewhere in the text.
Rescaling this per-nucleon ceiling $10^{-39}$~cm$^2$/GeV by $12^4$, we get the per-nucleus ceiling reduced cross section as about $10^{-35}$~cm$^2$/GeV.
The authors of Ref.~\cite{Graham:2018efk} cite the textbook in Ref.~\cite{text:kippenhahnweigert} for their values of $\rho_{\rm env}$ and $L_{\rm env}$.
In page 375 of this text, the WD density at the transition between the non-degenerate envelope and degenerate core is quoted as approximately $10^3$~g/cm$^3$, which roughly agrees with the density derived from the equations on that page for a 1.25 $M_\odot$ WD, which is about 710 g/cm$^3$.
This is clearly more than five orders of magnitude smaller than $\rho_{\rm WDc}$ as opposed to merely three.
This error may have perhaps arisen from a statement in page 375 of Ref.~\cite{text:kippenhahnweigert} to the effect that the central density of an average WD is $10^6$~g/cm$^3$, which would have indeed made $\rho_{\rm env} = 10^3$~g/cm$^3$ equal to $10^{-3} \rho_{\rm WDc}$.
But as just mentioned, the central density used in Ref.~\cite{Graham:2018efk} is 300 times larger.
Page 375 of Ref.~\cite{text:kippenhahnweigert} also states that $L_{\rm env}$ is ``1\% or less" of $R_{\rm WD}$, and Ref.~\cite{Graham:2018efk} perhaps took the upper end of this bound to get $L_{\rm env} = 50$~km.
In my estimate in Sec.~\ref{subsec:limitsclumps}, $\rho_{\rm env} = 100$~g/cm$^3$ and $L_{\rm env} = 5$~km as appropriate for a median-mass 1.05 $M_\odot$ WD, derived from Eqs.~(5.107), (5.108) and and (5.115) of Ref.~\cite{text:paddytheoryastroII}, also in agreement with Eqs.~(4.1.10) and (4.1.13) of Ref.~\cite{text:ShapiroTeukolsky}.
These values give a ceiling of $4 \times 10^{-32}~$cm$^2$/GeV.
In summary, it would appear that in Ref.~\cite{Graham:2018efk} a combination of taking an envelope density too high and the use of a thicker envelope conspire to give a discrepancy with my ceiling estimate by a factor of $\Oc(10^3-10^4)$. 
It is harder to trace the discrepancy between my ceiling estimate and Ref.~\cite{MACROSidhu2020}, as $\rho_{\rm env}$ and $L_{\rm env}$ are not given.
However, in Section III.C.1 (``Maximum constrained reduced cross section"), Ref.~\cite{MACROSidhu2020} states that the ceiling is at $10^{-16}$~cm$^2$/g for WDs and $10^{-12}$~cm$^2$/g for NSs. 
This would imply that the column density of a WD envelope is $10^4$ larger than that of an NS crust, which clearly cannot be true as it is well-known that NSs (even the outer layers) are generally much denser than WDs.
My best guess is that Ref.~\cite{MACROSidhu2020} took $\rho_{\rm env}$ and $L_{\rm env}$ from Ref.~\cite{Graham:2018efk}, before making refinements to the ceiling estimate so that it varies by a few decades as the DM mass is varied.
Yet another way to see that the ceiling estimates of Refs.~\cite{Graham:2018efk} and \cite{MACROSidhu2020} may be erroneous is to compare them with the geometric cross section of the WD core for particle capture.
This latter cross section $\pi R^2_{\rm WD} m_{\rm C}/M_{\rm WD}$ is about $10^{-39}$~cm$^2$. 
The ceiling from the non-degenerate envelope quoted by Ref.~\cite{Graham:2018efk}, if one were to extrapolate it to GeV mass DM, is only 10$^3$ times this value, and the one by Ref.~\cite{MACROSidhu2020} ranges between $10^{-42}-10^{-40}$~cm$^2$/GeV, both of which are clearly implausible as the column density of the WD envelope is smaller than that of the WD core by a factor of more than 10$^3$.

Several directions from here are ripe for inquiry.
I had considered elastic nuclear scatters of dark cluster and nugget constituents, but alternative possibilities are inelastic scatters -- such as baryon-absorption of Q-ball DM treated in Ref.~\cite{Graham:2018efk} -- and scattering on electrons by leptophilic long-range dark nuggets that deposit explosive energy.
The long-range potential of the dark nuggets need not be Yukawa, as treated here, but something else, e.g., Coulombic.
The energy stored in the long-range potential can also serve, through scattering alone, to heat planets (which has been done before with DM self-annihilations~\cite{GlobalWarming:Kawasaki:1991eu,GlobalWarming:Abbas:1996kk,GlobalWarming:Mack:2007xj,GlobalWarming:Hooper:2011dw,GlobalWarming:Garani:2019rcb,GlobalWarming:Bramante:2019fhi,GlobalWarming:Acevedo:2020gro,GlobalWarming:Leane:2020wob,GlobalWarming:Bramante:2022pmn}). 
On that note, they could also heat NSs colder than PSR J2144$-$3933, which may be observable in current and forthcoming telescopes in the ultraviolet, optical, and infrared~\cite{NSvIR:Baryakhtar:DKHNS,NSvIR:clumps2021,McKeen:2020oyr,McKeen:2021jbh,theluvoirteam2019luvoir,DES:2019rtl,Rubin1,Rubin2,JWST:Gardner:2006ky,ELT:neichel2018overview,TMT:2015pvw,green2012widefield}.
Thanks to their higher energies and fluxes than usually considered, long-range dark nuggets near or beyond the Planck mass ($\simeq$ 10 $\mu$g, contained in Fig.~\ref{fig:alphavM}) could be discovered in terrestrial set-ups, e.g. 
astroparticle detectors~\cite{MACRO:2000cdb,Bramante:2018qbc,Bramante:2018tos,Bramante:2019yss,Cappiello:2020lbk,BlobsGrabowska:2018lnd,Bai:2022nsv,Bhoonah:2020fys,xmaslights:DEAPCollaboration:2021raj,XENON:2023ysy,snowmass:Carney:2022gse},
ancient minerals~\cite{Snowden-Ifft:1995zgn,Baum:2018tfw,Drukier:2018pdy,Acevedo:2021tbl,Baum:2023cct} and other interesting systems~\cite{SinghSidhu:2018oqs,SinghSidhu:2019cpq,SinghSidhu:2019loh,SinghSidhu:2019znk,Dhakal:2022rwn}, by exploiting the large energy deposits and the peaked speed spectrum~\cite{Davoudiasl:2020ypv}.
See also the very recent Ref.~\cite{GWDuLeeWangZurek:2023dhk} for the projected reach of gravitational wave experiments.
Dark clusters and long-range dark nuggets making up a sub-component of DM (perhaps the only component with SM-scattering interactions) would shift my limits suitably while relaxing many others shown in this work.

In this work the class of Type I neutron star burst I have exclusively used (as did Ref.~\cite{MACROSidhu2020}) to set limits on dark  matter is the carbon-burning superburst, but there could potentially be limits from regular hydrogen/helium-burning bursts too. 
However, these limits are expected to be weaker than those from superbursts as they recur more frequently (hours to days, as opposed to months to years) while requiring a large energy deposition~\cite{supburst:underZanding:2017ugu}, both criteria restricting the range of clump masses that may be constrained.
A rarer form of Type I burst dubbed ``hyperburst", with a power about 100 times that of a superburst~\cite{hyperburstPage:2022ikz}, may also potentially help place limits on dark clumps.
However there is only one observation of such an event, and its physical cause, though thought to be unstable burning of neutron-rich isotopes of oxygen or neon, is even more unclear than superbursts~\cite{hyperburstPage:2022ikz}. 
For these reasons, I do not set limits on dark matter using this phenomenon.

I now make some remarks on the thermonuclear aspects of my work.
Recent experimental measurements of carbon fusion cross sections~\cite{FusionRatesTumino2018} using the
Trojan Horse method indicate the presence of resonances potentially enhancing the reaction rate by a factor of 25 compared to classical values~\cite{FusionRatesCaughlanFowler1988} 
at temperatures near 0.5 MeV, which was chosen in this work as the ignition temperature.
The enhancement in fusion would have the effect of reducing the actual ignition temperature, in turn facilitating accretion-induced collapse of white dwarfs into neutron stars during binary mergers and thereby reducing the rate of Type Ia supernovae via the so-called double-degenerate scenario~\cite{FusionRatesWD:Mori2018krw}.
As for my work, my constraints would appear to be conservative as I have chosen an ignition temperature that may be too high.
Similarly, the enhanced carbon fusion could reduce the critical density (or column depth) and recurrence times of superbursts by a factor of a few~\cite{FusionRatesSupburstDohi:2022gmm} (which was predicted in Ref.~\cite{FusionRatesWD:Cooper:2009ps} before the resonances were detected).
However, these conclusions have large theoretical uncertainties associated with them.
For instance, if the fusion rates are determined by a heavy-ion fusion hindrance model~\cite{FusionRatesWDHindrance:2007} the astrophysical $S$-factor can decrease in the energy range of interest; the effect of the hindrance on resonant reactions is unknown, so in principle the resonant enhancement may go away~\cite{FusionRatesWD:Mori2018krw}.
Similarly, a recent full microscopic
nuclear model in Ref.~\cite{FusionRatesFullMicroscopicModel:Taniguchi} (that appeared after the result in Ref.~\cite{FusionRatesTumino2018}), describing low-energy resonances using no adjustable parameter to tune the channel coupling and nuclear rotation, increases the superburst critical density and recurrence time to values close to those predicted by Ref.~\cite{FusionRatesCaughlanFowler1988} (see Table 3 of Ref.~\cite{FusionRatesSupburstDohi:2022gmm}).
It has also been suggested that without resonances in the carbon fusion rate, the NS ocean temperature is too low for ignition at the observationally inferred ignition column depth.
If this were the case, triggering by dark matter clumps may even explain superbursts, since the required higher ignition temperatures can be achieved. 
All said, given the above uncertainties, my choices of ignition temperature and critical density are warranted, but future studies on carbon fusion resonances may give rise to better-informed choices.

Some peculiar transients have been observed in sub-Chandrasekhar WDs in the outskirts of galaxies, with nebular spectra dominated by calcium~\cite{KasliwalCaRichTransients:2011se}.
These ``calcium-rich gap transient" supernovae may be potentially explained by DM clump transits through WDs in dwarf spheroidal galaxies that are expected to be located in the periphery of their parent DM halos, accounting for the distribution of these events~\cite{SmirnovLindenCaRichTransients:2022zip}. 
Another scenario that can prompt calcium-rich gap transients is that of DM in the form of charged massive particles (CHAMPs) that could be deflected by galactic magnetic fields to trigger these supernovae preferentially in galactic outer regions~\cite{FedderkeWDCHAMPs:2019jur}.

The accumulation in certain sub-Chandrasekhar WDs of charged massive particles (CHAMPs) making up DM, which might occur preferentially outside galaxies with magnetic fields that serve to deflect CHAMPs, could be an explanation of the distribution of calcium-rich gap transient WD supernovae [37] that do explode preferentially on the outskirts of galaxies 

Those of us interested in hunting dark matter via thermonuclear explosions have taken advantage of the trigger lengths (``flame widths") numerically estimated in Ref.~\cite{TimmesWoosley1992} by involving a large nuclear reaction network.
However, as mentioned before, these are only available for (i) a narrow range of densities, so that authors have had to rely on extrapolations at lower densities, (b) two different (50-50 C+O and 60-30-10 O+Ne+Mg) compositions, so that I had to simply adopt the C-O results for C-based superbursts. 
Given the significance of these computations for such fundamental questions as dark matter, I request experts in the nuclear astrophysics community to update the 31 year-old results of Ref.~\cite{TimmesWoosley1992}.
I also look forward to expansions in catalogues of white dwarfs and superbursts, the latter thanks to NICER~\cite{supburst:NICERearly} and AstroSat~\cite{supburst:Astrosat}-- or better yet, to the next Galactic supernova.
 
\section*{Acknowledgments}

I thank A.~Gopakumar for triggering a runaway by bringing my attention to low-mass x-ray binaries at the workshop {\em Neutron Stars: The Celestial Clocks That Probe Extreme Physics} at the Institute of Mathematical Sciences, where  I also had profitable discussions with M.~McLaughlin. 
I also thank 
J.~Bramante for pointing out that trigger-happy clumps could be smaller than the trigger volume,
C.~Cappiello for discussion on the unitarity limits of scattering cross sections,
C.~Dessert for mentioning helium fusion in RGB stars,
D.~Galloway for pointing me to Ref.~\cite{supburst:catalog2023},
V.~Lee for clarifying the maximum size of nuggets observable in pulsar timing arrays, and
H.~Verma for discussion on tidal effects.
The anonymous referees offered valuable comments, with Referee B pointing out the discrepancy between my estimate of the ceiling from the white dwarf non-degenerate envelope and that of previous literature. 
C.~Sinensis was of immense help with checking my calculations and typing up the manuscript.


\bibliography{refs}

\end{document}

%% file: universalnewcommands.tex
\newcommand{\gsim}{\gtrsim}
\newcommand{\lsim}{\lesssim}
\newcommand{\ra}{\rightarrow}

\def\Oc{\mathcal{O}}

\renewcommand{\tilde}{\widetilde} 

\newcommand{\beq}{\begin{equation}}
\newcommand{\eeq}{\end{equation}}
\newcommand{\bea}{\begin{eqnarray}}
\newcommand{\eea}{\end{eqnarray}}
\newcommand{\nn}{\nonumber}

\definecolor{rosy}{RGB}{230,235,252}
\definecolor{myframetitle}{RGB}{90,89,170}
\definecolor{myblocktitle}{RGB}{140,185,249}
\definecolor{mytitle}{RGB}{10,80,26}

\definecolor{darkgreen}{RGB}{27,130,45}
\definecolor{darkblue}{rgb}{0,0,0.3}
\definecolor{darkred}{rgb}{0.7,0,0}

\definecolor{light gray}{RGB}{220,220,220}
\definecolor{dark purple}{RGB}{108,0,217}
\definecolor{pink}{RGB}{190,20,100}
\definecolor{orang}{RGB}{193,63,0}
\definecolor{green}{RGB}{11,98,17}
\definecolor{darkpink}{RGB}{153,0,76}
\definecolor{bluegreen}{RGB}{0,102,102}
\definecolor{greenlagan}{RGB}{0,102,0}
\definecolor{redgreen}{RGB}{102,102,0}
\definecolor{Redgreen}{RGB}{153,76,0}
\definecolor{vividviolet}{rgb}{0.62, 0.0, 1.0}
\definecolor{amaranth}{rgb}{0.9, 0.17, 0.31}
\definecolor{palatinateblue}{rgb}{0.15, 0.23, 0.89}
\definecolor{brightpink}{rgb}{1.0, 0.0, 0.5}
\definecolor{cornflowerblue}{rgb}{0.39, 0.58, 0.93}
\definecolor{deepcarminepink}{rgb}{0.94, 0.19, 0.22}
\definecolor{radicalred}{rgb}{1.0, 0.21, 0.37}

%% file: main.bbl
\begin{thebibliography}{168}%
\makeatletter
\providecommand \@ifxundefined [1]{%
 \@ifx{#1\undefined}
}%
\providecommand \@ifnum [1]{%
 \ifnum #1\expandafter \@firstoftwo
 \else \expandafter \@secondoftwo
 \fi
}%
\providecommand \@ifx [1]{%
 \ifx #1\expandafter \@firstoftwo
 \else \expandafter \@secondoftwo
 \fi
}%
\providecommand \natexlab [1]{#1}%
\providecommand \enquote  [1]{``#1''}%
\providecommand \bibnamefont  [1]{#1}%
\providecommand \bibfnamefont [1]{#1}%
\providecommand \citenamefont [1]{#1}%
\providecommand \href@noop [0]{\@secondoftwo}%
\providecommand \href [0]{\begingroup \@sanitize@url \@href}%
\providecommand \@href[1]{\@@startlink{#1}\@@href}%
\providecommand \@@href[1]{\endgroup#1\@@endlink}%
\providecommand \@sanitize@url [0]{\catcode `\\12\catcode `\$12\catcode
  `\&12\catcode `\#12\catcode `\^12\catcode `\_12\catcode `\%12\relax}%
\providecommand \@@startlink[1]{}%
\providecommand \@@endlink[0]{}%
\providecommand \url  [0]{\begingroup\@sanitize@url \@url }%
\providecommand \@url [1]{\endgroup\@href {#1}{\urlprefix }}%
\providecommand \urlprefix  [0]{URL }%
\providecommand \Eprint [0]{\href }%
\providecommand \doibase [0]{http://dx.doi.org/}%
\providecommand \selectlanguage [0]{\@gobble}%
\providecommand \bibinfo  [0]{\@secondoftwo}%
\providecommand \bibfield  [0]{\@secondoftwo}%
\providecommand \translation [1]{[#1]}%
\providecommand \BibitemOpen [0]{}%
\providecommand \bibitemStop [0]{}%
\providecommand \bibitemNoStop [0]{.\EOS\space}%
\providecommand \EOS [0]{\spacefactor3000\relax}%
\providecommand \BibitemShut  [1]{\csname bibitem#1\endcsname}%
\let\auto@bib@innerbib\@empty
\bibitem [{\citenamefont {Bramante}\ \emph
  {et~al.}(2022{\natexlab{a}})\citenamefont {Bramante}, \citenamefont
  {Kavanagh},\ and\ \citenamefont {Raj}}]{NSvIR:clumps2021}%
  \BibitemOpen
  \bibfield  {author} {\bibinfo {author} {\bibfnamefont {J.}~\bibnamefont
  {Bramante}}, \bibinfo {author} {\bibfnamefont {B.~J.}\ \bibnamefont
  {Kavanagh}}, \ and\ \bibinfo {author} {\bibfnamefont {N.}~\bibnamefont
  {Raj}},\ }\href {\doibase 10.1103/PhysRevLett.128.231801} {\bibfield
  {journal} {\bibinfo  {journal} {Phys. Rev. Lett.}\ }\textbf {\bibinfo
  {volume} {128}},\ \bibinfo {pages} {231801} (\bibinfo {year}
  {2022}{\natexlab{a}})},\ \Eprint {http://arxiv.org/abs/2109.04582}
  {arXiv:2109.04582 [hep-ph]} \BibitemShut {NoStop}%
\bibitem [{\citenamefont {Zybin}\ \emph {et~al.}(1999)\citenamefont {Zybin},
  \citenamefont {Vysotsky},\ and\ \citenamefont {Gurevich}}]{Zybin:1999ic}%
  \BibitemOpen
  \bibfield  {author} {\bibinfo {author} {\bibfnamefont {K.~P.}\ \bibnamefont
  {Zybin}}, \bibinfo {author} {\bibfnamefont {M.~I.}\ \bibnamefont {Vysotsky}},
  \ and\ \bibinfo {author} {\bibfnamefont {A.~V.}\ \bibnamefont {Gurevich}},\
  }\href {\doibase 10.1016/S0375-9601(99)00434-X} {\bibfield  {journal}
  {\bibinfo  {journal} {Phys. Lett. A}\ }\textbf {\bibinfo {volume} {260}},\
  \bibinfo {pages} {262} (\bibinfo {year} {1999})}\BibitemShut {NoStop}%
\bibitem [{\citenamefont {Hofmann}\ \emph {et~al.}(2001)\citenamefont
  {Hofmann}, \citenamefont {Schwarz},\ and\ \citenamefont
  {Stoecker}}]{Hofmann:2001bi}%
  \BibitemOpen
  \bibfield  {author} {\bibinfo {author} {\bibfnamefont {S.}~\bibnamefont
  {Hofmann}}, \bibinfo {author} {\bibfnamefont {D.~J.}\ \bibnamefont
  {Schwarz}}, \ and\ \bibinfo {author} {\bibfnamefont {H.}~\bibnamefont
  {Stoecker}},\ }\href {\doibase 10.1103/PhysRevD.64.083507} {\bibfield
  {journal} {\bibinfo  {journal} {Phys. Rev. D}\ }\textbf {\bibinfo {volume}
  {64}},\ \bibinfo {pages} {083507} (\bibinfo {year} {2001})},\ \Eprint
  {http://arxiv.org/abs/astro-ph/0104173} {arXiv:astro-ph/0104173} \BibitemShut
  {NoStop}%
\bibitem [{\citenamefont {Berezinsky}\ \emph {et~al.}(2008)\citenamefont
  {Berezinsky}, \citenamefont {Dokuchaev},\ and\ \citenamefont
  {Eroshenko}}]{Berezinsky:2007qu}%
  \BibitemOpen
  \bibfield  {author} {\bibinfo {author} {\bibfnamefont {V.}~\bibnamefont
  {Berezinsky}}, \bibinfo {author} {\bibfnamefont {V.}~\bibnamefont
  {Dokuchaev}}, \ and\ \bibinfo {author} {\bibfnamefont {Y.}~\bibnamefont
  {Eroshenko}},\ }\href {\doibase 10.1103/PhysRevD.77.083519} {\bibfield
  {journal} {\bibinfo  {journal} {Phys. Rev. D}\ }\textbf {\bibinfo {volume}
  {77}},\ \bibinfo {pages} {083519} (\bibinfo {year} {2008})},\ \Eprint
  {http://arxiv.org/abs/0712.3499} {arXiv:0712.3499 [astro-ph]} \BibitemShut
  {NoStop}%
\bibitem [{\citenamefont {Bergstrom}\ \emph {et~al.}(1999)\citenamefont
  {Bergstrom}, \citenamefont {Edsjo}, \citenamefont {Gondolo},\ and\
  \citenamefont {Ullio}}]{BergstromGondolo:1998jj}%
  \BibitemOpen
  \bibfield  {author} {\bibinfo {author} {\bibfnamefont {L.}~\bibnamefont
  {Bergstrom}}, \bibinfo {author} {\bibfnamefont {J.}~\bibnamefont {Edsjo}},
  \bibinfo {author} {\bibfnamefont {P.}~\bibnamefont {Gondolo}}, \ and\
  \bibinfo {author} {\bibfnamefont {P.}~\bibnamefont {Ullio}},\ }\href
  {\doibase 10.1103/PhysRevD.59.043506} {\bibfield  {journal} {\bibinfo
  {journal} {Phys. Rev. D}\ }\textbf {\bibinfo {volume} {59}},\ \bibinfo
  {pages} {043506} (\bibinfo {year} {1999})},\ \Eprint
  {http://arxiv.org/abs/astro-ph/9806072} {arXiv:astro-ph/9806072} \BibitemShut
  {NoStop}%
\bibitem [{\citenamefont {van~den Bosch}\ \emph {et~al.}(2018)\citenamefont
  {van~den Bosch}, \citenamefont {Ogiya}, \citenamefont {Hahn},\ and\
  \citenamefont {Burkert}}]{vandenBosch:2017ynq}%
  \BibitemOpen
  \bibfield  {author} {\bibinfo {author} {\bibfnamefont {F.~C.}\ \bibnamefont
  {van~den Bosch}}, \bibinfo {author} {\bibfnamefont {G.}~\bibnamefont
  {Ogiya}}, \bibinfo {author} {\bibfnamefont {O.}~\bibnamefont {Hahn}}, \ and\
  \bibinfo {author} {\bibfnamefont {A.}~\bibnamefont {Burkert}},\ }\href
  {\doibase 10.1093/mnras/stx2956} {\bibfield  {journal} {\bibinfo  {journal}
  {Mon. Not. Roy. Astron. Soc.}\ }\textbf {\bibinfo {volume} {474}},\ \bibinfo
  {pages} {3043} (\bibinfo {year} {2018})},\ \Eprint
  {http://arxiv.org/abs/1711.05276} {arXiv:1711.05276 [astro-ph.GA]}
  \BibitemShut {NoStop}%
\bibitem [{\citenamefont {van~den Bosch}\ and\ \citenamefont
  {Ogiya}(2018)}]{vandenBosch:2018tyt}%
  \BibitemOpen
  \bibfield  {author} {\bibinfo {author} {\bibfnamefont {F.~C.}\ \bibnamefont
  {van~den Bosch}}\ and\ \bibinfo {author} {\bibfnamefont {G.}~\bibnamefont
  {Ogiya}},\ }\href {\doibase 10.1093/mnras/sty084} {\bibfield  {journal}
  {\bibinfo  {journal} {Mon. Not. Roy. Astron. Soc.}\ }\textbf {\bibinfo
  {volume} {475}},\ \bibinfo {pages} {4066} (\bibinfo {year} {2018})},\ \Eprint
  {http://arxiv.org/abs/1801.05427} {arXiv:1801.05427 [astro-ph.GA]}
  \BibitemShut {NoStop}%
\bibitem [{\citenamefont {Erickcek}\ and\ \citenamefont
  {Sigurdson}(2011)}]{ErickcekSigurdson}%
  \BibitemOpen
  \bibfield  {author} {\bibinfo {author} {\bibfnamefont {A.~L.}\ \bibnamefont
  {Erickcek}}\ and\ \bibinfo {author} {\bibfnamefont {K.}~\bibnamefont
  {Sigurdson}},\ }\href {\doibase 10.1103/PhysRevD.84.083503} {\bibfield
  {journal} {\bibinfo  {journal} {Phys. Rev. D}\ }\textbf {\bibinfo {volume}
  {84}},\ \bibinfo {pages} {083503} (\bibinfo {year} {2011})}\BibitemShut
  {NoStop}%
\bibitem [{\citenamefont {Barenboim}\ and\ \citenamefont
  {Rasero}(2014)}]{Barenboim:2013gya}%
  \BibitemOpen
  \bibfield  {author} {\bibinfo {author} {\bibfnamefont {G.}~\bibnamefont
  {Barenboim}}\ and\ \bibinfo {author} {\bibfnamefont {J.}~\bibnamefont
  {Rasero}},\ }\href {\doibase 10.1007/JHEP04(2014)138} {\bibfield  {journal}
  {\bibinfo  {journal} {JHEP}\ }\textbf {\bibinfo {volume} {04}},\ \bibinfo
  {pages} {138} (\bibinfo {year} {2014})},\ \Eprint
  {http://arxiv.org/abs/1311.4034} {arXiv:1311.4034 [hep-ph]} \BibitemShut
  {NoStop}%
\bibitem [{\citenamefont {Fan}\ \emph {et~al.}(2014)\citenamefont {Fan},
  \citenamefont {\"Ozsoy},\ and\ \citenamefont {Watson}}]{FanWatson}%
  \BibitemOpen
  \bibfield  {author} {\bibinfo {author} {\bibfnamefont {J.}~\bibnamefont
  {Fan}}, \bibinfo {author} {\bibfnamefont {O.}~\bibnamefont {\"Ozsoy}}, \ and\
  \bibinfo {author} {\bibfnamefont {S.}~\bibnamefont {Watson}},\ }\href
  {\doibase 10.1103/PhysRevD.90.043536} {\bibfield  {journal} {\bibinfo
  {journal} {Phys. Rev. D}\ }\textbf {\bibinfo {volume} {90}},\ \bibinfo
  {pages} {043536} (\bibinfo {year} {2014})}\BibitemShut {NoStop}%
\bibitem [{\citenamefont {Dror}\ \emph {et~al.}(2018)\citenamefont {Dror},
  \citenamefont {Kuflik}, \citenamefont {Melcher},\ and\ \citenamefont
  {Watson}}]{drorcodecay}%
  \BibitemOpen
  \bibfield  {author} {\bibinfo {author} {\bibfnamefont {J.~A.}\ \bibnamefont
  {Dror}}, \bibinfo {author} {\bibfnamefont {E.}~\bibnamefont {Kuflik}},
  \bibinfo {author} {\bibfnamefont {B.}~\bibnamefont {Melcher}}, \ and\
  \bibinfo {author} {\bibfnamefont {S.}~\bibnamefont {Watson}},\ }\href
  {\doibase 10.1103/PhysRevD.97.063524} {\bibfield  {journal} {\bibinfo
  {journal} {Phys. Rev. D}\ }\textbf {\bibinfo {volume} {97}},\ \bibinfo
  {pages} {063524} (\bibinfo {year} {2018})}\BibitemShut {NoStop}%
\bibitem [{\citenamefont {Graham}\ \emph {et~al.}(2016)\citenamefont {Graham},
  \citenamefont {Mardon},\ and\ \citenamefont {Rajendran}}]{inflatflucs}%
  \BibitemOpen
  \bibfield  {author} {\bibinfo {author} {\bibfnamefont {P.~W.}\ \bibnamefont
  {Graham}}, \bibinfo {author} {\bibfnamefont {J.}~\bibnamefont {Mardon}}, \
  and\ \bibinfo {author} {\bibfnamefont {S.}~\bibnamefont {Rajendran}},\ }\href
  {\doibase 10.1103/PhysRevD.93.103520} {\bibfield  {journal} {\bibinfo
  {journal} {Phys. Rev. D}\ }\textbf {\bibinfo {volume} {93}},\ \bibinfo
  {pages} {103520} (\bibinfo {year} {2016})}\BibitemShut {NoStop}%
\bibitem [{\citenamefont {Buckley}\ and\ \citenamefont
  {DiFranzo}(2018)}]{Buckley:2017ttd}%
  \BibitemOpen
  \bibfield  {author} {\bibinfo {author} {\bibfnamefont {M.~R.}\ \bibnamefont
  {Buckley}}\ and\ \bibinfo {author} {\bibfnamefont {A.}~\bibnamefont
  {DiFranzo}},\ }\href {\doibase 10.1103/PhysRevLett.120.051102} {\bibfield
  {journal} {\bibinfo  {journal} {Phys. Rev. Lett.}\ }\textbf {\bibinfo
  {volume} {120}},\ \bibinfo {pages} {051102} (\bibinfo {year} {2018})},\
  \Eprint {http://arxiv.org/abs/1707.03829} {arXiv:1707.03829 [hep-ph]}
  \BibitemShut {NoStop}%
\bibitem [{\citenamefont {Nussinov}\ and\ \citenamefont
  {Zhang}(2020)}]{nussinovcluster}%
  \BibitemOpen
  \bibfield  {author} {\bibinfo {author} {\bibfnamefont {S.}~\bibnamefont
  {Nussinov}}\ and\ \bibinfo {author} {\bibfnamefont {Y.}~\bibnamefont
  {Zhang}},\ }\href {\doibase 10.1007/JHEP03(2020)133} {\bibfield  {journal}
  {\bibinfo  {journal} {JHEP}\ }\textbf {\bibinfo {volume} {03}},\ \bibinfo
  {pages} {133} (\bibinfo {year} {2020})},\ \Eprint
  {http://arxiv.org/abs/1807.00846} {arXiv:1807.00846 [hep-ph]} \BibitemShut
  {NoStop}%
\bibitem [{\citenamefont {Barenboim}\ \emph {et~al.}(2021)\citenamefont
  {Barenboim}, \citenamefont {Blinov},\ and\ \citenamefont
  {Stebbins}}]{Barenboim:2021swl}%
  \BibitemOpen
  \bibfield  {author} {\bibinfo {author} {\bibfnamefont {G.}~\bibnamefont
  {Barenboim}}, \bibinfo {author} {\bibfnamefont {N.}~\bibnamefont {Blinov}}, \
  and\ \bibinfo {author} {\bibfnamefont {A.}~\bibnamefont {Stebbins}},\
  }\href@noop {} {\  (\bibinfo {year} {2021})},\ \Eprint
  {http://arxiv.org/abs/2107.10293} {arXiv:2107.10293 [astro-ph.CO]}
  \BibitemShut {NoStop}%
\bibitem [{\citenamefont {Dom\`enech}\ \emph {et~al.}(2023)\citenamefont
  {Dom\`enech}, \citenamefont {Inman}, \citenamefont {Kusenko},\ and\
  \citenamefont {Sasaki}}]{Domenech:2023afs}%
  \BibitemOpen
  \bibfield  {author} {\bibinfo {author} {\bibfnamefont {G.}~\bibnamefont
  {Dom\`enech}}, \bibinfo {author} {\bibfnamefont {D.}~\bibnamefont {Inman}},
  \bibinfo {author} {\bibfnamefont {A.}~\bibnamefont {Kusenko}}, \ and\
  \bibinfo {author} {\bibfnamefont {M.}~\bibnamefont {Sasaki}},\ }\href@noop {}
  {\  (\bibinfo {year} {2023})},\ \Eprint {http://arxiv.org/abs/2304.13053}
  {arXiv:2304.13053 [astro-ph.CO]} \BibitemShut {NoStop}%
\bibitem [{\citenamefont {Graham}\ \emph {et~al.}(2015)\citenamefont {Graham},
  \citenamefont {Rajendran},\ and\ \citenamefont
  {Varela}}]{GrahamRajendVarelaPBHWD}%
  \BibitemOpen
  \bibfield  {author} {\bibinfo {author} {\bibfnamefont {P.~W.}\ \bibnamefont
  {Graham}}, \bibinfo {author} {\bibfnamefont {S.}~\bibnamefont {Rajendran}}, \
  and\ \bibinfo {author} {\bibfnamefont {J.}~\bibnamefont {Varela}},\ }\href
  {\doibase 10.1103/PhysRevD.92.063007} {\bibfield  {journal} {\bibinfo
  {journal} {Phys. Rev. D}\ }\textbf {\bibinfo {volume} {92}},\ \bibinfo
  {pages} {063007} (\bibinfo {year} {2015})}\BibitemShut {NoStop}%
\bibitem [{\citenamefont {Montero-Camacho}\ \emph {et~al.}(2019)\citenamefont
  {Montero-Camacho}, \citenamefont {Fang}, \citenamefont {Vasquez},
  \citenamefont {Silva},\ and\ \citenamefont
  {Hirata}}]{MonteroCamachoHirata2019}%
  \BibitemOpen
  \bibfield  {author} {\bibinfo {author} {\bibfnamefont {P.}~\bibnamefont
  {Montero-Camacho}}, \bibinfo {author} {\bibfnamefont {X.}~\bibnamefont
  {Fang}}, \bibinfo {author} {\bibfnamefont {G.}~\bibnamefont {Vasquez}},
  \bibinfo {author} {\bibfnamefont {M.}~\bibnamefont {Silva}}, \ and\ \bibinfo
  {author} {\bibfnamefont {C.~M.}\ \bibnamefont {Hirata}},\ }\href {\doibase
  10.1088/1475-7516/2019/08/031} {\bibfield  {journal} {\bibinfo  {journal}
  {Journal of Cosmology and Astroparticle Physics}\ }\textbf {\bibinfo {volume}
  {2019}},\ \bibinfo {pages} {031} (\bibinfo {year} {2019})}\BibitemShut
  {NoStop}%
\bibitem [{\citenamefont {Steigerwald}\ and\ \citenamefont
  {Tejeda}(2021)}]{SteigerwaldDetonationPBH:2021vgi}%
  \BibitemOpen
  \bibfield  {author} {\bibinfo {author} {\bibfnamefont {H.}~\bibnamefont
  {Steigerwald}}\ and\ \bibinfo {author} {\bibfnamefont {E.}~\bibnamefont
  {Tejeda}},\ }\href {\doibase 10.1103/PhysRevLett.127.011101} {\bibfield
  {journal} {\bibinfo  {journal} {Phys. Rev. Lett.}\ }\textbf {\bibinfo
  {volume} {127}},\ \bibinfo {pages} {011101} (\bibinfo {year} {2021})},\
  \Eprint {http://arxiv.org/abs/2104.07066} {arXiv:2104.07066 [astro-ph.HE]}
  \BibitemShut {NoStop}%
\bibitem [{\citenamefont {Leung}\ \emph {et~al.}(2013)\citenamefont {Leung},
  \citenamefont {Chu}, \citenamefont {Lin},\ and\ \citenamefont
  {Wong}}]{Leung:2013pra}%
  \BibitemOpen
  \bibfield  {author} {\bibinfo {author} {\bibfnamefont {S.~C.}\ \bibnamefont
  {Leung}}, \bibinfo {author} {\bibfnamefont {M.~C.}\ \bibnamefont {Chu}},
  \bibinfo {author} {\bibfnamefont {L.~M.}\ \bibnamefont {Lin}}, \ and\
  \bibinfo {author} {\bibfnamefont {K.~W.}\ \bibnamefont {Wong}},\ }\href
  {\doibase 10.1103/PhysRevD.87.123506} {\bibfield  {journal} {\bibinfo
  {journal} {Phys. Rev. D}\ }\textbf {\bibinfo {volume} {87}},\ \bibinfo
  {pages} {123506} (\bibinfo {year} {2013})},\ \Eprint
  {http://arxiv.org/abs/1305.6142} {arXiv:1305.6142 [astro-ph.CO]} \BibitemShut
  {NoStop}%
\bibitem [{\citenamefont {Bramante}(2015)}]{Bramante:2015cua}%
  \BibitemOpen
  \bibfield  {author} {\bibinfo {author} {\bibfnamefont {J.}~\bibnamefont
  {Bramante}},\ }\href {\doibase 10.1103/PhysRevLett.115.141301} {\bibfield
  {journal} {\bibinfo  {journal} {Phys. Rev. Lett.}\ }\textbf {\bibinfo
  {volume} {115}},\ \bibinfo {pages} {141301} (\bibinfo {year} {2015})},\
  \Eprint {http://arxiv.org/abs/1505.07464} {arXiv:1505.07464 [hep-ph]}
  \BibitemShut {NoStop}%
\bibitem [{\citenamefont {Fedderke}\ \emph {et~al.}(2020)\citenamefont
  {Fedderke}, \citenamefont {Graham},\ and\ \citenamefont
  {Rajendran}}]{FedderkeWDCHAMPs:2019jur}%
  \BibitemOpen
  \bibfield  {author} {\bibinfo {author} {\bibfnamefont {M.~A.}\ \bibnamefont
  {Fedderke}}, \bibinfo {author} {\bibfnamefont {P.~W.}\ \bibnamefont
  {Graham}}, \ and\ \bibinfo {author} {\bibfnamefont {S.}~\bibnamefont
  {Rajendran}},\ }\href {\doibase 10.1103/PhysRevD.101.115021} {\bibfield
  {journal} {\bibinfo  {journal} {Phys. Rev. D}\ }\textbf {\bibinfo {volume}
  {101}},\ \bibinfo {pages} {115021} (\bibinfo {year} {2020})},\ \Eprint
  {http://arxiv.org/abs/1911.08883} {arXiv:1911.08883 [hep-ph]} \BibitemShut
  {NoStop}%
\bibitem [{\citenamefont {Steigerwald}\ \emph {et~al.}(2022)\citenamefont
  {Steigerwald}, \citenamefont {Marra},\ and\ \citenamefont
  {Profumo}}]{SteigerwaldProfumo:2022pjo}%
  \BibitemOpen
  \bibfield  {author} {\bibinfo {author} {\bibfnamefont {H.}~\bibnamefont
  {Steigerwald}}, \bibinfo {author} {\bibfnamefont {V.}~\bibnamefont {Marra}},
  \ and\ \bibinfo {author} {\bibfnamefont {S.}~\bibnamefont {Profumo}},\ }\href
  {\doibase 10.1103/PhysRevD.105.083507} {\bibfield  {journal} {\bibinfo
  {journal} {Phys. Rev. D}\ }\textbf {\bibinfo {volume} {105}},\ \bibinfo
  {pages} {083507} (\bibinfo {year} {2022})},\ \Eprint
  {http://arxiv.org/abs/2203.09054} {arXiv:2203.09054 [astro-ph.CO]}
  \BibitemShut {NoStop}%
\bibitem [{\citenamefont {Graham}\ \emph {et~al.}(2018)\citenamefont {Graham},
  \citenamefont {Janish}, \citenamefont {Narayan}, \citenamefont {Rajendran},\
  and\ \citenamefont {Riggins}}]{Graham:2018efk}%
  \BibitemOpen
  \bibfield  {author} {\bibinfo {author} {\bibfnamefont {P.~W.}\ \bibnamefont
  {Graham}}, \bibinfo {author} {\bibfnamefont {R.}~\bibnamefont {Janish}},
  \bibinfo {author} {\bibfnamefont {V.}~\bibnamefont {Narayan}}, \bibinfo
  {author} {\bibfnamefont {S.}~\bibnamefont {Rajendran}}, \ and\ \bibinfo
  {author} {\bibfnamefont {P.}~\bibnamefont {Riggins}},\ }\href {\doibase
  10.1103/PhysRevD.98.115027} {\bibfield  {journal} {\bibinfo  {journal} {Phys.
  Rev. D}\ }\textbf {\bibinfo {volume} {98}},\ \bibinfo {pages} {115027}
  (\bibinfo {year} {2018})},\ \Eprint {http://arxiv.org/abs/1805.07381}
  {arXiv:1805.07381 [hep-ph]} \BibitemShut {NoStop}%
\bibitem [{\citenamefont {Sidhu}\ and\ \citenamefont
  {Starkman}(2020)}]{MACROSidhu2020}%
  \BibitemOpen
  \bibfield  {author} {\bibinfo {author} {\bibfnamefont {J.~S.}\ \bibnamefont
  {Sidhu}}\ and\ \bibinfo {author} {\bibfnamefont {G.~D.}\ \bibnamefont
  {Starkman}},\ }\href {\doibase 10.1103/physrevd.101.083503} {\bibfield
  {journal} {\bibinfo  {journal} {Physical Review D}\ }\textbf {\bibinfo
  {volume} {101}} (\bibinfo {year} {2020}),\
  10.1103/physrevd.101.083503}\BibitemShut {NoStop}%
\bibitem [{\citenamefont {Acevedo}\ \emph
  {et~al.}(2021{\natexlab{a}})\citenamefont {Acevedo}, \citenamefont
  {Bramante},\ and\ \citenamefont {Goodman}}]{Acevedo:2020avd}%
  \BibitemOpen
  \bibfield  {author} {\bibinfo {author} {\bibfnamefont {J.~F.}\ \bibnamefont
  {Acevedo}}, \bibinfo {author} {\bibfnamefont {J.}~\bibnamefont {Bramante}}, \
  and\ \bibinfo {author} {\bibfnamefont {A.}~\bibnamefont {Goodman}},\ }\href
  {\doibase 10.1103/PhysRevD.103.123022} {\bibfield  {journal} {\bibinfo
  {journal} {Phys. Rev. D}\ }\textbf {\bibinfo {volume} {103}},\ \bibinfo
  {pages} {123022} (\bibinfo {year} {2021}{\natexlab{a}})},\ \Eprint
  {http://arxiv.org/abs/2012.10998} {arXiv:2012.10998 [hep-ph]} \BibitemShut
  {NoStop}%
\bibitem [{\citenamefont {Acevedo}\ and\ \citenamefont
  {Bramante}(2019)}]{Acevedo:2019gre}%
  \BibitemOpen
  \bibfield  {author} {\bibinfo {author} {\bibfnamefont {J.~F.}\ \bibnamefont
  {Acevedo}}\ and\ \bibinfo {author} {\bibfnamefont {J.}~\bibnamefont
  {Bramante}},\ }\href {\doibase 10.1103/PhysRevD.100.043020} {\bibfield
  {journal} {\bibinfo  {journal} {Phys. Rev. D}\ }\textbf {\bibinfo {volume}
  {100}},\ \bibinfo {pages} {043020} (\bibinfo {year} {2019})},\ \Eprint
  {http://arxiv.org/abs/1904.11993} {arXiv:1904.11993 [hep-ph]} \BibitemShut
  {NoStop}%
\bibitem [{\citenamefont {Janish}\ \emph {et~al.}(2019)\citenamefont {Janish},
  \citenamefont {Narayan},\ and\ \citenamefont {Riggins}}]{Janish:2019nkk}%
  \BibitemOpen
  \bibfield  {author} {\bibinfo {author} {\bibfnamefont {R.}~\bibnamefont
  {Janish}}, \bibinfo {author} {\bibfnamefont {V.}~\bibnamefont {Narayan}}, \
  and\ \bibinfo {author} {\bibfnamefont {P.}~\bibnamefont {Riggins}},\ }\href
  {\doibase 10.1103/PhysRevD.100.035008} {\bibfield  {journal} {\bibinfo
  {journal} {Phys. Rev. D}\ }\textbf {\bibinfo {volume} {100}},\ \bibinfo
  {pages} {035008} (\bibinfo {year} {2019})},\ \Eprint
  {http://arxiv.org/abs/1905.00395} {arXiv:1905.00395 [hep-ph]} \BibitemShut
  {NoStop}%
\bibitem [{\citenamefont {Das}\ \emph {et~al.}(2022)\citenamefont {Das},
  \citenamefont {Ellis}, \citenamefont {Schuster},\ and\ \citenamefont
  {Zhou}}]{StellarShocksDas2022}%
  \BibitemOpen
  \bibfield  {author} {\bibinfo {author} {\bibfnamefont {A.}~\bibnamefont
  {Das}}, \bibinfo {author} {\bibfnamefont {S.~A.}\ \bibnamefont {Ellis}},
  \bibinfo {author} {\bibfnamefont {P.~C.}\ \bibnamefont {Schuster}}, \ and\
  \bibinfo {author} {\bibfnamefont {K.}~\bibnamefont {Zhou}},\ }\href {\doibase
  10.1103/physrevlett.128.021101} {\bibfield  {journal} {\bibinfo  {journal}
  {Physical Review Letters}\ }\textbf {\bibinfo {volume} {128}} (\bibinfo
  {year} {2022}),\ 10.1103/physrevlett.128.021101}\BibitemShut {NoStop}%
\bibitem [{\citenamefont {Galloway}\ \emph {et~al.}(2020)\citenamefont
  {Galloway}, \citenamefont {in~'t Zand}, \citenamefont {Chenevez},
  \citenamefont {W\"orpel}, \citenamefont {Keek}, \citenamefont {Ootes},
  \citenamefont {Watts}, \citenamefont {Gisler}, \citenamefont
  {Sanchez-Fernandez},\ and\ \citenamefont
  {Kuulkers}}]{supburst:MINBARcatalog:2020}%
  \BibitemOpen
  \bibfield  {author} {\bibinfo {author} {\bibfnamefont {D.~K.}\ \bibnamefont
  {Galloway}}, \bibinfo {author} {\bibfnamefont {J.}~\bibnamefont {in~'t
  Zand}}, \bibinfo {author} {\bibfnamefont {J.}~\bibnamefont {Chenevez}},
  \bibinfo {author} {\bibfnamefont {H.}~\bibnamefont {W\"orpel}}, \bibinfo
  {author} {\bibfnamefont {L.}~\bibnamefont {Keek}}, \bibinfo {author}
  {\bibfnamefont {L.}~\bibnamefont {Ootes}}, \bibinfo {author} {\bibfnamefont
  {A.~L.}\ \bibnamefont {Watts}}, \bibinfo {author} {\bibfnamefont
  {L.}~\bibnamefont {Gisler}}, \bibinfo {author} {\bibfnamefont
  {C.}~\bibnamefont {Sanchez-Fernandez}}, \ and\ \bibinfo {author}
  {\bibfnamefont {E.}~\bibnamefont {Kuulkers}},\ }\href {\doibase
  10.3847/1538-4365/ab9f2e} {\bibfield  {journal} {\bibinfo  {journal}
  {Astrophys. J. Suppl.}\ }\textbf {\bibinfo {volume} {249}},\ \bibinfo {pages}
  {32} (\bibinfo {year} {2020})},\ \Eprint {http://arxiv.org/abs/2003.00685}
  {arXiv:2003.00685 [astro-ph.HE]} \BibitemShut {NoStop}%
\bibitem [{\citenamefont {Alizai}(2023)}]{supburst:catalog2023}%
  \BibitemOpen
  \bibfield  {author} {\bibinfo {author} {\bibfnamefont {e.~a.}\ \bibnamefont
  {Alizai}, \bibfnamefont {K}},\ }\href {\doibase 10.1093/mnras/stad374}
  {\bibfield  {journal} {\bibinfo  {journal} {Monthly Notices of the Royal
  Astronomical Society}\ }\textbf {\bibinfo {volume} {521}},\ \bibinfo {pages}
  {3608} (\bibinfo {year} {2023})},\ \Eprint
  {http://arxiv.org/abs/https://academic.oup.com/mnras/article-pdf/521/3/3608/49630407/stad374.pdf}
  {https://academic.oup.com/mnras/article-pdf/521/3/3608/49630407/stad374.pdf}
  \BibitemShut {NoStop}%
\bibitem [{\citenamefont {Cumming}\ and\ \citenamefont
  {Bildsten}(2001)}]{supburst:Cflashes:Cumming:2001wg}%
  \BibitemOpen
  \bibfield  {author} {\bibinfo {author} {\bibfnamefont {A.}~\bibnamefont
  {Cumming}}\ and\ \bibinfo {author} {\bibfnamefont {L.}~\bibnamefont
  {Bildsten}},\ }\href {\doibase 10.1086/323937} {\bibfield  {journal}
  {\bibinfo  {journal} {Astrophys. J. Lett.}\ }\textbf {\bibinfo {volume}
  {559}},\ \bibinfo {pages} {L127} (\bibinfo {year} {2001})},\ \Eprint
  {http://arxiv.org/abs/astro-ph/0107213} {arXiv:astro-ph/0107213} \BibitemShut
  {NoStop}%
\bibitem [{\citenamefont {in~'t Zand}(2017)}]{supburst:underZanding:2017ugu}%
  \BibitemOpen
  \bibfield  {author} {\bibinfo {author} {\bibfnamefont {J.}~\bibnamefont
  {in~'t Zand}}\ }(\bibinfo {year} {2017})\ \Eprint
  {http://arxiv.org/abs/1702.04899} {arXiv:1702.04899 [astro-ph.HE]}
  \BibitemShut {NoStop}%
\bibitem [{\citenamefont {Gresham}\ \emph {et~al.}(2023)\citenamefont
  {Gresham}, \citenamefont {Lee},\ and\ \citenamefont
  {Zurek}}]{GreshamLeeZurek:2022biw}%
  \BibitemOpen
  \bibfield  {author} {\bibinfo {author} {\bibfnamefont {M.~I.}\ \bibnamefont
  {Gresham}}, \bibinfo {author} {\bibfnamefont {V.~S.~H.}\ \bibnamefont {Lee}},
  \ and\ \bibinfo {author} {\bibfnamefont {K.~M.}\ \bibnamefont {Zurek}},\
  }\href {\doibase 10.1088/1475-7516/2023/02/048} {\bibfield  {journal}
  {\bibinfo  {journal} {JCAP}\ }\textbf {\bibinfo {volume} {02}},\ \bibinfo
  {pages} {048} (\bibinfo {year} {2023})},\ \Eprint
  {http://arxiv.org/abs/2209.03963} {arXiv:2209.03963 [astro-ph.HE]}
  \BibitemShut {NoStop}%
\bibitem [{\citenamefont {Witten}(1984)}]{nuggetsWitten:1984rs}%
  \BibitemOpen
  \bibfield  {author} {\bibinfo {author} {\bibfnamefont {E.}~\bibnamefont
  {Witten}},\ }\href {\doibase 10.1103/PhysRevD.30.272} {\bibfield  {journal}
  {\bibinfo  {journal} {Phys. Rev. D}\ }\textbf {\bibinfo {volume} {30}},\
  \bibinfo {pages} {272} (\bibinfo {year} {1984})}\BibitemShut {NoStop}%
\bibitem [{\citenamefont {Krnjaic}\ and\ \citenamefont
  {Sigurdson}(2015)}]{nucleiKrnjaic:2014xza}%
  \BibitemOpen
  \bibfield  {author} {\bibinfo {author} {\bibfnamefont {G.}~\bibnamefont
  {Krnjaic}}\ and\ \bibinfo {author} {\bibfnamefont {K.}~\bibnamefont
  {Sigurdson}},\ }\href {\doibase 10.1016/j.physletb.2015.11.001} {\bibfield
  {journal} {\bibinfo  {journal} {Phys. Lett. B}\ }\textbf {\bibinfo {volume}
  {751}},\ \bibinfo {pages} {464} (\bibinfo {year} {2015})},\ \Eprint
  {http://arxiv.org/abs/1406.1171} {arXiv:1406.1171 [hep-ph]} \BibitemShut
  {NoStop}%
\bibitem [{\citenamefont {Detmold}\ \emph {et~al.}(2014)\citenamefont
  {Detmold}, \citenamefont {McCullough},\ and\ \citenamefont
  {Pochinsky}}]{nucleiDetmoldMcCullough:2014qqa}%
  \BibitemOpen
  \bibfield  {author} {\bibinfo {author} {\bibfnamefont {W.}~\bibnamefont
  {Detmold}}, \bibinfo {author} {\bibfnamefont {M.}~\bibnamefont {McCullough}},
  \ and\ \bibinfo {author} {\bibfnamefont {A.}~\bibnamefont {Pochinsky}},\
  }\href {\doibase 10.1103/PhysRevD.90.115013} {\bibfield  {journal} {\bibinfo
  {journal} {Phys. Rev. D}\ }\textbf {\bibinfo {volume} {90}},\ \bibinfo
  {pages} {115013} (\bibinfo {year} {2014})},\ \Eprint
  {http://arxiv.org/abs/1406.2276} {arXiv:1406.2276 [hep-ph]} \BibitemShut
  {NoStop}%
\bibitem [{\citenamefont {Hardy}\ \emph
  {et~al.}(2015{\natexlab{a}})\citenamefont {Hardy}, \citenamefont {Lasenby},
  \citenamefont {March-Russell},\ and\ \citenamefont
  {West}}]{nucleiHardy:2014mqa}%
  \BibitemOpen
  \bibfield  {author} {\bibinfo {author} {\bibfnamefont {E.}~\bibnamefont
  {Hardy}}, \bibinfo {author} {\bibfnamefont {R.}~\bibnamefont {Lasenby}},
  \bibinfo {author} {\bibfnamefont {J.}~\bibnamefont {March-Russell}}, \ and\
  \bibinfo {author} {\bibfnamefont {S.~M.}\ \bibnamefont {West}},\ }\href
  {\doibase 10.1007/JHEP06(2015)011} {\bibfield  {journal} {\bibinfo  {journal}
  {JHEP}\ }\textbf {\bibinfo {volume} {06}},\ \bibinfo {pages} {011} (\bibinfo
  {year} {2015}{\natexlab{a}})},\ \Eprint {http://arxiv.org/abs/1411.3739}
  {arXiv:1411.3739 [hep-ph]} \BibitemShut {NoStop}%
\bibitem [{\citenamefont {Wise}\ and\ \citenamefont
  {Zhang}(2014)}]{boundstateWise:2014jva}%
  \BibitemOpen
  \bibfield  {author} {\bibinfo {author} {\bibfnamefont {M.~B.}\ \bibnamefont
  {Wise}}\ and\ \bibinfo {author} {\bibfnamefont {Y.}~\bibnamefont {Zhang}},\
  }\href {\doibase 10.1103/PhysRevD.90.055030} {\bibfield  {journal} {\bibinfo
  {journal} {Phys. Rev. D}\ }\textbf {\bibinfo {volume} {90}},\ \bibinfo
  {pages} {055030} (\bibinfo {year} {2014})},\ \bibinfo {note} {[Erratum:
  Phys.Rev.D 91, 039907 (2015)]},\ \Eprint {http://arxiv.org/abs/1407.4121}
  {arXiv:1407.4121 [hep-ph]} \BibitemShut {NoStop}%
\bibitem [{\citenamefont {Wise}\ and\ \citenamefont
  {Zhang}(2015)}]{boundstateWise:2014ola}%
  \BibitemOpen
  \bibfield  {author} {\bibinfo {author} {\bibfnamefont {M.~B.}\ \bibnamefont
  {Wise}}\ and\ \bibinfo {author} {\bibfnamefont {Y.}~\bibnamefont {Zhang}},\
  }\href {\doibase 10.1007/JHEP02(2015)023} {\bibfield  {journal} {\bibinfo
  {journal} {JHEP}\ }\textbf {\bibinfo {volume} {02}},\ \bibinfo {pages} {023}
  (\bibinfo {year} {2015})},\ \bibinfo {note} {[Erratum: JHEP 10, 165
  (2015)]},\ \Eprint {http://arxiv.org/abs/1411.1772} {arXiv:1411.1772
  [hep-ph]} \BibitemShut {NoStop}%
\bibitem [{\citenamefont {Hardy}\ \emph
  {et~al.}(2015{\natexlab{b}})\citenamefont {Hardy}, \citenamefont {Lasenby},
  \citenamefont {March-Russell},\ and\ \citenamefont
  {West}}]{nucleiHardy:2015boa}%
  \BibitemOpen
  \bibfield  {author} {\bibinfo {author} {\bibfnamefont {E.}~\bibnamefont
  {Hardy}}, \bibinfo {author} {\bibfnamefont {R.}~\bibnamefont {Lasenby}},
  \bibinfo {author} {\bibfnamefont {J.}~\bibnamefont {March-Russell}}, \ and\
  \bibinfo {author} {\bibfnamefont {S.~M.}\ \bibnamefont {West}},\ }\href
  {\doibase 10.1007/JHEP07(2015)133} {\bibfield  {journal} {\bibinfo  {journal}
  {JHEP}\ }\textbf {\bibinfo {volume} {07}},\ \bibinfo {pages} {133} (\bibinfo
  {year} {2015}{\natexlab{b}})},\ \Eprint {http://arxiv.org/abs/1504.05419}
  {arXiv:1504.05419 [hep-ph]} \BibitemShut {NoStop}%
\bibitem [{\citenamefont {Gresham}\ \emph {et~al.}(2017)\citenamefont
  {Gresham}, \citenamefont {Lou},\ and\ \citenamefont
  {Zurek}}]{nucleiGresham:2017zqi}%
  \BibitemOpen
  \bibfield  {author} {\bibinfo {author} {\bibfnamefont {M.~I.}\ \bibnamefont
  {Gresham}}, \bibinfo {author} {\bibfnamefont {H.~K.}\ \bibnamefont {Lou}}, \
  and\ \bibinfo {author} {\bibfnamefont {K.~M.}\ \bibnamefont {Zurek}},\ }\href
  {\doibase 10.1103/PhysRevD.96.096012} {\bibfield  {journal} {\bibinfo
  {journal} {Phys. Rev. D}\ }\textbf {\bibinfo {volume} {96}},\ \bibinfo
  {pages} {096012} (\bibinfo {year} {2017})},\ \Eprint
  {http://arxiv.org/abs/1707.02313} {arXiv:1707.02313 [hep-ph]} \BibitemShut
  {NoStop}%
\bibitem [{\citenamefont {Gresham}\ \emph
  {et~al.}(2018{\natexlab{a}})\citenamefont {Gresham}, \citenamefont {Lou},\
  and\ \citenamefont {Zurek}}]{nuggetsGresham:2017cvl}%
  \BibitemOpen
  \bibfield  {author} {\bibinfo {author} {\bibfnamefont {M.~I.}\ \bibnamefont
  {Gresham}}, \bibinfo {author} {\bibfnamefont {H.~K.}\ \bibnamefont {Lou}}, \
  and\ \bibinfo {author} {\bibfnamefont {K.~M.}\ \bibnamefont {Zurek}},\ }\href
  {\doibase 10.1103/PhysRevD.97.036003} {\bibfield  {journal} {\bibinfo
  {journal} {Phys. Rev. D}\ }\textbf {\bibinfo {volume} {97}},\ \bibinfo
  {pages} {036003} (\bibinfo {year} {2018}{\natexlab{a}})},\ \Eprint
  {http://arxiv.org/abs/1707.02316} {arXiv:1707.02316 [hep-ph]} \BibitemShut
  {NoStop}%
\bibitem [{\citenamefont {McDermott}(2018)}]{nucleiMcDermott:2017vyk}%
  \BibitemOpen
  \bibfield  {author} {\bibinfo {author} {\bibfnamefont {S.~D.}\ \bibnamefont
  {McDermott}},\ }\href {\doibase 10.1103/PhysRevLett.120.221806} {\bibfield
  {journal} {\bibinfo  {journal} {Phys. Rev. Lett.}\ }\textbf {\bibinfo
  {volume} {120}},\ \bibinfo {pages} {221806} (\bibinfo {year} {2018})},\
  \Eprint {http://arxiv.org/abs/1711.00857} {arXiv:1711.00857 [hep-ph]}
  \BibitemShut {NoStop}%
\bibitem [{\citenamefont {Gresham}\ \emph
  {et~al.}(2018{\natexlab{b}})\citenamefont {Gresham}, \citenamefont {Lou},\
  and\ \citenamefont {Zurek}}]{nuggetsGresham:2018anj}%
  \BibitemOpen
  \bibfield  {author} {\bibinfo {author} {\bibfnamefont {M.~I.}\ \bibnamefont
  {Gresham}}, \bibinfo {author} {\bibfnamefont {H.~K.}\ \bibnamefont {Lou}}, \
  and\ \bibinfo {author} {\bibfnamefont {K.~M.}\ \bibnamefont {Zurek}},\ }\href
  {\doibase 10.1103/PhysRevD.98.096001} {\bibfield  {journal} {\bibinfo
  {journal} {Phys. Rev. D}\ }\textbf {\bibinfo {volume} {98}},\ \bibinfo
  {pages} {096001} (\bibinfo {year} {2018}{\natexlab{b}})},\ \Eprint
  {http://arxiv.org/abs/1805.04512} {arXiv:1805.04512 [hep-ph]} \BibitemShut
  {NoStop}%
\bibitem [{\citenamefont {Bai}\ \emph {et~al.}(2019)\citenamefont {Bai},
  \citenamefont {Long},\ and\ \citenamefont {Lu}}]{BaiLongLunuggets:2018dxf}%
  \BibitemOpen
  \bibfield  {author} {\bibinfo {author} {\bibfnamefont {Y.}~\bibnamefont
  {Bai}}, \bibinfo {author} {\bibfnamefont {A.~J.}\ \bibnamefont {Long}}, \
  and\ \bibinfo {author} {\bibfnamefont {S.}~\bibnamefont {Lu}},\ }\href
  {\doibase 10.1103/PhysRevD.99.055047} {\bibfield  {journal} {\bibinfo
  {journal} {Phys. Rev. D}\ }\textbf {\bibinfo {volume} {99}},\ \bibinfo
  {pages} {055047} (\bibinfo {year} {2019})},\ \Eprint
  {http://arxiv.org/abs/1810.04360} {arXiv:1810.04360 [hep-ph]} \BibitemShut
  {NoStop}%
\bibitem [{\citenamefont {Grabowska}\ \emph {et~al.}(2018)\citenamefont
  {Grabowska}, \citenamefont {Melia},\ and\ \citenamefont
  {Rajendran}}]{BlobsGrabowska:2018lnd}%
  \BibitemOpen
  \bibfield  {author} {\bibinfo {author} {\bibfnamefont {D.~M.}\ \bibnamefont
  {Grabowska}}, \bibinfo {author} {\bibfnamefont {T.}~\bibnamefont {Melia}}, \
  and\ \bibinfo {author} {\bibfnamefont {S.}~\bibnamefont {Rajendran}},\ }\href
  {\doibase 10.1103/PhysRevD.98.115020} {\bibfield  {journal} {\bibinfo
  {journal} {Phys. Rev. D}\ }\textbf {\bibinfo {volume} {98}},\ \bibinfo
  {pages} {115020} (\bibinfo {year} {2018})},\ \Eprint
  {http://arxiv.org/abs/1807.03788} {arXiv:1807.03788 [hep-ph]} \BibitemShut
  {NoStop}%
\bibitem [{\citenamefont {Croon}\ \emph
  {et~al.}(2020{\natexlab{a}})\citenamefont {Croon}, \citenamefont {McKeen},
  \citenamefont {Raj},\ and\ \citenamefont {Wang}}]{ECOlocation1}%
  \BibitemOpen
  \bibfield  {author} {\bibinfo {author} {\bibfnamefont {D.}~\bibnamefont
  {Croon}}, \bibinfo {author} {\bibfnamefont {D.}~\bibnamefont {McKeen}},
  \bibinfo {author} {\bibfnamefont {N.}~\bibnamefont {Raj}}, \ and\ \bibinfo
  {author} {\bibfnamefont {Z.}~\bibnamefont {Wang}},\ }\href {\doibase
  10.1103/PhysRevD.102.083021} {\bibfield  {journal} {\bibinfo  {journal}
  {Phys. Rev. D}\ }\textbf {\bibinfo {volume} {102}},\ \bibinfo {pages}
  {083021} (\bibinfo {year} {2020}{\natexlab{a}})},\ \Eprint
  {http://arxiv.org/abs/2007.12697} {arXiv:2007.12697 [astro-ph.CO]}
  \BibitemShut {NoStop}%
\bibitem [{\citenamefont {Croon}\ \emph
  {et~al.}(2020{\natexlab{b}})\citenamefont {Croon}, \citenamefont {McKeen},\
  and\ \citenamefont {Raj}}]{ECOlocation2}%
  \BibitemOpen
  \bibfield  {author} {\bibinfo {author} {\bibfnamefont {D.}~\bibnamefont
  {Croon}}, \bibinfo {author} {\bibfnamefont {D.}~\bibnamefont {McKeen}}, \
  and\ \bibinfo {author} {\bibfnamefont {N.}~\bibnamefont {Raj}},\ }\href
  {\doibase 10.1103/PhysRevD.101.083013} {\bibfield  {journal} {\bibinfo
  {journal} {Phys. Rev. D}\ }\textbf {\bibinfo {volume} {101}},\ \bibinfo
  {pages} {083013} (\bibinfo {year} {2020}{\natexlab{b}})},\ \Eprint
  {http://arxiv.org/abs/2002.08962} {arXiv:2002.08962 [astro-ph.CO]}
  \BibitemShut {NoStop}%
\bibitem [{\citenamefont {Dror}\ \emph {et~al.}(2019)\citenamefont {Dror},
  \citenamefont {Ramani}, \citenamefont {Trickle},\ and\ \citenamefont
  {Zurek}}]{PTA:Dror:2019twh}%
  \BibitemOpen
  \bibfield  {author} {\bibinfo {author} {\bibfnamefont {J.~A.}\ \bibnamefont
  {Dror}}, \bibinfo {author} {\bibfnamefont {H.}~\bibnamefont {Ramani}},
  \bibinfo {author} {\bibfnamefont {T.}~\bibnamefont {Trickle}}, \ and\
  \bibinfo {author} {\bibfnamefont {K.~M.}\ \bibnamefont {Zurek}},\ }\href
  {\doibase 10.1103/PhysRevD.100.023003} {\bibfield  {journal} {\bibinfo
  {journal} {Phys. Rev. D}\ }\textbf {\bibinfo {volume} {100}},\ \bibinfo
  {pages} {023003} (\bibinfo {year} {2019})},\ \Eprint
  {http://arxiv.org/abs/1901.04490} {arXiv:1901.04490 [astro-ph.CO]}
  \BibitemShut {NoStop}%
\bibitem [{\citenamefont {Lee}\ \emph {et~al.}(2021)\citenamefont {Lee},
  \citenamefont {Taylor}, \citenamefont {Trickle},\ and\ \citenamefont
  {Zurek}}]{PTALee:2021zqw}%
  \BibitemOpen
  \bibfield  {author} {\bibinfo {author} {\bibfnamefont {V.~S.~H.}\
  \bibnamefont {Lee}}, \bibinfo {author} {\bibfnamefont {S.~R.}\ \bibnamefont
  {Taylor}}, \bibinfo {author} {\bibfnamefont {T.}~\bibnamefont {Trickle}}, \
  and\ \bibinfo {author} {\bibfnamefont {K.~M.}\ \bibnamefont {Zurek}},\ }\href
  {\doibase 10.1088/1475-7516/2021/08/025} {\bibfield  {journal} {\bibinfo
  {journal} {JCAP}\ }\textbf {\bibinfo {volume} {08}},\ \bibinfo {pages} {025}
  (\bibinfo {year} {2021})},\ \Eprint {http://arxiv.org/abs/2104.05717}
  {arXiv:2104.05717 [astro-ph.CO]} \BibitemShut {NoStop}%
\bibitem [{\citenamefont {Bai}\ \emph {et~al.}(2020)\citenamefont {Bai},
  \citenamefont {Long},\ and\ \citenamefont {Lu}}]{dMACHOS:Bai:2020jfm}%
  \BibitemOpen
  \bibfield  {author} {\bibinfo {author} {\bibfnamefont {Y.}~\bibnamefont
  {Bai}}, \bibinfo {author} {\bibfnamefont {A.~J.}\ \bibnamefont {Long}}, \
  and\ \bibinfo {author} {\bibfnamefont {S.}~\bibnamefont {Lu}},\ }\href
  {\doibase 10.1088/1475-7516/2020/09/044} {\bibfield  {journal} {\bibinfo
  {journal} {JCAP}\ }\textbf {\bibinfo {volume} {09}},\ \bibinfo {pages} {044}
  (\bibinfo {year} {2020})},\ \Eprint {http://arxiv.org/abs/2003.13182}
  {arXiv:2003.13182 [astro-ph.CO]} \BibitemShut {NoStop}%
\bibitem [{\citenamefont {Wadekar}\ and\ \citenamefont
  {Wang}(2023)}]{Wadekar:2022ymq}%
  \BibitemOpen
  \bibfield  {author} {\bibinfo {author} {\bibfnamefont {D.}~\bibnamefont
  {Wadekar}}\ and\ \bibinfo {author} {\bibfnamefont {Z.}~\bibnamefont {Wang}},\
  }\href {\doibase 10.1103/PhysRevD.107.083011} {\bibfield  {journal} {\bibinfo
   {journal} {Phys. Rev. D}\ }\textbf {\bibinfo {volume} {107}},\ \bibinfo
  {pages} {083011} (\bibinfo {year} {2023})},\ \Eprint
  {http://arxiv.org/abs/2211.07668} {arXiv:2211.07668 [hep-ph]} \BibitemShut
  {NoStop}%
\bibitem [{\citenamefont {Croon}\ \emph {et~al.}(2023)\citenamefont {Croon},
  \citenamefont {Ipek},\ and\ \citenamefont {McKeen}}]{Croon:2022tmr}%
  \BibitemOpen
  \bibfield  {author} {\bibinfo {author} {\bibfnamefont {D.}~\bibnamefont
  {Croon}}, \bibinfo {author} {\bibfnamefont {S.}~\bibnamefont {Ipek}}, \ and\
  \bibinfo {author} {\bibfnamefont {D.}~\bibnamefont {McKeen}},\ }\href
  {\doibase 10.1103/PhysRevD.107.063012} {\bibfield  {journal} {\bibinfo
  {journal} {Phys. Rev. D}\ }\textbf {\bibinfo {volume} {107}},\ \bibinfo
  {pages} {063012} (\bibinfo {year} {2023})},\ \Eprint
  {http://arxiv.org/abs/2205.15396} {arXiv:2205.15396 [astro-ph.CO]}
  \BibitemShut {NoStop}%
\bibitem [{\citenamefont {Stiff}\ \emph {et~al.}(2001)\citenamefont {Stiff},
  \citenamefont {Widrow},\ and\ \citenamefont
  {Frieman}}]{DDclumps:WidrowStiff:2001dq}%
  \BibitemOpen
  \bibfield  {author} {\bibinfo {author} {\bibfnamefont {D.}~\bibnamefont
  {Stiff}}, \bibinfo {author} {\bibfnamefont {L.~M.}\ \bibnamefont {Widrow}}, \
  and\ \bibinfo {author} {\bibfnamefont {J.}~\bibnamefont {Frieman}},\ }\href
  {\doibase 10.1103/PhysRevD.64.083516} {\bibfield  {journal} {\bibinfo
  {journal} {Phys. Rev. D}\ }\textbf {\bibinfo {volume} {64}},\ \bibinfo
  {pages} {083516} (\bibinfo {year} {2001})},\ \Eprint
  {http://arxiv.org/abs/astro-ph/0106048} {arXiv:astro-ph/0106048} \BibitemShut
  {NoStop}%
\bibitem [{\citenamefont {Kamionkowski}\ and\ \citenamefont
  {Koushiappas}(2008)}]{DDclumps:Kamionkowski:2008vw}%
  \BibitemOpen
  \bibfield  {author} {\bibinfo {author} {\bibfnamefont {M.}~\bibnamefont
  {Kamionkowski}}\ and\ \bibinfo {author} {\bibfnamefont {S.~M.}\ \bibnamefont
  {Koushiappas}},\ }\href {\doibase 10.1103/PhysRevD.77.103509} {\bibfield
  {journal} {\bibinfo  {journal} {Phys. Rev. D}\ }\textbf {\bibinfo {volume}
  {77}},\ \bibinfo {pages} {103509} (\bibinfo {year} {2008})},\ \Eprint
  {http://arxiv.org/abs/0801.3269} {arXiv:0801.3269 [astro-ph]} \BibitemShut
  {NoStop}%
\bibitem [{\citenamefont {Ibarra}\ \emph {et~al.}(2019)\citenamefont {Ibarra},
  \citenamefont {Kavanagh},\ and\ \citenamefont {Rappelt}}]{bradleyclumpSun}%
  \BibitemOpen
  \bibfield  {author} {\bibinfo {author} {\bibfnamefont {A.}~\bibnamefont
  {Ibarra}}, \bibinfo {author} {\bibfnamefont {B.~J.}\ \bibnamefont
  {Kavanagh}}, \ and\ \bibinfo {author} {\bibfnamefont {A.}~\bibnamefont
  {Rappelt}},\ }\href {\doibase 10.1088/1475-7516/2019/12/013} {\bibfield
  {journal} {\bibinfo  {journal} {JCAP}\ }\textbf {\bibinfo {volume} {12}},\
  \bibinfo {pages} {013} (\bibinfo {year} {2019})},\ \Eprint
  {http://arxiv.org/abs/1908.00747} {arXiv:1908.00747 [hep-ph]} \BibitemShut
  {NoStop}%
\bibitem [{\citenamefont {Bramante}\ \emph
  {et~al.}(2018{\natexlab{a}})\citenamefont {Bramante}, \citenamefont
  {Broerman}, \citenamefont {Lang},\ and\ \citenamefont
  {Raj}}]{xmaslights:Bramante:2018qbc}%
  \BibitemOpen
  \bibfield  {author} {\bibinfo {author} {\bibfnamefont {J.}~\bibnamefont
  {Bramante}}, \bibinfo {author} {\bibfnamefont {B.}~\bibnamefont {Broerman}},
  \bibinfo {author} {\bibfnamefont {R.~F.}\ \bibnamefont {Lang}}, \ and\
  \bibinfo {author} {\bibfnamefont {N.}~\bibnamefont {Raj}},\ }\href {\doibase
  10.1103/PhysRevD.98.083516} {\bibfield  {journal} {\bibinfo  {journal} {Phys.
  Rev. D}\ }\textbf {\bibinfo {volume} {98}},\ \bibinfo {pages} {083516}
  (\bibinfo {year} {2018}{\natexlab{a}})},\ \Eprint
  {http://arxiv.org/abs/1803.08044} {arXiv:1803.08044 [hep-ph]} \BibitemShut
  {NoStop}%
\bibitem [{\citenamefont {Bramante}\ \emph
  {et~al.}(2019{\natexlab{a}})\citenamefont {Bramante}, \citenamefont
  {Broerman}, \citenamefont {Kumar}, \citenamefont {Lang}, \citenamefont
  {Pospelov},\ and\ \citenamefont {Raj}}]{xmaslights:Bramante:2018tos}%
  \BibitemOpen
  \bibfield  {author} {\bibinfo {author} {\bibfnamefont {J.}~\bibnamefont
  {Bramante}}, \bibinfo {author} {\bibfnamefont {B.}~\bibnamefont {Broerman}},
  \bibinfo {author} {\bibfnamefont {J.}~\bibnamefont {Kumar}}, \bibinfo
  {author} {\bibfnamefont {R.~F.}\ \bibnamefont {Lang}}, \bibinfo {author}
  {\bibfnamefont {M.}~\bibnamefont {Pospelov}}, \ and\ \bibinfo {author}
  {\bibfnamefont {N.}~\bibnamefont {Raj}},\ }\href {\doibase
  10.1103/PhysRevD.99.083010} {\bibfield  {journal} {\bibinfo  {journal} {Phys.
  Rev. D}\ }\textbf {\bibinfo {volume} {99}},\ \bibinfo {pages} {083010}
  (\bibinfo {year} {2019}{\natexlab{a}})},\ \Eprint
  {http://arxiv.org/abs/1812.09325} {arXiv:1812.09325 [hep-ph]} \BibitemShut
  {NoStop}%
\bibitem [{\citenamefont {Bramante}\ \emph
  {et~al.}(2019{\natexlab{b}})\citenamefont {Bramante}, \citenamefont {Kumar},\
  and\ \citenamefont {Raj}}]{xmaslights:Bramante:2019yss}%
  \BibitemOpen
  \bibfield  {author} {\bibinfo {author} {\bibfnamefont {J.}~\bibnamefont
  {Bramante}}, \bibinfo {author} {\bibfnamefont {J.}~\bibnamefont {Kumar}}, \
  and\ \bibinfo {author} {\bibfnamefont {N.}~\bibnamefont {Raj}},\ }\href
  {\doibase 10.1103/PhysRevD.100.123016} {\bibfield  {journal} {\bibinfo
  {journal} {Phys. Rev. D}\ }\textbf {\bibinfo {volume} {100}},\ \bibinfo
  {pages} {123016} (\bibinfo {year} {2019}{\natexlab{b}})},\ \Eprint
  {http://arxiv.org/abs/1910.05380} {arXiv:1910.05380 [hep-ph]} \BibitemShut
  {NoStop}%
\bibitem [{\citenamefont {Adhikari}\ \emph {et~al.}(2022)\citenamefont
  {Adhikari} \emph {et~al.}}]{xmaslights:DEAPCollaboration:2021raj}%
  \BibitemOpen
  \bibfield  {author} {\bibinfo {author} {\bibfnamefont {P.}~\bibnamefont
  {Adhikari}} \emph {et~al.} (\bibinfo {collaboration} {(DEAP
  Collaboration)\textdaggerdbl{}, DEAP}),\ }\href {\doibase
  10.1103/PhysRevLett.128.011801} {\bibfield  {journal} {\bibinfo  {journal}
  {Phys. Rev. Lett.}\ }\textbf {\bibinfo {volume} {128}},\ \bibinfo {pages}
  {011801} (\bibinfo {year} {2022})},\ \Eprint
  {http://arxiv.org/abs/2108.09405} {arXiv:2108.09405 [astro-ph.CO]}
  \BibitemShut {NoStop}%
\bibitem [{\citenamefont {Acevedo}\ \emph
  {et~al.}(2021{\natexlab{b}})\citenamefont {Acevedo}, \citenamefont
  {Bramante},\ and\ \citenamefont {Goodman}}]{Acevedo:2021kly}%
  \BibitemOpen
  \bibfield  {author} {\bibinfo {author} {\bibfnamefont {J.~F.}\ \bibnamefont
  {Acevedo}}, \bibinfo {author} {\bibfnamefont {J.}~\bibnamefont {Bramante}}, \
  and\ \bibinfo {author} {\bibfnamefont {A.}~\bibnamefont {Goodman}},\
  }\href@noop {} {\  (\bibinfo {year} {2021}{\natexlab{b}})},\ \Eprint
  {http://arxiv.org/abs/2108.10889} {arXiv:2108.10889 [hep-ph]} \BibitemShut
  {NoStop}%
\bibitem [{\citenamefont {Carney}\ \emph {et~al.}(2022)\citenamefont {Carney}
  \emph {et~al.}}]{snowmass:Carney:2022gse}%
  \BibitemOpen
  \bibfield  {author} {\bibinfo {author} {\bibfnamefont {D.}~\bibnamefont
  {Carney}} \emph {et~al.},\ }\href@noop {} {\  (\bibinfo {year} {2022})},\
  \Eprint {http://arxiv.org/abs/2203.06508} {arXiv:2203.06508 [hep-ph]}
  \BibitemShut {NoStop}%
\bibitem [{\citenamefont {Jacobs}\ \emph {et~al.}(2015)\citenamefont {Jacobs},
  \citenamefont {Starkman},\ and\ \citenamefont {Lynn}}]{Jacobs:2014yca}%
  \BibitemOpen
  \bibfield  {author} {\bibinfo {author} {\bibfnamefont {D.~M.}\ \bibnamefont
  {Jacobs}}, \bibinfo {author} {\bibfnamefont {G.~D.}\ \bibnamefont
  {Starkman}}, \ and\ \bibinfo {author} {\bibfnamefont {B.~W.}\ \bibnamefont
  {Lynn}},\ }\href {\doibase 10.1093/mnras/stv774} {\bibfield  {journal}
  {\bibinfo  {journal} {Mon. Not. Roy. Astron. Soc.}\ }\textbf {\bibinfo
  {volume} {450}},\ \bibinfo {pages} {3418} (\bibinfo {year} {2015})},\ \Eprint
  {http://arxiv.org/abs/1410.2236} {arXiv:1410.2236 [astro-ph.CO]} \BibitemShut
  {NoStop}%
\bibitem [{\citenamefont {Ebadi}\ \emph {et~al.}(2021)\citenamefont {Ebadi}
  \emph {et~al.}}]{Ebadi:2021cte}%
  \BibitemOpen
  \bibfield  {author} {\bibinfo {author} {\bibfnamefont {R.}~\bibnamefont
  {Ebadi}} \emph {et~al.},\ }\href {\doibase 10.1103/PhysRevD.104.015041}
  {\bibfield  {journal} {\bibinfo  {journal} {Phys. Rev. D}\ }\textbf {\bibinfo
  {volume} {104}},\ \bibinfo {pages} {015041} (\bibinfo {year} {2021})},\
  \Eprint {http://arxiv.org/abs/2105.03998} {arXiv:2105.03998 [hep-ph]}
  \BibitemShut {NoStop}%
\bibitem [{\citenamefont {Acevedo}\ \emph
  {et~al.}(2021{\natexlab{c}})\citenamefont {Acevedo}, \citenamefont
  {Bramante},\ and\ \citenamefont {Goodman}}]{Acevedo:2021tbl}%
  \BibitemOpen
  \bibfield  {author} {\bibinfo {author} {\bibfnamefont {J.~F.}\ \bibnamefont
  {Acevedo}}, \bibinfo {author} {\bibfnamefont {J.}~\bibnamefont {Bramante}}, \
  and\ \bibinfo {author} {\bibfnamefont {A.}~\bibnamefont {Goodman}},\
  }\href@noop {} {\  (\bibinfo {year} {2021}{\natexlab{c}})},\ \Eprint
  {http://arxiv.org/abs/2105.06473} {arXiv:2105.06473 [hep-ph]} \BibitemShut
  {NoStop}%
\bibitem [{\citenamefont {Dessert}\ and\ \citenamefont
  {Johnson}(2022)}]{Dessert:2021wjx}%
  \BibitemOpen
  \bibfield  {author} {\bibinfo {author} {\bibfnamefont {C.}~\bibnamefont
  {Dessert}}\ and\ \bibinfo {author} {\bibfnamefont {Z.}~\bibnamefont
  {Johnson}},\ }\href {\doibase 10.1103/PhysRevD.106.103034} {\bibfield
  {journal} {\bibinfo  {journal} {Phys. Rev. D}\ }\textbf {\bibinfo {volume}
  {106}},\ \bibinfo {pages} {103034} (\bibinfo {year} {2022})},\ \Eprint
  {http://arxiv.org/abs/2112.06949} {arXiv:2112.06949 [hep-ph]} \BibitemShut
  {NoStop}%
\bibitem [{\citenamefont {Baryakhtar}\ \emph {et~al.}(2017)\citenamefont
  {Baryakhtar}, \citenamefont {Bramante}, \citenamefont {Li}, \citenamefont
  {Linden},\ and\ \citenamefont {Raj}}]{NSvIR:Baryakhtar:DKHNS}%
  \BibitemOpen
  \bibfield  {author} {\bibinfo {author} {\bibfnamefont {M.}~\bibnamefont
  {Baryakhtar}}, \bibinfo {author} {\bibfnamefont {J.}~\bibnamefont
  {Bramante}}, \bibinfo {author} {\bibfnamefont {S.~W.}\ \bibnamefont {Li}},
  \bibinfo {author} {\bibfnamefont {T.}~\bibnamefont {Linden}}, \ and\ \bibinfo
  {author} {\bibfnamefont {N.}~\bibnamefont {Raj}},\ }\href {\doibase
  10.1103/PhysRevLett.119.131801} {\bibfield  {journal} {\bibinfo  {journal}
  {Phys. Rev. Lett.}\ }\textbf {\bibinfo {volume} {119}},\ \bibinfo {pages}
  {131801} (\bibinfo {year} {2017})},\ \Eprint
  {http://arxiv.org/abs/1704.01577} {arXiv:1704.01577 [hep-ph]} \BibitemShut
  {NoStop}%
\bibitem [{\citenamefont {Raj}\ \emph {et~al.}(2018)\citenamefont {Raj},
  \citenamefont {Tanedo},\ and\ \citenamefont {Yu}}]{NSvIR:Raj:DKHNSOps}%
  \BibitemOpen
  \bibfield  {author} {\bibinfo {author} {\bibfnamefont {N.}~\bibnamefont
  {Raj}}, \bibinfo {author} {\bibfnamefont {P.}~\bibnamefont {Tanedo}}, \ and\
  \bibinfo {author} {\bibfnamefont {H.-B.}\ \bibnamefont {Yu}},\ }\href
  {\doibase 10.1103/PhysRevD.97.043006} {\bibfield  {journal} {\bibinfo
  {journal} {Phys. Rev. D}\ }\textbf {\bibinfo {volume} {97}},\ \bibinfo
  {pages} {043006} (\bibinfo {year} {2018})},\ \Eprint
  {http://arxiv.org/abs/1707.09442} {arXiv:1707.09442 [hep-ph]} \BibitemShut
  {NoStop}%
\bibitem [{\citenamefont {Chen}\ and\ \citenamefont
  {Lin}(2018)}]{NSvIR:SelfIntDM}%
  \BibitemOpen
  \bibfield  {author} {\bibinfo {author} {\bibfnamefont {C.-S.}\ \bibnamefont
  {Chen}}\ and\ \bibinfo {author} {\bibfnamefont {Y.-H.}\ \bibnamefont {Lin}},\
  }\href {\doibase 10.1007/JHEP08(2018)069} {\bibfield  {journal} {\bibinfo
  {journal} {JHEP}\ }\textbf {\bibinfo {volume} {08}},\ \bibinfo {pages} {069}
  (\bibinfo {year} {2018})},\ \Eprint {http://arxiv.org/abs/1804.03409}
  {arXiv:1804.03409 [hep-ph]} \BibitemShut {NoStop}%
\bibitem [{\citenamefont {Bell}\ \emph {et~al.}(2018)\citenamefont {Bell},
  \citenamefont {Busoni},\ and\ \citenamefont
  {Robles}}]{NSvIR:Bell2018:Inelastic}%
  \BibitemOpen
  \bibfield  {author} {\bibinfo {author} {\bibfnamefont {N.~F.}\ \bibnamefont
  {Bell}}, \bibinfo {author} {\bibfnamefont {G.}~\bibnamefont {Busoni}}, \ and\
  \bibinfo {author} {\bibfnamefont {S.}~\bibnamefont {Robles}},\ }\href
  {\doibase 10.1088/1475-7516/2018/09/018} {\bibfield  {journal} {\bibinfo
  {journal} {JCAP}\ }\textbf {\bibinfo {volume} {1809}},\ \bibinfo {pages}
  {018} (\bibinfo {year} {2018})},\ \Eprint {http://arxiv.org/abs/1807.02840}
  {arXiv:1807.02840 [hep-ph]} \BibitemShut {NoStop}%
\bibitem [{\citenamefont {Garani}\ \emph {et~al.}(2019)\citenamefont {Garani},
  \citenamefont {Genolini},\ and\ \citenamefont
  {Hambye}}]{NSvIR:GaraniGenoliniHambye}%
  \BibitemOpen
  \bibfield  {author} {\bibinfo {author} {\bibfnamefont {R.}~\bibnamefont
  {Garani}}, \bibinfo {author} {\bibfnamefont {Y.}~\bibnamefont {Genolini}}, \
  and\ \bibinfo {author} {\bibfnamefont {T.}~\bibnamefont {Hambye}},\ }\href
  {\doibase 10.1088/1475-7516/2019/05/035} {\bibfield  {journal} {\bibinfo
  {journal} {JCAP}\ }\textbf {\bibinfo {volume} {1905}},\ \bibinfo {pages}
  {035} (\bibinfo {year} {2019})},\ \Eprint {http://arxiv.org/abs/1812.08773}
  {arXiv:1812.08773 [hep-ph]} \BibitemShut {NoStop}%
\bibitem [{\citenamefont {Camargo}\ \emph {et~al.}(2019)\citenamefont
  {Camargo}, \citenamefont {Queiroz},\ and\ \citenamefont
  {Sturani}}]{NSvIR:Queiroz:Spectroscopy}%
  \BibitemOpen
  \bibfield  {author} {\bibinfo {author} {\bibfnamefont {D.~A.}\ \bibnamefont
  {Camargo}}, \bibinfo {author} {\bibfnamefont {F.~S.}\ \bibnamefont
  {Queiroz}}, \ and\ \bibinfo {author} {\bibfnamefont {R.}~\bibnamefont
  {Sturani}},\ }\href {\doibase 10.1088/1475-7516/2019/09/051} {\bibfield
  {journal} {\bibinfo  {journal} {JCAP}\ }\textbf {\bibinfo {volume} {1909}},\
  \bibinfo {pages} {051} (\bibinfo {year} {2019})},\ \Eprint
  {http://arxiv.org/abs/1901.05474} {arXiv:1901.05474 [hep-ph]} \BibitemShut
  {NoStop}%
\bibitem [{\citenamefont {Bell}\ \emph {et~al.}(2019)\citenamefont {Bell},
  \citenamefont {Busoni},\ and\ \citenamefont
  {Robles}}]{NSvIR:Bell2019:Leptophilic}%
  \BibitemOpen
  \bibfield  {author} {\bibinfo {author} {\bibfnamefont {N.~F.}\ \bibnamefont
  {Bell}}, \bibinfo {author} {\bibfnamefont {G.}~\bibnamefont {Busoni}}, \ and\
  \bibinfo {author} {\bibfnamefont {S.}~\bibnamefont {Robles}},\ }\href
  {\doibase 10.1088/1475-7516/2019/06/054} {\bibfield  {journal} {\bibinfo
  {journal} {JCAP}\ }\textbf {\bibinfo {volume} {1906}},\ \bibinfo {pages}
  {054} (\bibinfo {year} {2019})},\ \Eprint {http://arxiv.org/abs/1904.09803}
  {arXiv:1904.09803 [hep-ph]} \BibitemShut {NoStop}%
\bibitem [{\citenamefont {Hamaguchi}\ \emph {et~al.}(2019)\citenamefont
  {Hamaguchi}, \citenamefont {Nagata},\ and\ \citenamefont
  {Yanagi}}]{NSvIR:Hamaguchi:RotochemicalvDM2019}%
  \BibitemOpen
  \bibfield  {author} {\bibinfo {author} {\bibfnamefont {K.}~\bibnamefont
  {Hamaguchi}}, \bibinfo {author} {\bibfnamefont {N.}~\bibnamefont {Nagata}}, \
  and\ \bibinfo {author} {\bibfnamefont {K.}~\bibnamefont {Yanagi}},\ }\href
  {\doibase 10.1016/j.physletb.2019.06.060} {\bibfield  {journal} {\bibinfo
  {journal} {Phys. Lett.}\ }\textbf {\bibinfo {volume} {B795}},\ \bibinfo
  {pages} {484} (\bibinfo {year} {2019})},\ \Eprint
  {http://arxiv.org/abs/1905.02991} {arXiv:1905.02991 [hep-ph]} \BibitemShut
  {NoStop}%
\bibitem [{\citenamefont {Garani}\ and\ \citenamefont
  {Heeck}(2019)}]{NSvIR:GaraniHeeck:Muophilic}%
  \BibitemOpen
  \bibfield  {author} {\bibinfo {author} {\bibfnamefont {R.}~\bibnamefont
  {Garani}}\ and\ \bibinfo {author} {\bibfnamefont {J.}~\bibnamefont {Heeck}},\
  }\href {\doibase 10.1103/PhysRevD.100.035039} {\bibfield  {journal} {\bibinfo
   {journal} {Phys. Rev.}\ }\textbf {\bibinfo {volume} {D100}},\ \bibinfo
  {pages} {035039} (\bibinfo {year} {2019})},\ \Eprint
  {http://arxiv.org/abs/1906.10145} {arXiv:1906.10145 [hep-ph]} \BibitemShut
  {NoStop}%
\bibitem [{\citenamefont {Acevedo}\ \emph {et~al.}(2020)\citenamefont
  {Acevedo}, \citenamefont {Bramante}, \citenamefont {Leane},\ and\
  \citenamefont {Raj}}]{NSvIR:Pasta}%
  \BibitemOpen
  \bibfield  {author} {\bibinfo {author} {\bibfnamefont {J.~F.}\ \bibnamefont
  {Acevedo}}, \bibinfo {author} {\bibfnamefont {J.}~\bibnamefont {Bramante}},
  \bibinfo {author} {\bibfnamefont {R.~K.}\ \bibnamefont {Leane}}, \ and\
  \bibinfo {author} {\bibfnamefont {N.}~\bibnamefont {Raj}},\ }\href {\doibase
  10.1088/1475-7516/2020/03/038} {\bibfield  {journal} {\bibinfo  {journal}
  {JCAP}\ }\textbf {\bibinfo {volume} {03}},\ \bibinfo {pages} {038} (\bibinfo
  {year} {2020})},\ \Eprint {http://arxiv.org/abs/1911.06334} {arXiv:1911.06334
  [hep-ph]} \BibitemShut {NoStop}%
\bibitem [{\citenamefont {Joglekar}\ \emph {et~al.}(2019)\citenamefont
  {Joglekar}, \citenamefont {Raj}, \citenamefont {Tanedo},\ and\ \citenamefont
  {Yu}}]{NSvIR:Riverside:LeptophilicShort}%
  \BibitemOpen
  \bibfield  {author} {\bibinfo {author} {\bibfnamefont {A.}~\bibnamefont
  {Joglekar}}, \bibinfo {author} {\bibfnamefont {N.}~\bibnamefont {Raj}},
  \bibinfo {author} {\bibfnamefont {P.}~\bibnamefont {Tanedo}}, \ and\ \bibinfo
  {author} {\bibfnamefont {H.-B.}\ \bibnamefont {Yu}},\ }\href@noop {} {\
  (\bibinfo {year} {2019})},\ \Eprint {http://arxiv.org/abs/1911.13293}
  {arXiv:1911.13293 [hep-ph]} \BibitemShut {NoStop}%
\bibitem [{\citenamefont {Keung}\ \emph {et~al.}(2020)\citenamefont {Keung},
  \citenamefont {Marfatia},\ and\ \citenamefont
  {Tseng}}]{NSvIR:Marfatia:DarkBaryon}%
  \BibitemOpen
  \bibfield  {author} {\bibinfo {author} {\bibfnamefont {W.-Y.}\ \bibnamefont
  {Keung}}, \bibinfo {author} {\bibfnamefont {D.}~\bibnamefont {Marfatia}}, \
  and\ \bibinfo {author} {\bibfnamefont {P.-Y.}\ \bibnamefont {Tseng}},\ }\href
  {\doibase 10.1007/JHEP07(2020)181} {\bibfield  {journal} {\bibinfo  {journal}
  {JHEP}\ }\textbf {\bibinfo {volume} {07}},\ \bibinfo {pages} {181} (\bibinfo
  {year} {2020})},\ \Eprint {http://arxiv.org/abs/2001.09140} {arXiv:2001.09140
  [hep-ph]} \BibitemShut {NoStop}%
\bibitem [{\citenamefont {Joglekar}\ \emph {et~al.}(2020)\citenamefont
  {Joglekar}, \citenamefont {Raj}, \citenamefont {Tanedo},\ and\ \citenamefont
  {Yu}}]{NSvIR:Riverside:LeptophilicLong}%
  \BibitemOpen
  \bibfield  {author} {\bibinfo {author} {\bibfnamefont {A.}~\bibnamefont
  {Joglekar}}, \bibinfo {author} {\bibfnamefont {N.}~\bibnamefont {Raj}},
  \bibinfo {author} {\bibfnamefont {P.}~\bibnamefont {Tanedo}}, \ and\ \bibinfo
  {author} {\bibfnamefont {H.-B.}\ \bibnamefont {Yu}},\ }\href@noop {} {\
  (\bibinfo {year} {2020})},\ \Eprint {http://arxiv.org/abs/2004.09539}
  {arXiv:2004.09539 [hep-ph]} \BibitemShut {NoStop}%
\bibitem [{\citenamefont {Bell}\ \emph
  {et~al.}(2020{\natexlab{a}})\citenamefont {Bell}, \citenamefont {Busoni},
  \citenamefont {Robles},\ and\ \citenamefont {Virgato}}]{NSvIR:Bell:Improved}%
  \BibitemOpen
  \bibfield  {author} {\bibinfo {author} {\bibfnamefont {N.~F.}\ \bibnamefont
  {Bell}}, \bibinfo {author} {\bibfnamefont {G.}~\bibnamefont {Busoni}},
  \bibinfo {author} {\bibfnamefont {S.}~\bibnamefont {Robles}}, \ and\ \bibinfo
  {author} {\bibfnamefont {M.}~\bibnamefont {Virgato}},\ }\href@noop {} {\
  (\bibinfo {year} {2020}{\natexlab{a}})},\ \Eprint
  {http://arxiv.org/abs/2004.14888} {arXiv:2004.14888 [hep-ph]} \BibitemShut
  {NoStop}%
\bibitem [{\citenamefont {Dasgupta}\ \emph {et~al.}(2020)\citenamefont
  {Dasgupta}, \citenamefont {Gupta},\ and\ \citenamefont
  {Ray}}]{NSvIR:DasguptaGuptaRay:LightMed}%
  \BibitemOpen
  \bibfield  {author} {\bibinfo {author} {\bibfnamefont {B.}~\bibnamefont
  {Dasgupta}}, \bibinfo {author} {\bibfnamefont {A.}~\bibnamefont {Gupta}}, \
  and\ \bibinfo {author} {\bibfnamefont {A.}~\bibnamefont {Ray}},\ }\href
  {\doibase 10.1088/1475-7516/2020/10/023} {\bibfield  {journal} {\bibinfo
  {journal} {JCAP}\ }\textbf {\bibinfo {volume} {10}},\ \bibinfo {pages} {023}
  (\bibinfo {year} {2020})},\ \Eprint {http://arxiv.org/abs/2006.10773}
  {arXiv:2006.10773 [hep-ph]} \BibitemShut {NoStop}%
\bibitem [{\citenamefont {Garani}\ \emph {et~al.}(2021)\citenamefont {Garani},
  \citenamefont {Gupta},\ and\ \citenamefont
  {Raj}}]{NSvIR:GaraniGuptaRaj:Thermalizn}%
  \BibitemOpen
  \bibfield  {author} {\bibinfo {author} {\bibfnamefont {R.}~\bibnamefont
  {Garani}}, \bibinfo {author} {\bibfnamefont {A.}~\bibnamefont {Gupta}}, \
  and\ \bibinfo {author} {\bibfnamefont {N.}~\bibnamefont {Raj}},\ }\href
  {\doibase 10.1103/PhysRevD.103.043019} {\bibfield  {journal} {\bibinfo
  {journal} {Phys. Rev. D}\ }\textbf {\bibinfo {volume} {103}},\ \bibinfo
  {pages} {043019} (\bibinfo {year} {2021})},\ \Eprint
  {http://arxiv.org/abs/2009.10728} {arXiv:2009.10728 [hep-ph]} \BibitemShut
  {NoStop}%
\bibitem [{\citenamefont {Bell}\ \emph
  {et~al.}(2021{\natexlab{a}})\citenamefont {Bell}, \citenamefont {Busoni},
  \citenamefont {Robles},\ and\ \citenamefont
  {Virgato}}]{NSvIR:Bell:ImprovedLepton}%
  \BibitemOpen
  \bibfield  {author} {\bibinfo {author} {\bibfnamefont {N.~F.}\ \bibnamefont
  {Bell}}, \bibinfo {author} {\bibfnamefont {G.}~\bibnamefont {Busoni}},
  \bibinfo {author} {\bibfnamefont {S.}~\bibnamefont {Robles}}, \ and\ \bibinfo
  {author} {\bibfnamefont {M.}~\bibnamefont {Virgato}},\ }\href {\doibase
  10.1088/1475-7516/2021/03/086} {\bibfield  {journal} {\bibinfo  {journal}
  {JCAP}\ }\textbf {\bibinfo {volume} {03}},\ \bibinfo {pages} {086} (\bibinfo
  {year} {2021}{\natexlab{a}})},\ \Eprint {http://arxiv.org/abs/2010.13257}
  {arXiv:2010.13257 [hep-ph]} \BibitemShut {NoStop}%
\bibitem [{\citenamefont {Bell}\ \emph
  {et~al.}(2020{\natexlab{b}})\citenamefont {Bell}, \citenamefont {Busoni},
  \citenamefont {Motta}, \citenamefont {Robles}, \citenamefont {Thomas},\ and\
  \citenamefont {Virgato}}]{NSvIR:Bell2020improved}%
  \BibitemOpen
  \bibfield  {author} {\bibinfo {author} {\bibfnamefont {N.~F.}\ \bibnamefont
  {Bell}}, \bibinfo {author} {\bibfnamefont {G.}~\bibnamefont {Busoni}},
  \bibinfo {author} {\bibfnamefont {T.~F.}\ \bibnamefont {Motta}}, \bibinfo
  {author} {\bibfnamefont {S.}~\bibnamefont {Robles}}, \bibinfo {author}
  {\bibfnamefont {A.~W.}\ \bibnamefont {Thomas}}, \ and\ \bibinfo {author}
  {\bibfnamefont {M.}~\bibnamefont {Virgato}},\ }\href@noop {} {\  (\bibinfo
  {year} {2020}{\natexlab{b}})},\ \Eprint {http://arxiv.org/abs/2012.08918}
  {arXiv:2012.08918 [hep-ph]} \BibitemShut {NoStop}%
\bibitem [{\citenamefont {Maity}\ and\ \citenamefont
  {Queiroz}(2021)}]{NSvIR:Queiroz:BosonDM}%
  \BibitemOpen
  \bibfield  {author} {\bibinfo {author} {\bibfnamefont {T.~N.}\ \bibnamefont
  {Maity}}\ and\ \bibinfo {author} {\bibfnamefont {F.~S.}\ \bibnamefont
  {Queiroz}},\ }\href@noop {} {\  (\bibinfo {year} {2021})},\ \Eprint
  {http://arxiv.org/abs/2104.02700} {arXiv:2104.02700 [hep-ph]} \BibitemShut
  {NoStop}%
\bibitem [{\citenamefont {Anzuini}\ \emph {et~al.}(2021)\citenamefont
  {Anzuini}, \citenamefont {Bell}, \citenamefont {Busoni}, \citenamefont
  {Motta}, \citenamefont {Robles}, \citenamefont {Thomas},\ and\ \citenamefont
  {Virgato}}]{NSvIR:anzuiniBell2021improved}%
  \BibitemOpen
  \bibfield  {author} {\bibinfo {author} {\bibfnamefont {F.}~\bibnamefont
  {Anzuini}}, \bibinfo {author} {\bibfnamefont {N.~F.}\ \bibnamefont {Bell}},
  \bibinfo {author} {\bibfnamefont {G.}~\bibnamefont {Busoni}}, \bibinfo
  {author} {\bibfnamefont {T.~F.}\ \bibnamefont {Motta}}, \bibinfo {author}
  {\bibfnamefont {S.}~\bibnamefont {Robles}}, \bibinfo {author} {\bibfnamefont
  {A.~W.}\ \bibnamefont {Thomas}}, \ and\ \bibinfo {author} {\bibfnamefont
  {M.}~\bibnamefont {Virgato}},\ }\href@noop {} {\  (\bibinfo {year} {2021})},\
  \Eprint {http://arxiv.org/abs/2108.02525} {arXiv:2108.02525 [hep-ph]}
  \BibitemShut {NoStop}%
\bibitem [{\citenamefont {Lin}\ and\ \citenamefont
  {Lin}(2021)}]{NSvIR:Lin2021:spin1med}%
  \BibitemOpen
  \bibfield  {author} {\bibinfo {author} {\bibfnamefont {G.-L.}\ \bibnamefont
  {Lin}}\ and\ \bibinfo {author} {\bibfnamefont {Y.-H.}\ \bibnamefont {Lin}},\
  }\href {\doibase 10.1103/PhysRevD.104.063021} {\bibfield  {journal} {\bibinfo
   {journal} {Phys. Rev. D}\ }\textbf {\bibinfo {volume} {104}},\ \bibinfo
  {pages} {063021} (\bibinfo {year} {2021})},\ \Eprint
  {http://arxiv.org/abs/2102.11151} {arXiv:2102.11151 [hep-ph]} \BibitemShut
  {NoStop}%
\bibitem [{\citenamefont {Zeng}\ \emph {et~al.}(2022)\citenamefont {Zeng},
  \citenamefont {Xiao},\ and\ \citenamefont {Wang}}]{NSvIR:zeng2021PNGBDM}%
  \BibitemOpen
  \bibfield  {author} {\bibinfo {author} {\bibfnamefont {Y.-P.}\ \bibnamefont
  {Zeng}}, \bibinfo {author} {\bibfnamefont {X.}~\bibnamefont {Xiao}}, \ and\
  \bibinfo {author} {\bibfnamefont {W.}~\bibnamefont {Wang}},\ }\href {\doibase
  10.1016/j.physletb.2021.136822} {\bibfield  {journal} {\bibinfo  {journal}
  {Phys. Lett. B}\ }\textbf {\bibinfo {volume} {824}},\ \bibinfo {pages}
  {136822} (\bibinfo {year} {2022})},\ \Eprint
  {http://arxiv.org/abs/2108.11381} {arXiv:2108.11381 [hep-ph]} \BibitemShut
  {NoStop}%
\bibitem [{\citenamefont {Fujiwara}\ \emph {et~al.}(2022)\citenamefont
  {Fujiwara}, \citenamefont {Hamaguchi}, \citenamefont {Nagata},\ and\
  \citenamefont {Zheng}}]{NSvIR:HamaguchiEWmultiplet:2022uiq}%
  \BibitemOpen
  \bibfield  {author} {\bibinfo {author} {\bibfnamefont {M.}~\bibnamefont
  {Fujiwara}}, \bibinfo {author} {\bibfnamefont {K.}~\bibnamefont {Hamaguchi}},
  \bibinfo {author} {\bibfnamefont {N.}~\bibnamefont {Nagata}}, \ and\ \bibinfo
  {author} {\bibfnamefont {J.}~\bibnamefont {Zheng}},\ }\href@noop {} {\
  (\bibinfo {year} {2022})},\ \Eprint {http://arxiv.org/abs/2204.02238}
  {arXiv:2204.02238 [hep-ph]} \BibitemShut {NoStop}%
\bibitem [{\citenamefont {Hamaguchi}\ \emph {et~al.}(2022)\citenamefont
  {Hamaguchi}, \citenamefont {Nagata},\ and\ \citenamefont
  {Ramirez-Quezada}}]{NsvIR:HamaguchiMug-2:2022wpz}%
  \BibitemOpen
  \bibfield  {author} {\bibinfo {author} {\bibfnamefont {K.}~\bibnamefont
  {Hamaguchi}}, \bibinfo {author} {\bibfnamefont {N.}~\bibnamefont {Nagata}}, \
  and\ \bibinfo {author} {\bibfnamefont {M.~E.}\ \bibnamefont
  {Ramirez-Quezada}},\ }\href@noop {} {\  (\bibinfo {year} {2022})},\ \Eprint
  {http://arxiv.org/abs/2204.02413} {arXiv:2204.02413 [hep-ph]} \BibitemShut
  {NoStop}%
\bibitem [{\citenamefont {Chatterjee}\ \emph {et~al.}(2022)\citenamefont
  {Chatterjee}, \citenamefont {Garani}, \citenamefont {Jain}, \citenamefont
  {Kanodia}, \citenamefont {Kumar},\ and\ \citenamefont
  {Vempati}}]{NSvIR:IISc2022}%
  \BibitemOpen
  \bibfield  {author} {\bibinfo {author} {\bibfnamefont {S.}~\bibnamefont
  {Chatterjee}}, \bibinfo {author} {\bibfnamefont {R.}~\bibnamefont {Garani}},
  \bibinfo {author} {\bibfnamefont {R.~K.}\ \bibnamefont {Jain}}, \bibinfo
  {author} {\bibfnamefont {B.}~\bibnamefont {Kanodia}}, \bibinfo {author}
  {\bibfnamefont {M.~S.~N.}\ \bibnamefont {Kumar}}, \ and\ \bibinfo {author}
  {\bibfnamefont {S.~K.}\ \bibnamefont {Vempati}},\ }\href {\doibase
  10.48550/ARXIV.2205.05048} {\  (\bibinfo {year} {2022}),\
  10.48550/ARXIV.2205.05048},\ \Eprint {http://arxiv.org/abs/2205.05048}
  {2205.05048} \BibitemShut {NoStop}%
\bibitem [{\citenamefont {Coffey}\ \emph {et~al.}(2022)\citenamefont {Coffey},
  \citenamefont {McKeen}, \citenamefont {Morrissey},\ and\ \citenamefont
  {Raj}}]{NSvIR:PseudoscaTRIUMF:2022eav}%
  \BibitemOpen
  \bibfield  {author} {\bibinfo {author} {\bibfnamefont {J.}~\bibnamefont
  {Coffey}}, \bibinfo {author} {\bibfnamefont {D.}~\bibnamefont {McKeen}},
  \bibinfo {author} {\bibfnamefont {D.~E.}\ \bibnamefont {Morrissey}}, \ and\
  \bibinfo {author} {\bibfnamefont {N.}~\bibnamefont {Raj}},\ }\href {\doibase
  10.1103/PhysRevD.106.115019} {\bibfield  {journal} {\bibinfo  {journal}
  {Phys. Rev. D}\ }\textbf {\bibinfo {volume} {106}},\ \bibinfo {pages}
  {115019} (\bibinfo {year} {2022})},\ \Eprint
  {http://arxiv.org/abs/2207.02221} {arXiv:2207.02221 [hep-ph]} \BibitemShut
  {NoStop}%
\bibitem [{\citenamefont {Alvarez}\ \emph {et~al.}(2023)\citenamefont
  {Alvarez}, \citenamefont {Joglekar}, \citenamefont {Phoroutan-Mehr},\ and\
  \citenamefont {Yu}}]{NSvIR:InelasticJoglekarYu:2023fjj}%
  \BibitemOpen
  \bibfield  {author} {\bibinfo {author} {\bibfnamefont {G.}~\bibnamefont
  {Alvarez}}, \bibinfo {author} {\bibfnamefont {A.}~\bibnamefont {Joglekar}},
  \bibinfo {author} {\bibfnamefont {M.}~\bibnamefont {Phoroutan-Mehr}}, \ and\
  \bibinfo {author} {\bibfnamefont {H.-B.}\ \bibnamefont {Yu}},\ }\href@noop {}
  {\  (\bibinfo {year} {2023})},\ \Eprint {http://arxiv.org/abs/2301.08767}
  {arXiv:2301.08767 [hep-ph]} \BibitemShut {NoStop}%
\bibitem [{\citenamefont {Garani}\ \emph {et~al.}(2023)\citenamefont {Garani},
  \citenamefont {Raj},\ and\ \citenamefont
  {Reynoso-Cordova}}]{globularGaraniRajReynosoC:2023esk}%
  \BibitemOpen
  \bibfield  {author} {\bibinfo {author} {\bibfnamefont {R.}~\bibnamefont
  {Garani}}, \bibinfo {author} {\bibfnamefont {N.}~\bibnamefont {Raj}}, \ and\
  \bibinfo {author} {\bibfnamefont {J.}~\bibnamefont {Reynoso-Cordova}},\
  }\href@noop {} {\  (\bibinfo {year} {2023})},\ \Eprint
  {http://arxiv.org/abs/2303.18009} {arXiv:2303.18009 [astro-ph.HE]}
  \BibitemShut {NoStop}%
\bibitem [{\citenamefont {Berti}\ \emph {et~al.}(2022)\citenamefont {Berti}
  \emph {et~al.}}]{snowmass:Berti:2022rwn}%
  \BibitemOpen
  \bibfield  {author} {\bibinfo {author} {\bibfnamefont {E.}~\bibnamefont
  {Berti}} \emph {et~al.},\ }in\ \href@noop {} {\emph {\bibinfo {booktitle}
  {{2022 Snowmass Summer Study}}}}\ (\bibinfo {year} {2022})\ \Eprint
  {http://arxiv.org/abs/2203.07984} {arXiv:2203.07984 [hep-ph]} \BibitemShut
  {NoStop}%
\bibitem [{\citenamefont {Bramante}\ and\ \citenamefont
  {Raj}(2024)}]{reviewdarkincompact}%
  \BibitemOpen
  \bibfield  {author} {\bibinfo {author} {\bibfnamefont {J.}~\bibnamefont
  {Bramante}}\ and\ \bibinfo {author} {\bibfnamefont {N.}~\bibnamefont {Raj}},\
  }\href {\doibase 10.1016/j.physrep.2023.12.001} {\bibfield  {journal}
  {\bibinfo  {journal} {Phys. Rept.}\ }\textbf {\bibinfo {volume} {1052}},\
  \bibinfo {pages} {1} (\bibinfo {year} {2024})},\ \Eprint
  {http://arxiv.org/abs/2307.14435} {arXiv:2307.14435 [hep-ph]} \BibitemShut
  {NoStop}%
\bibitem [{\citenamefont {{Timmes}}\ and\ \citenamefont
  {{Woosley}}(1992)}]{TimmesWoosley1992}%
  \BibitemOpen
  \bibfield  {author} {\bibinfo {author} {\bibfnamefont {F.~X.}\ \bibnamefont
  {{Timmes}}}\ and\ \bibinfo {author} {\bibfnamefont {S.~E.}\ \bibnamefont
  {{Woosley}}},\ }\href {\doibase 10.1086/171746} {\bibfield  {journal}
  {\bibinfo  {journal} {APJ}\ }\textbf {\bibinfo {volume} {396}},\ \bibinfo
  {pages} {649} (\bibinfo {year} {1992})}\BibitemShut {NoStop}%
\bibitem [{\citenamefont {Prakash}\ \emph {et~al.}(1988)\citenamefont
  {Prakash}, \citenamefont {Ainsworth},\ and\ \citenamefont
  {Lattimer}}]{EoSPrakash:1988md}%
  \BibitemOpen
  \bibfield  {author} {\bibinfo {author} {\bibfnamefont {M.}~\bibnamefont
  {Prakash}}, \bibinfo {author} {\bibfnamefont {T.~L.}\ \bibnamefont
  {Ainsworth}}, \ and\ \bibinfo {author} {\bibfnamefont {J.~M.}\ \bibnamefont
  {Lattimer}},\ }\href {\doibase 10.1103/PhysRevLett.61.2518} {\bibfield
  {journal} {\bibinfo  {journal} {Phys. Rev. Lett.}\ }\textbf {\bibinfo
  {volume} {61}},\ \bibinfo {pages} {2518} (\bibinfo {year}
  {1988})}\BibitemShut {NoStop}%
\bibitem [{\citenamefont {{Datta}}\ \emph {et~al.}(1995)\citenamefont
  {{Datta}}, \citenamefont {{Thampan}},\ and\ \citenamefont
  {{Bhattacharya}}}]{crustdensityprofileRRIIIA}%
  \BibitemOpen
  \bibfield  {author} {\bibinfo {author} {\bibfnamefont {B.}~\bibnamefont
  {{Datta}}}, \bibinfo {author} {\bibfnamefont {A.~V.}\ \bibnamefont
  {{Thampan}}}, \ and\ \bibinfo {author} {\bibfnamefont {D.}~\bibnamefont
  {{Bhattacharya}}},\ }\href {\doibase 10.1007/BF02715610} {\bibfield
  {journal} {\bibinfo  {journal} {Journal of Astrophysics and Astronomy}\
  }\textbf {\bibinfo {volume} {16}},\ \bibinfo {pages} {375} (\bibinfo {year}
  {1995})}\BibitemShut {NoStop}%
\bibitem [{\citenamefont {Brayeur}\ and\ \citenamefont
  {Tinyakov}(2012)}]{Brayeur:2011yw}%
  \BibitemOpen
  \bibfield  {author} {\bibinfo {author} {\bibfnamefont {L.}~\bibnamefont
  {Brayeur}}\ and\ \bibinfo {author} {\bibfnamefont {P.}~\bibnamefont
  {Tinyakov}},\ }\href {\doibase 10.1103/PhysRevLett.109.061301} {\bibfield
  {journal} {\bibinfo  {journal} {Phys. Rev. Lett.}\ }\textbf {\bibinfo
  {volume} {109}},\ \bibinfo {pages} {061301} (\bibinfo {year} {2012})},\
  \Eprint {http://arxiv.org/abs/1111.3205} {arXiv:1111.3205 [astro-ph.CO]}
  \BibitemShut {NoStop}%
\bibitem [{\citenamefont {Bell}\ \emph
  {et~al.}(2021{\natexlab{b}})\citenamefont {Bell}, \citenamefont {Busoni},
  \citenamefont {Ramirez-Quezada}, \citenamefont {Robles},\ and\ \citenamefont
  {Virgato}}]{BellImprovedWD:2021fye}%
  \BibitemOpen
  \bibfield  {author} {\bibinfo {author} {\bibfnamefont {N.~F.}\ \bibnamefont
  {Bell}}, \bibinfo {author} {\bibfnamefont {G.}~\bibnamefont {Busoni}},
  \bibinfo {author} {\bibfnamefont {M.~E.}\ \bibnamefont {Ramirez-Quezada}},
  \bibinfo {author} {\bibfnamefont {S.}~\bibnamefont {Robles}}, \ and\ \bibinfo
  {author} {\bibfnamefont {M.}~\bibnamefont {Virgato}},\ }\href {\doibase
  10.1088/1475-7516/2021/10/083} {\bibfield  {journal} {\bibinfo  {journal}
  {JCAP}\ }\textbf {\bibinfo {volume} {10}},\ \bibinfo {pages} {083} (\bibinfo
  {year} {2021}{\natexlab{b}})},\ \Eprint {http://arxiv.org/abs/2104.14367}
  {arXiv:2104.14367 [hep-ph]} \BibitemShut {NoStop}%
\bibitem [{\citenamefont {Agashe}\ \emph {et~al.}(2016)\citenamefont {Agashe},
  \citenamefont {Cui}, \citenamefont {Necib},\ and\ \citenamefont
  {Thaler}}]{Agashe:2015xkj}%
  \BibitemOpen
  \bibfield  {author} {\bibinfo {author} {\bibfnamefont {K.}~\bibnamefont
  {Agashe}}, \bibinfo {author} {\bibfnamefont {Y.}~\bibnamefont {Cui}},
  \bibinfo {author} {\bibfnamefont {L.}~\bibnamefont {Necib}}, \ and\ \bibinfo
  {author} {\bibfnamefont {J.}~\bibnamefont {Thaler}},\ }\href {\doibase
  10.1088/1742-6596/718/4/042041} {\bibfield  {journal} {\bibinfo  {journal}
  {J. Phys. Conf. Ser.}\ }\textbf {\bibinfo {volume} {718}},\ \bibinfo {pages}
  {042041} (\bibinfo {year} {2016})},\ \Eprint
  {http://arxiv.org/abs/1512.03782} {arXiv:1512.03782 [hep-ph]} \BibitemShut
  {NoStop}%
\bibitem [{\citenamefont {Cumming}\ \emph {et~al.}(2006)\citenamefont
  {Cumming}, \citenamefont {Macbeth}, \citenamefont {in~'t Zand},\ and\
  \citenamefont {Page}}]{supburst:trecur:CummingZandPage2005}%
  \BibitemOpen
  \bibfield  {author} {\bibinfo {author} {\bibfnamefont {A.}~\bibnamefont
  {Cumming}}, \bibinfo {author} {\bibfnamefont {J.}~\bibnamefont {Macbeth}},
  \bibinfo {author} {\bibfnamefont {J.~J.~M.}\ \bibnamefont {in~'t Zand}}, \
  and\ \bibinfo {author} {\bibfnamefont {D.}~\bibnamefont {Page}},\ }\href
  {\doibase 10.1086/504698} {\bibfield  {journal} {\bibinfo  {journal}
  {Astrophys. J.}\ }\textbf {\bibinfo {volume} {646}},\ \bibinfo {pages} {429}
  (\bibinfo {year} {2006})},\ \Eprint {http://arxiv.org/abs/astro-ph/0508432}
  {arXiv:astro-ph/0508432} \BibitemShut {NoStop}%
\bibitem [{\citenamefont {{Wenger}}\ \emph {et~al.}(2000)\citenamefont
  {{Wenger}}, \citenamefont {{Ochsenbein}}, \citenamefont {{Egret}},
  \citenamefont {{Dubois}}, \citenamefont {{Bonnarel}}, \citenamefont
  {{Borde}}, \citenamefont {{Genova}}, \citenamefont {{Jasniewicz}},
  \citenamefont {{Lalo{\"e}}}, \citenamefont {{Lesteven}},\ and\ \citenamefont
  {{Monier}}}]{SIMBAD}%
  \BibitemOpen
  \bibfield  {author} {\bibinfo {author} {\bibfnamefont {M.}~\bibnamefont
  {{Wenger}}}, \bibinfo {author} {\bibfnamefont {F.}~\bibnamefont
  {{Ochsenbein}}}, \bibinfo {author} {\bibfnamefont {D.}~\bibnamefont
  {{Egret}}}, \bibinfo {author} {\bibfnamefont {P.}~\bibnamefont {{Dubois}}},
  \bibinfo {author} {\bibfnamefont {F.}~\bibnamefont {{Bonnarel}}}, \bibinfo
  {author} {\bibfnamefont {S.}~\bibnamefont {{Borde}}}, \bibinfo {author}
  {\bibfnamefont {F.}~\bibnamefont {{Genova}}}, \bibinfo {author}
  {\bibfnamefont {G.}~\bibnamefont {{Jasniewicz}}}, \bibinfo {author}
  {\bibfnamefont {S.}~\bibnamefont {{Lalo{\"e}}}}, \bibinfo {author}
  {\bibfnamefont {S.}~\bibnamefont {{Lesteven}}}, \ and\ \bibinfo {author}
  {\bibfnamefont {R.}~\bibnamefont {{Monier}}},\ }\href {\doibase
  10.1051/aas:2000332} {\bibfield  {journal} {\bibinfo  {journal} {AAPS}\
  }\textbf {\bibinfo {volume} {143}},\ \bibinfo {pages} {9} (\bibinfo {year}
  {2000})},\ \Eprint {http://arxiv.org/abs/astro-ph/0002110}
  {arXiv:astro-ph/0002110 [astro-ph]} \BibitemShut {NoStop}%
\bibitem [{\citenamefont {{Dufour}}\ \emph {et~al.}(2017)\citenamefont
  {{Dufour}}, \citenamefont {{Blouin}}, \citenamefont {{Coutu}}, \citenamefont
  {{Fortin-Archambault}}, \citenamefont {{Thibeault}}, \citenamefont
  {{Bergeron}},\ and\ \citenamefont {{Fontaine}}}]{MontrealWDDatabase}%
  \BibitemOpen
  \bibfield  {author} {\bibinfo {author} {\bibfnamefont {P.}~\bibnamefont
  {{Dufour}}}, \bibinfo {author} {\bibfnamefont {S.}~\bibnamefont {{Blouin}}},
  \bibinfo {author} {\bibfnamefont {S.}~\bibnamefont {{Coutu}}}, \bibinfo
  {author} {\bibfnamefont {M.}~\bibnamefont {{Fortin-Archambault}}}, \bibinfo
  {author} {\bibfnamefont {C.}~\bibnamefont {{Thibeault}}}, \bibinfo {author}
  {\bibfnamefont {P.}~\bibnamefont {{Bergeron}}}, \ and\ \bibinfo {author}
  {\bibfnamefont {G.}~\bibnamefont {{Fontaine}}},\ }in\ \href {\doibase
  10.48550/arXiv.1610.00986} {\emph {\bibinfo {booktitle} {20th European White
  Dwarf Workshop}}},\ \bibinfo {series} {Astronomical Society of the Pacific
  Conference Series}, Vol.\ \bibinfo {volume} {509},\ \bibinfo {editor} {edited
  by\ \bibinfo {editor} {\bibfnamefont {P.~E.}\ \bibnamefont {{Tremblay}}},
  \bibinfo {editor} {\bibfnamefont {B.}~\bibnamefont {{Gaensicke}}}, \ and\
  \bibinfo {editor} {\bibfnamefont {T.}~\bibnamefont {{Marsh}}}}\ (\bibinfo
  {year} {2017})\ p.~\bibinfo {pages} {3},\ \Eprint
  {http://arxiv.org/abs/1610.00986} {arXiv:1610.00986 [astro-ph.SR]}
  \BibitemShut {NoStop}%
\bibitem [{\citenamefont {Cirelli}\ \emph {et~al.}(2011)\citenamefont
  {Cirelli}, \citenamefont {Corcella}, \citenamefont {Hektor}, \citenamefont
  {Hutsi}, \citenamefont {Kadastik}, \citenamefont {Panci}, \citenamefont
  {Raidal}, \citenamefont {Sala},\ and\ \citenamefont
  {Strumia}}]{PPPCookbook:2010xx}%
  \BibitemOpen
  \bibfield  {author} {\bibinfo {author} {\bibfnamefont {M.}~\bibnamefont
  {Cirelli}}, \bibinfo {author} {\bibfnamefont {G.}~\bibnamefont {Corcella}},
  \bibinfo {author} {\bibfnamefont {A.}~\bibnamefont {Hektor}}, \bibinfo
  {author} {\bibfnamefont {G.}~\bibnamefont {Hutsi}}, \bibinfo {author}
  {\bibfnamefont {M.}~\bibnamefont {Kadastik}}, \bibinfo {author}
  {\bibfnamefont {P.}~\bibnamefont {Panci}}, \bibinfo {author} {\bibfnamefont
  {M.}~\bibnamefont {Raidal}}, \bibinfo {author} {\bibfnamefont
  {F.}~\bibnamefont {Sala}}, \ and\ \bibinfo {author} {\bibfnamefont
  {A.}~\bibnamefont {Strumia}},\ }\href {\doibase
  10.1088/1475-7516/2012/10/E01} {\bibfield  {journal} {\bibinfo  {journal}
  {JCAP}\ }\textbf {\bibinfo {volume} {03}},\ \bibinfo {pages} {051} (\bibinfo
  {year} {2011})},\ \bibinfo {note} {[Erratum: JCAP 10, E01 (2012)]},\ \Eprint
  {http://arxiv.org/abs/1012.4515} {arXiv:1012.4515 [hep-ph]} \BibitemShut
  {NoStop}%
\bibitem [{\citenamefont {Padmanabhan}(1990)}]{text:paddytheoryastroII}%
  \BibitemOpen
  \bibfield  {author} {\bibinfo {author} {\bibfnamefont {T.}~\bibnamefont
  {Padmanabhan}},\ }\href
  {https://link.springer.com/book/10.1007/978-3-642-61523-8} {\emph {\bibinfo
  {title} {Stellar Structure and Evolution}}}\ (\bibinfo  {publisher}
  {Springer},\ \bibinfo {year} {1990})\BibitemShut {NoStop}%
\bibitem [{\citenamefont {Bramante}\ \emph
  {et~al.}(2018{\natexlab{b}})\citenamefont {Bramante}, \citenamefont
  {Broerman}, \citenamefont {Lang},\ and\ \citenamefont
  {Raj}}]{Bramante:2018qbc}%
  \BibitemOpen
  \bibfield  {author} {\bibinfo {author} {\bibfnamefont {J.}~\bibnamefont
  {Bramante}}, \bibinfo {author} {\bibfnamefont {B.}~\bibnamefont {Broerman}},
  \bibinfo {author} {\bibfnamefont {R.~F.}\ \bibnamefont {Lang}}, \ and\
  \bibinfo {author} {\bibfnamefont {N.}~\bibnamefont {Raj}},\ }\href {\doibase
  10.1103/PhysRevD.98.083516} {\bibfield  {journal} {\bibinfo  {journal} {Phys.
  Rev. D}\ }\textbf {\bibinfo {volume} {98}},\ \bibinfo {pages} {083516}
  (\bibinfo {year} {2018}{\natexlab{b}})},\ \Eprint
  {http://arxiv.org/abs/1803.08044} {arXiv:1803.08044 [hep-ph]} \BibitemShut
  {NoStop}%
\bibitem [{\citenamefont {Goodman}\ and\ \citenamefont
  {Witten}(1985)}]{GoodmanWitten:1984dc}%
  \BibitemOpen
  \bibfield  {author} {\bibinfo {author} {\bibfnamefont {M.~W.}\ \bibnamefont
  {Goodman}}\ and\ \bibinfo {author} {\bibfnamefont {E.}~\bibnamefont
  {Witten}},\ }\href {\doibase 10.1103/PhysRevD.31.3059} {\bibfield  {journal}
  {\bibinfo  {journal} {Phys. Rev. D}\ }\textbf {\bibinfo {volume} {31}},\
  \bibinfo {pages} {3059} (\bibinfo {year} {1985})}\BibitemShut {NoStop}%
\bibitem [{\citenamefont {Digman}\ \emph {et~al.}(2019)\citenamefont {Digman},
  \citenamefont {Cappiello}, \citenamefont {Beacom}, \citenamefont {Hirata},\
  and\ \citenamefont {Peter}}]{DigmanBarn:2019wdm}%
  \BibitemOpen
  \bibfield  {author} {\bibinfo {author} {\bibfnamefont {M.~C.}\ \bibnamefont
  {Digman}}, \bibinfo {author} {\bibfnamefont {C.~V.}\ \bibnamefont
  {Cappiello}}, \bibinfo {author} {\bibfnamefont {J.~F.}\ \bibnamefont
  {Beacom}}, \bibinfo {author} {\bibfnamefont {C.~M.}\ \bibnamefont {Hirata}},
  \ and\ \bibinfo {author} {\bibfnamefont {A.~H.~G.}\ \bibnamefont {Peter}},\
  }\href {\doibase 10.1103/PhysRevD.100.063013} {\bibfield  {journal} {\bibinfo
   {journal} {Phys. Rev. D}\ }\textbf {\bibinfo {volume} {100}},\ \bibinfo
  {pages} {063013} (\bibinfo {year} {2019})},\ \bibinfo {note} {[Erratum:
  Phys.Rev.D 106, 089902 (2022)]},\ \Eprint {http://arxiv.org/abs/1907.10618}
  {arXiv:1907.10618 [hep-ph]} \BibitemShut {NoStop}%
\bibitem [{\citenamefont {Xu}\ and\ \citenamefont
  {Farrar}(2023)}]{resonantXuFarrar:2020qjk}%
  \BibitemOpen
  \bibfield  {author} {\bibinfo {author} {\bibfnamefont {X.}~\bibnamefont
  {Xu}}\ and\ \bibinfo {author} {\bibfnamefont {G.~R.}\ \bibnamefont
  {Farrar}},\ }\href {\doibase 10.1103/PhysRevD.107.095028} {\bibfield
  {journal} {\bibinfo  {journal} {Phys. Rev. D}\ }\textbf {\bibinfo {volume}
  {107}},\ \bibinfo {pages} {095028} (\bibinfo {year} {2023})},\ \Eprint
  {http://arxiv.org/abs/2101.00142} {arXiv:2101.00142 [hep-ph]} \BibitemShut
  {NoStop}%
\bibitem [{\citenamefont {Bhoonah}\ \emph {et~al.}(2021)\citenamefont
  {Bhoonah}, \citenamefont {Bramante}, \citenamefont {Courtman},\ and\
  \citenamefont {Song}}]{Bhoonah:2020fys}%
  \BibitemOpen
  \bibfield  {author} {\bibinfo {author} {\bibfnamefont {A.}~\bibnamefont
  {Bhoonah}}, \bibinfo {author} {\bibfnamefont {J.}~\bibnamefont {Bramante}},
  \bibinfo {author} {\bibfnamefont {B.}~\bibnamefont {Courtman}}, \ and\
  \bibinfo {author} {\bibfnamefont {N.}~\bibnamefont {Song}},\ }\href {\doibase
  10.1103/PhysRevD.103.103001} {\bibfield  {journal} {\bibinfo  {journal}
  {Phys. Rev. D}\ }\textbf {\bibinfo {volume} {103}},\ \bibinfo {pages}
  {103001} (\bibinfo {year} {2021})},\ \Eprint
  {http://arxiv.org/abs/2012.13406} {arXiv:2012.13406 [hep-ph]} \BibitemShut
  {NoStop}%
\bibitem [{\citenamefont {Bhoonah}\ \emph {et~al.}(2019)\citenamefont
  {Bhoonah}, \citenamefont {Bramante}, \citenamefont {Elahi},\ and\
  \citenamefont {Schon}}]{gasclouds:Bhoonah:2018gjb}%
  \BibitemOpen
  \bibfield  {author} {\bibinfo {author} {\bibfnamefont {A.}~\bibnamefont
  {Bhoonah}}, \bibinfo {author} {\bibfnamefont {J.}~\bibnamefont {Bramante}},
  \bibinfo {author} {\bibfnamefont {F.}~\bibnamefont {Elahi}}, \ and\ \bibinfo
  {author} {\bibfnamefont {S.}~\bibnamefont {Schon}},\ }\href {\doibase
  10.1103/PhysRevD.100.023001} {\bibfield  {journal} {\bibinfo  {journal}
  {Phys. Rev. D}\ }\textbf {\bibinfo {volume} {100}},\ \bibinfo {pages}
  {023001} (\bibinfo {year} {2019})},\ \Eprint
  {http://arxiv.org/abs/1812.10919} {arXiv:1812.10919 [hep-ph]} \BibitemShut
  {NoStop}%
\bibitem [{\citenamefont {{Napiwotzki}}(2009)}]{WDdistribs:Napiwotzki}%
  \BibitemOpen
  \bibfield  {author} {\bibinfo {author} {\bibfnamefont {R.}~\bibnamefont
  {{Napiwotzki}}},\ }in\ \href {\doibase 10.1088/1742-6596/172/1/012004} {\emph
  {\bibinfo {booktitle} {Journal of Physics Conference Series}}},\ \bibinfo
  {series} {Journal of Physics Conference Series}, Vol.\ \bibinfo {volume}
  {172}\ (\bibinfo {year} {2009})\ p.\ \bibinfo {pages} {012004},\ \Eprint
  {http://arxiv.org/abs/0903.2159} {arXiv:0903.2159 [astro-ph.SR]} \BibitemShut
  {NoStop}%
\bibitem [{\citenamefont {{Blaes}}\ and\ \citenamefont
  {{Madau}}(1993)}]{NSdistribs:BlaesMadau}%
  \BibitemOpen
  \bibfield  {author} {\bibinfo {author} {\bibfnamefont {O.}~\bibnamefont
  {{Blaes}}}\ and\ \bibinfo {author} {\bibfnamefont {P.}~\bibnamefont
  {{Madau}}},\ }\href {\doibase 10.1086/172240} {\bibfield  {journal} {\bibinfo
   {journal} {APJ}\ }\textbf {\bibinfo {volume} {403}},\ \bibinfo {pages} {690}
  (\bibinfo {year} {1993})}\BibitemShut {NoStop}%
\bibitem [{\citenamefont {Sartore}\ \emph {et~al.}(2010)\citenamefont
  {Sartore}, \citenamefont {Ripamonti}, \citenamefont {Treves},\ and\
  \citenamefont {Turolla}}]{NSdistribs:Sartore2010}%
  \BibitemOpen
  \bibfield  {author} {\bibinfo {author} {\bibfnamefont {N.}~\bibnamefont
  {Sartore}}, \bibinfo {author} {\bibfnamefont {E.}~\bibnamefont {Ripamonti}},
  \bibinfo {author} {\bibfnamefont {A.}~\bibnamefont {Treves}}, \ and\ \bibinfo
  {author} {\bibfnamefont {R.}~\bibnamefont {Turolla}},\ }\href {\doibase
  10.1051/0004-6361/200912222} {\bibfield  {journal} {\bibinfo  {journal}
  {Astronomy and Astrophysics}\ }\textbf {\bibinfo {volume} {510}},\ \bibinfo
  {pages} {A23} (\bibinfo {year} {2010})}\BibitemShut {NoStop}%
\bibitem [{\citenamefont {Guillot}\ \emph {et~al.}(2019)\citenamefont
  {Guillot}, \citenamefont {Pavlov}, \citenamefont {Reyes}, \citenamefont
  {Reisenegger}, \citenamefont {Rodriguez}, \citenamefont {Rangelov},\ and\
  \citenamefont {Kargaltsev}}]{coldestNSHST}%
  \BibitemOpen
  \bibfield  {author} {\bibinfo {author} {\bibfnamefont {S.}~\bibnamefont
  {Guillot}}, \bibinfo {author} {\bibfnamefont {G.~G.}\ \bibnamefont {Pavlov}},
  \bibinfo {author} {\bibfnamefont {C.}~\bibnamefont {Reyes}}, \bibinfo
  {author} {\bibfnamefont {A.}~\bibnamefont {Reisenegger}}, \bibinfo {author}
  {\bibfnamefont {L.}~\bibnamefont {Rodriguez}}, \bibinfo {author}
  {\bibfnamefont {B.}~\bibnamefont {Rangelov}}, \ and\ \bibinfo {author}
  {\bibfnamefont {O.}~\bibnamefont {Kargaltsev}},\ }\href {\doibase
  10.3847/1538-4357/ab0f38} {\bibfield  {journal} {\bibinfo  {journal}
  {Astrophys. J.}\ }\textbf {\bibinfo {volume} {874}},\ \bibinfo {pages} {175}
  (\bibinfo {year} {2019})},\ \Eprint {http://arxiv.org/abs/1901.07998}
  {arXiv:1901.07998 [astro-ph.HE]} \BibitemShut {NoStop}%
\bibitem [{\citenamefont {Kippenhahn}\ and\ \citenamefont
  {Weigert}(2010)}]{text:kippenhahnweigert}%
  \BibitemOpen
  \bibfield  {author} {\bibinfo {author} {\bibfnamefont {R.}~\bibnamefont
  {Kippenhahn}}\ and\ \bibinfo {author} {\bibfnamefont {A.}~\bibnamefont
  {Weigert}},\ }\href
  {https://books.google.co.in/books/about/Theoretical_Astrophysics_Volume_2_Stars.html?id=TOjwtYYb63cC&redir_esc=y}
  {\emph {\bibinfo {title} {Theoretical Astrophysics II: Stars and Stellar
  Systems}}}\ (\bibinfo  {publisher} {Cambridge University Press},\ \bibinfo
  {year} {2010})\BibitemShut {NoStop}%
\bibitem [{\citenamefont {{Shapiro}}\ and\ \citenamefont
  {{Teukolsky}}(1983)}]{text:ShapiroTeukolsky}%
  \BibitemOpen
  \bibfield  {author} {\bibinfo {author} {\bibfnamefont {S.~L.}\ \bibnamefont
  {{Shapiro}}}\ and\ \bibinfo {author} {\bibfnamefont {S.~A.}\ \bibnamefont
  {{Teukolsky}}},\ }\href@noop {} {\emph {\bibinfo {title} {{Black holes, white
  dwarfs, and neutron stars : the physics of compact objects}}}}\ (\bibinfo
  {year} {1983})\BibitemShut {NoStop}%
\bibitem [{\citenamefont {Kawasaki}\ \emph {et~al.}(1992)\citenamefont
  {Kawasaki}, \citenamefont {Murayama},\ and\ \citenamefont
  {Yanagida}}]{GlobalWarming:Kawasaki:1991eu}%
  \BibitemOpen
  \bibfield  {author} {\bibinfo {author} {\bibfnamefont {M.}~\bibnamefont
  {Kawasaki}}, \bibinfo {author} {\bibfnamefont {H.}~\bibnamefont {Murayama}},
  \ and\ \bibinfo {author} {\bibfnamefont {T.}~\bibnamefont {Yanagida}},\
  }\href {\doibase 10.1143/PTP.87.685} {\bibfield  {journal} {\bibinfo
  {journal} {Prog. Theor. Phys.}\ }\textbf {\bibinfo {volume} {87}},\ \bibinfo
  {pages} {685} (\bibinfo {year} {1992})}\BibitemShut {NoStop}%
\bibitem [{\citenamefont {Abbas}\ and\ \citenamefont
  {Abbas}(1998)}]{GlobalWarming:Abbas:1996kk}%
  \BibitemOpen
  \bibfield  {author} {\bibinfo {author} {\bibfnamefont {S.}~\bibnamefont
  {Abbas}}\ and\ \bibinfo {author} {\bibfnamefont {A.}~\bibnamefont {Abbas}},\
  }\href {\doibase 10.1016/S0927-6505(97)00051-0} {\bibfield  {journal}
  {\bibinfo  {journal} {Astropart. Phys.}\ }\textbf {\bibinfo {volume} {8}},\
  \bibinfo {pages} {317} (\bibinfo {year} {1998})},\ \Eprint
  {http://arxiv.org/abs/astro-ph/9612214} {arXiv:astro-ph/9612214} \BibitemShut
  {NoStop}%
\bibitem [{\citenamefont {Mack}\ \emph {et~al.}(2007)\citenamefont {Mack},
  \citenamefont {Beacom},\ and\ \citenamefont
  {Bertone}}]{GlobalWarming:Mack:2007xj}%
  \BibitemOpen
  \bibfield  {author} {\bibinfo {author} {\bibfnamefont {G.~D.}\ \bibnamefont
  {Mack}}, \bibinfo {author} {\bibfnamefont {J.~F.}\ \bibnamefont {Beacom}}, \
  and\ \bibinfo {author} {\bibfnamefont {G.}~\bibnamefont {Bertone}},\ }\href
  {\doibase 10.1103/PhysRevD.76.043523} {\bibfield  {journal} {\bibinfo
  {journal} {Phys. Rev. D}\ }\textbf {\bibinfo {volume} {76}},\ \bibinfo
  {pages} {043523} (\bibinfo {year} {2007})},\ \Eprint
  {http://arxiv.org/abs/0705.4298} {arXiv:0705.4298 [astro-ph]} \BibitemShut
  {NoStop}%
\bibitem [{\citenamefont {Hooper}\ and\ \citenamefont
  {Steffen}(2012)}]{GlobalWarming:Hooper:2011dw}%
  \BibitemOpen
  \bibfield  {author} {\bibinfo {author} {\bibfnamefont {D.}~\bibnamefont
  {Hooper}}\ and\ \bibinfo {author} {\bibfnamefont {J.~H.}\ \bibnamefont
  {Steffen}},\ }\href {\doibase 10.1088/1475-7516/2012/07/046} {\bibfield
  {journal} {\bibinfo  {journal} {JCAP}\ }\textbf {\bibinfo {volume} {07}},\
  \bibinfo {pages} {046} (\bibinfo {year} {2012})},\ \Eprint
  {http://arxiv.org/abs/1103.5086} {arXiv:1103.5086 [astro-ph.EP]} \BibitemShut
  {NoStop}%
\bibitem [{\citenamefont {Garani}\ and\ \citenamefont
  {Tinyakov}(2020)}]{GlobalWarming:Garani:2019rcb}%
  \BibitemOpen
  \bibfield  {author} {\bibinfo {author} {\bibfnamefont {R.}~\bibnamefont
  {Garani}}\ and\ \bibinfo {author} {\bibfnamefont {P.}~\bibnamefont
  {Tinyakov}},\ }\href {\doibase 10.1016/j.physletb.2020.135403} {\bibfield
  {journal} {\bibinfo  {journal} {Phys. Lett. B}\ }\textbf {\bibinfo {volume}
  {804}},\ \bibinfo {pages} {135403} (\bibinfo {year} {2020})},\ \Eprint
  {http://arxiv.org/abs/1912.00443} {arXiv:1912.00443 [hep-ph]} \BibitemShut
  {NoStop}%
\bibitem [{\citenamefont {Bramante}\ \emph {et~al.}(2020)\citenamefont
  {Bramante}, \citenamefont {Buchanan}, \citenamefont {Goodman},\ and\
  \citenamefont {Lodhi}}]{GlobalWarming:Bramante:2019fhi}%
  \BibitemOpen
  \bibfield  {author} {\bibinfo {author} {\bibfnamefont {J.}~\bibnamefont
  {Bramante}}, \bibinfo {author} {\bibfnamefont {A.}~\bibnamefont {Buchanan}},
  \bibinfo {author} {\bibfnamefont {A.}~\bibnamefont {Goodman}}, \ and\
  \bibinfo {author} {\bibfnamefont {E.}~\bibnamefont {Lodhi}},\ }\href
  {\doibase 10.1103/PhysRevD.101.043001} {\bibfield  {journal} {\bibinfo
  {journal} {Phys. Rev. D}\ }\textbf {\bibinfo {volume} {101}},\ \bibinfo
  {pages} {043001} (\bibinfo {year} {2020})},\ \Eprint
  {http://arxiv.org/abs/1909.11683} {arXiv:1909.11683 [hep-ph]} \BibitemShut
  {NoStop}%
\bibitem [{\citenamefont {Acevedo}\ \emph
  {et~al.}(2021{\natexlab{d}})\citenamefont {Acevedo}, \citenamefont
  {Bramante}, \citenamefont {Goodman}, \citenamefont {Kopp},\ and\
  \citenamefont {Opferkuch}}]{GlobalWarming:Acevedo:2020gro}%
  \BibitemOpen
  \bibfield  {author} {\bibinfo {author} {\bibfnamefont {J.~F.}\ \bibnamefont
  {Acevedo}}, \bibinfo {author} {\bibfnamefont {J.}~\bibnamefont {Bramante}},
  \bibinfo {author} {\bibfnamefont {A.}~\bibnamefont {Goodman}}, \bibinfo
  {author} {\bibfnamefont {J.}~\bibnamefont {Kopp}}, \ and\ \bibinfo {author}
  {\bibfnamefont {T.}~\bibnamefont {Opferkuch}},\ }\href {\doibase
  10.1088/1475-7516/2021/04/026} {\bibfield  {journal} {\bibinfo  {journal}
  {JCAP}\ }\textbf {\bibinfo {volume} {04}},\ \bibinfo {pages} {026} (\bibinfo
  {year} {2021}{\natexlab{d}})},\ \Eprint {http://arxiv.org/abs/2012.09176}
  {arXiv:2012.09176 [hep-ph]} \BibitemShut {NoStop}%
\bibitem [{\citenamefont {Leane}\ and\ \citenamefont
  {Smirnov}(2021)}]{GlobalWarming:Leane:2020wob}%
  \BibitemOpen
  \bibfield  {author} {\bibinfo {author} {\bibfnamefont {R.~K.}\ \bibnamefont
  {Leane}}\ and\ \bibinfo {author} {\bibfnamefont {J.}~\bibnamefont
  {Smirnov}},\ }\href {\doibase 10.1103/PhysRevLett.126.161101} {\bibfield
  {journal} {\bibinfo  {journal} {Phys. Rev. Lett.}\ }\textbf {\bibinfo
  {volume} {126}},\ \bibinfo {pages} {161101} (\bibinfo {year} {2021})},\
  \Eprint {http://arxiv.org/abs/2010.00015} {arXiv:2010.00015 [hep-ph]}
  \BibitemShut {NoStop}%
\bibitem [{\citenamefont {Bramante}\ \emph
  {et~al.}(2022{\natexlab{b}})\citenamefont {Bramante}, \citenamefont {Kumar},
  \citenamefont {Mohlabeng}, \citenamefont {Raj},\ and\ \citenamefont
  {Song}}]{GlobalWarming:Bramante:2022pmn}%
  \BibitemOpen
  \bibfield  {author} {\bibinfo {author} {\bibfnamefont {J.}~\bibnamefont
  {Bramante}}, \bibinfo {author} {\bibfnamefont {J.}~\bibnamefont {Kumar}},
  \bibinfo {author} {\bibfnamefont {G.}~\bibnamefont {Mohlabeng}}, \bibinfo
  {author} {\bibfnamefont {N.}~\bibnamefont {Raj}}, \ and\ \bibinfo {author}
  {\bibfnamefont {N.}~\bibnamefont {Song}},\ }\href@noop {} {\  (\bibinfo
  {year} {2022}{\natexlab{b}})},\ \Eprint {http://arxiv.org/abs/2210.01812}
  {arXiv:2210.01812 [hep-ph]} \BibitemShut {NoStop}%
\bibitem [{\citenamefont {McKeen}\ \emph
  {et~al.}(2021{\natexlab{a}})\citenamefont {McKeen}, \citenamefont
  {Pospelov},\ and\ \citenamefont {Raj}}]{McKeen:2020oyr}%
  \BibitemOpen
  \bibfield  {author} {\bibinfo {author} {\bibfnamefont {D.}~\bibnamefont
  {McKeen}}, \bibinfo {author} {\bibfnamefont {M.}~\bibnamefont {Pospelov}}, \
  and\ \bibinfo {author} {\bibfnamefont {N.}~\bibnamefont {Raj}},\ }\href
  {\doibase 10.1103/PhysRevD.103.115002} {\bibfield  {journal} {\bibinfo
  {journal} {Phys. Rev. D}\ }\textbf {\bibinfo {volume} {103}},\ \bibinfo
  {pages} {115002} (\bibinfo {year} {2021}{\natexlab{a}})},\ \Eprint
  {http://arxiv.org/abs/2012.09865} {arXiv:2012.09865 [hep-ph]} \BibitemShut
  {NoStop}%
\bibitem [{\citenamefont {McKeen}\ \emph
  {et~al.}(2021{\natexlab{b}})\citenamefont {McKeen}, \citenamefont
  {Pospelov},\ and\ \citenamefont {Raj}}]{McKeen:2021jbh}%
  \BibitemOpen
  \bibfield  {author} {\bibinfo {author} {\bibfnamefont {D.}~\bibnamefont
  {McKeen}}, \bibinfo {author} {\bibfnamefont {M.}~\bibnamefont {Pospelov}}, \
  and\ \bibinfo {author} {\bibfnamefont {N.}~\bibnamefont {Raj}},\ }\href
  {\doibase 10.1103/PhysRevLett.127.061805} {\bibfield  {journal} {\bibinfo
  {journal} {Phys. Rev. Lett.}\ }\textbf {\bibinfo {volume} {127}},\ \bibinfo
  {pages} {061805} (\bibinfo {year} {2021}{\natexlab{b}})},\ \Eprint
  {http://arxiv.org/abs/2105.09951} {arXiv:2105.09951 [hep-ph]} \BibitemShut
  {NoStop}%
\bibitem [{\citenamefont {Team}(2019)}]{theluvoirteam2019luvoir}%
  \BibitemOpen
  \bibfield  {author} {\bibinfo {author} {\bibfnamefont {T.~L.}\ \bibnamefont
  {Team}},\ }\href@noop {} {\enquote {\bibinfo {title} {The luvoir mission
  concept study final report},}\ } (\bibinfo {year} {2019}),\ \Eprint
  {http://arxiv.org/abs/1912.06219} {arXiv:1912.06219 [astro-ph.IM]}
  \BibitemShut {NoStop}%
\bibitem [{\citenamefont {Diehl}\ \emph {et~al.}(2019)\citenamefont {Diehl}
  \emph {et~al.}}]{DES:2019rtl}%
  \BibitemOpen
  \bibfield  {author} {\bibinfo {author} {\bibfnamefont {H.~T.}\ \bibnamefont
  {Diehl}} \emph {et~al.} (\bibinfo {collaboration} {DES}),\ }\href {\doibase
  10.2172/1596042} {\  (\bibinfo {year} {2019}),\ 10.2172/1596042}\BibitemShut
  {NoStop}%
\bibitem [{\citenamefont {Peterson}(2016)}]{Rubin1}%
  \BibitemOpen
  \bibfield  {author} {\bibinfo {author} {\bibfnamefont {J.~R.}\ \bibnamefont
  {Peterson}},\ }\href {\doibase 10.2172/1272167} {\  (\bibinfo {year}
  {2016}),\ 10.2172/1272167}\BibitemShut {NoStop}%
\bibitem [{\citenamefont {Ivezi\'c}\ \emph {et~al.}(2019)\citenamefont
  {Ivezi\'c} \emph {et~al.}}]{Rubin2}%
  \BibitemOpen
  \bibfield  {author} {\bibinfo {author} {\bibfnamefont {v.}~\bibnamefont
  {Ivezi\'c}} \emph {et~al.} (\bibinfo {collaboration} {LSST}),\ }\href
  {\doibase 10.3847/1538-4357/ab042c} {\bibfield  {journal} {\bibinfo
  {journal} {Astrophys. J.}\ }\textbf {\bibinfo {volume} {873}},\ \bibinfo
  {pages} {111} (\bibinfo {year} {2019})},\ \Eprint
  {http://arxiv.org/abs/0805.2366} {arXiv:0805.2366 [astro-ph]} \BibitemShut
  {NoStop}%
\bibitem [{\citenamefont {Gardner}\ \emph {et~al.}(2006)\citenamefont {Gardner}
  \emph {et~al.}}]{JWST:Gardner:2006ky}%
  \BibitemOpen
  \bibfield  {author} {\bibinfo {author} {\bibfnamefont {J.~P.}\ \bibnamefont
  {Gardner}} \emph {et~al.},\ }\href {\doibase 10.1007/s11214-006-8315-7}
  {\bibfield  {journal} {\bibinfo  {journal} {Space Sci. Rev.}\ }\textbf
  {\bibinfo {volume} {123}},\ \bibinfo {pages} {485} (\bibinfo {year}
  {2006})},\ \Eprint {http://arxiv.org/abs/astro-ph/0606175}
  {arXiv:astro-ph/0606175} \BibitemShut {NoStop}%
\bibitem [{\citenamefont {Neichel}\ \emph {et~al.}(2018)\citenamefont
  {Neichel}, \citenamefont {Mouillet}, \citenamefont {Gendron}, \citenamefont
  {Correia}, \citenamefont {Sauvage},\ and\ \citenamefont
  {Fusco}}]{ELT:neichel2018overview}%
  \BibitemOpen
  \bibfield  {author} {\bibinfo {author} {\bibfnamefont {B.}~\bibnamefont
  {Neichel}}, \bibinfo {author} {\bibfnamefont {D.}~\bibnamefont {Mouillet}},
  \bibinfo {author} {\bibfnamefont {E.}~\bibnamefont {Gendron}}, \bibinfo
  {author} {\bibfnamefont {C.}~\bibnamefont {Correia}}, \bibinfo {author}
  {\bibfnamefont {J.~F.}\ \bibnamefont {Sauvage}}, \ and\ \bibinfo {author}
  {\bibfnamefont {T.}~\bibnamefont {Fusco}},\ }\href@noop {} {\enquote
  {\bibinfo {title} {Overview of the european extremely large telescope and its
  instrument suite},}\ } (\bibinfo {year} {2018}),\ \Eprint
  {http://arxiv.org/abs/1812.06639} {arXiv:1812.06639 [astro-ph.IM]}
  \BibitemShut {NoStop}%
\bibitem [{\citenamefont {Skidmore}\ \emph {et~al.}(2015)\citenamefont
  {Skidmore} \emph {et~al.}}]{TMT:2015pvw}%
  \BibitemOpen
  \bibfield  {author} {\bibinfo {author} {\bibfnamefont {W.}~\bibnamefont
  {Skidmore}} \emph {et~al.} (\bibinfo {collaboration} {TMT International
  Science Development Teams \& TMT Science Advisory Committee}),\ }\href
  {\doibase 10.1088/1674-4527/15/12/001} {\bibfield  {journal} {\bibinfo
  {journal} {Res. Astron. Astrophys.}\ }\textbf {\bibinfo {volume} {15}},\
  \bibinfo {pages} {1945} (\bibinfo {year} {2015})},\ \Eprint
  {http://arxiv.org/abs/1505.01195} {arXiv:1505.01195 [astro-ph.IM]}
  \BibitemShut {NoStop}%
\bibitem [{\citenamefont {J.~Green}(2012)}]{green2012widefield}%
  \BibitemOpen
  \bibfield  {author} {\bibinfo {author} {\bibfnamefont {e.~a.}\ \bibnamefont
  {J.~Green}},\ }\href@noop {} {\enquote {\bibinfo {title} {Wide-field infrared
  survey telescope (wfirst) final report},}\ } (\bibinfo {year} {2012}),\
  \Eprint {http://arxiv.org/abs/1208.4012} {arXiv:1208.4012 [astro-ph.IM]}
  \BibitemShut {NoStop}%
\bibitem [{\citenamefont {De~Mitri}\ \emph {et~al.}(2002)\citenamefont
  {De~Mitri} \emph {et~al.}}]{MACRO:2000cdb}%
  \BibitemOpen
  \bibfield  {author} {\bibinfo {author} {\bibfnamefont {I.}~\bibnamefont
  {De~Mitri}} \emph {et~al.} (\bibinfo {collaboration} {MACRO}),\ }\href
  {\doibase 10.1016/S0920-5632(02)01479-2} {\bibfield  {journal} {\bibinfo
  {journal} {Nucl. Phys. B Proc. Suppl.}\ }\textbf {\bibinfo {volume} {110}},\
  \bibinfo {pages} {186} (\bibinfo {year} {2002})},\ \Eprint
  {http://arxiv.org/abs/hep-ex/0009002} {arXiv:hep-ex/0009002} \BibitemShut
  {NoStop}%
\bibitem [{\citenamefont {Bramante}\ \emph
  {et~al.}(2019{\natexlab{c}})\citenamefont {Bramante}, \citenamefont
  {Broerman}, \citenamefont {Kumar}, \citenamefont {Lang}, \citenamefont
  {Pospelov},\ and\ \citenamefont {Raj}}]{Bramante:2018tos}%
  \BibitemOpen
  \bibfield  {author} {\bibinfo {author} {\bibfnamefont {J.}~\bibnamefont
  {Bramante}}, \bibinfo {author} {\bibfnamefont {B.}~\bibnamefont {Broerman}},
  \bibinfo {author} {\bibfnamefont {J.}~\bibnamefont {Kumar}}, \bibinfo
  {author} {\bibfnamefont {R.~F.}\ \bibnamefont {Lang}}, \bibinfo {author}
  {\bibfnamefont {M.}~\bibnamefont {Pospelov}}, \ and\ \bibinfo {author}
  {\bibfnamefont {N.}~\bibnamefont {Raj}},\ }\href {\doibase
  10.1103/PhysRevD.99.083010} {\bibfield  {journal} {\bibinfo  {journal} {Phys.
  Rev. D}\ }\textbf {\bibinfo {volume} {99}},\ \bibinfo {pages} {083010}
  (\bibinfo {year} {2019}{\natexlab{c}})},\ \Eprint
  {http://arxiv.org/abs/1812.09325} {arXiv:1812.09325 [hep-ph]} \BibitemShut
  {NoStop}%
\bibitem [{\citenamefont {Bramante}\ \emph
  {et~al.}(2019{\natexlab{d}})\citenamefont {Bramante}, \citenamefont {Kumar},\
  and\ \citenamefont {Raj}}]{Bramante:2019yss}%
  \BibitemOpen
  \bibfield  {author} {\bibinfo {author} {\bibfnamefont {J.}~\bibnamefont
  {Bramante}}, \bibinfo {author} {\bibfnamefont {J.}~\bibnamefont {Kumar}}, \
  and\ \bibinfo {author} {\bibfnamefont {N.}~\bibnamefont {Raj}},\ }\href
  {\doibase 10.1103/PhysRevD.100.123016} {\bibfield  {journal} {\bibinfo
  {journal} {Phys. Rev. D}\ }\textbf {\bibinfo {volume} {100}},\ \bibinfo
  {pages} {123016} (\bibinfo {year} {2019}{\natexlab{d}})},\ \Eprint
  {http://arxiv.org/abs/1910.05380} {arXiv:1910.05380 [hep-ph]} \BibitemShut
  {NoStop}%
\bibitem [{\citenamefont {Cappiello}\ \emph {et~al.}(2021)\citenamefont
  {Cappiello}, \citenamefont {Collar},\ and\ \citenamefont
  {Beacom}}]{Cappiello:2020lbk}%
  \BibitemOpen
  \bibfield  {author} {\bibinfo {author} {\bibfnamefont {C.~V.}\ \bibnamefont
  {Cappiello}}, \bibinfo {author} {\bibfnamefont {J.~I.}\ \bibnamefont
  {Collar}}, \ and\ \bibinfo {author} {\bibfnamefont {J.~F.}\ \bibnamefont
  {Beacom}},\ }\href {\doibase 10.1103/PhysRevD.103.023019} {\bibfield
  {journal} {\bibinfo  {journal} {Phys. Rev. D}\ }\textbf {\bibinfo {volume}
  {103}},\ \bibinfo {pages} {023019} (\bibinfo {year} {2021})},\ \Eprint
  {http://arxiv.org/abs/2008.10646} {arXiv:2008.10646 [hep-ex]} \BibitemShut
  {NoStop}%
\bibitem [{\citenamefont {Bai}\ \emph {et~al.}(2022)\citenamefont {Bai},
  \citenamefont {Berger},\ and\ \citenamefont {Korwar}}]{Bai:2022nsv}%
  \BibitemOpen
  \bibfield  {author} {\bibinfo {author} {\bibfnamefont {Y.}~\bibnamefont
  {Bai}}, \bibinfo {author} {\bibfnamefont {J.}~\bibnamefont {Berger}}, \ and\
  \bibinfo {author} {\bibfnamefont {M.}~\bibnamefont {Korwar}},\ }\href
  {\doibase 10.1007/JHEP11(2022)079} {\bibfield  {journal} {\bibinfo  {journal}
  {JHEP}\ }\textbf {\bibinfo {volume} {11}},\ \bibinfo {pages} {079} (\bibinfo
  {year} {2022})},\ \Eprint {http://arxiv.org/abs/2206.07928} {arXiv:2206.07928
  [hep-ph]} \BibitemShut {NoStop}%
\bibitem [{\citenamefont {Aprile}\ \emph {et~al.}(2023)\citenamefont {Aprile}
  \emph {et~al.}}]{XENON:2023ysy}%
  \BibitemOpen
  \bibfield  {author} {\bibinfo {author} {\bibfnamefont {E.}~\bibnamefont
  {Aprile}} \emph {et~al.} (\bibinfo {collaboration} {XENON}),\ }\href@noop {}
  {\  (\bibinfo {year} {2023})},\ \Eprint {http://arxiv.org/abs/2304.10931}
  {arXiv:2304.10931 [hep-ex]} \BibitemShut {NoStop}%
\bibitem [{\citenamefont {Snowden-Ifft}\ \emph {et~al.}(1995)\citenamefont
  {Snowden-Ifft}, \citenamefont {Freeman},\ and\ \citenamefont
  {Price}}]{Snowden-Ifft:1995zgn}%
  \BibitemOpen
  \bibfield  {author} {\bibinfo {author} {\bibfnamefont {D.~P.}\ \bibnamefont
  {Snowden-Ifft}}, \bibinfo {author} {\bibfnamefont {E.~S.}\ \bibnamefont
  {Freeman}}, \ and\ \bibinfo {author} {\bibfnamefont {P.~B.}\ \bibnamefont
  {Price}},\ }\href {\doibase 10.1103/PhysRevLett.74.4133} {\bibfield
  {journal} {\bibinfo  {journal} {Phys. Rev. Lett.}\ }\textbf {\bibinfo
  {volume} {74}},\ \bibinfo {pages} {4133} (\bibinfo {year}
  {1995})}\BibitemShut {NoStop}%
\bibitem [{\citenamefont {Baum}\ \emph {et~al.}(2020)\citenamefont {Baum},
  \citenamefont {Drukier}, \citenamefont {Freese}, \citenamefont {G\'orski},\
  and\ \citenamefont {Stengel}}]{Baum:2018tfw}%
  \BibitemOpen
  \bibfield  {author} {\bibinfo {author} {\bibfnamefont {S.}~\bibnamefont
  {Baum}}, \bibinfo {author} {\bibfnamefont {A.~K.}\ \bibnamefont {Drukier}},
  \bibinfo {author} {\bibfnamefont {K.}~\bibnamefont {Freese}}, \bibinfo
  {author} {\bibfnamefont {M.}~\bibnamefont {G\'orski}}, \ and\ \bibinfo
  {author} {\bibfnamefont {P.}~\bibnamefont {Stengel}},\ }\href {\doibase
  10.1016/j.physletb.2020.135325} {\bibfield  {journal} {\bibinfo  {journal}
  {Phys. Lett. B}\ }\textbf {\bibinfo {volume} {803}},\ \bibinfo {pages}
  {135325} (\bibinfo {year} {2020})},\ \Eprint
  {http://arxiv.org/abs/1806.05991} {arXiv:1806.05991 [astro-ph.CO]}
  \BibitemShut {NoStop}%
\bibitem [{\citenamefont {Drukier}\ \emph {et~al.}(2019)\citenamefont
  {Drukier}, \citenamefont {Baum}, \citenamefont {Freese}, \citenamefont
  {G\'orski},\ and\ \citenamefont {Stengel}}]{Drukier:2018pdy}%
  \BibitemOpen
  \bibfield  {author} {\bibinfo {author} {\bibfnamefont {A.~K.}\ \bibnamefont
  {Drukier}}, \bibinfo {author} {\bibfnamefont {S.}~\bibnamefont {Baum}},
  \bibinfo {author} {\bibfnamefont {K.}~\bibnamefont {Freese}}, \bibinfo
  {author} {\bibfnamefont {M.}~\bibnamefont {G\'orski}}, \ and\ \bibinfo
  {author} {\bibfnamefont {P.}~\bibnamefont {Stengel}},\ }\href {\doibase
  10.1103/PhysRevD.99.043014} {\bibfield  {journal} {\bibinfo  {journal} {Phys.
  Rev. D}\ }\textbf {\bibinfo {volume} {99}},\ \bibinfo {pages} {043014}
  (\bibinfo {year} {2019})},\ \Eprint {http://arxiv.org/abs/1811.06844}
  {arXiv:1811.06844 [astro-ph.CO]} \BibitemShut {NoStop}%
\bibitem [{\citenamefont {Baum}\ \emph {et~al.}(2023)\citenamefont {Baum} \emph
  {et~al.}}]{Baum:2023cct}%
  \BibitemOpen
  \bibfield  {author} {\bibinfo {author} {\bibfnamefont {S.}~\bibnamefont
  {Baum}} \emph {et~al.},\ }\href {\doibase 10.1016/j.dark.2023.101245}
  {\bibfield  {journal} {\bibinfo  {journal} {Phys. Dark Univ.}\ }\textbf
  {\bibinfo {volume} {41}},\ \bibinfo {pages} {101245} (\bibinfo {year}
  {2023})},\ \Eprint {http://arxiv.org/abs/2301.07118} {arXiv:2301.07118
  [astro-ph.IM]} \BibitemShut {NoStop}%
\bibitem [{\citenamefont {Singh~Sidhu}\ \emph
  {et~al.}(2019{\natexlab{a}})\citenamefont {Singh~Sidhu}, \citenamefont
  {Abraham}, \citenamefont {Covault},\ and\ \citenamefont
  {Starkman}}]{SinghSidhu:2018oqs}%
  \BibitemOpen
  \bibfield  {author} {\bibinfo {author} {\bibfnamefont {J.}~\bibnamefont
  {Singh~Sidhu}}, \bibinfo {author} {\bibfnamefont {R.~M.}\ \bibnamefont
  {Abraham}}, \bibinfo {author} {\bibfnamefont {C.}~\bibnamefont {Covault}}, \
  and\ \bibinfo {author} {\bibfnamefont {G.}~\bibnamefont {Starkman}},\ }\href
  {\doibase 10.1088/1475-7516/2019/02/037} {\bibfield  {journal} {\bibinfo
  {journal} {JCAP}\ }\textbf {\bibinfo {volume} {02}},\ \bibinfo {pages} {037}
  (\bibinfo {year} {2019}{\natexlab{a}})},\ \Eprint
  {http://arxiv.org/abs/1808.06978} {arXiv:1808.06978 [astro-ph.HE]}
  \BibitemShut {NoStop}%
\bibitem [{\citenamefont {Singh~Sidhu}\ and\ \citenamefont
  {Starkman}(2019)}]{SinghSidhu:2019cpq}%
  \BibitemOpen
  \bibfield  {author} {\bibinfo {author} {\bibfnamefont {J.}~\bibnamefont
  {Singh~Sidhu}}\ and\ \bibinfo {author} {\bibfnamefont {G.}~\bibnamefont
  {Starkman}},\ }\href {\doibase 10.1103/PhysRevD.100.123008} {\bibfield
  {journal} {\bibinfo  {journal} {Phys. Rev. D}\ }\textbf {\bibinfo {volume}
  {100}},\ \bibinfo {pages} {123008} (\bibinfo {year} {2019})},\ \Eprint
  {http://arxiv.org/abs/1908.00557} {arXiv:1908.00557 [astro-ph.CO]}
  \BibitemShut {NoStop}%
\bibitem [{\citenamefont {Singh~Sidhu}\ \emph {et~al.}(2020)\citenamefont
  {Singh~Sidhu}, \citenamefont {Scherrer},\ and\ \citenamefont
  {Starkman}}]{SinghSidhu:2019loh}%
  \BibitemOpen
  \bibfield  {author} {\bibinfo {author} {\bibfnamefont {J.}~\bibnamefont
  {Singh~Sidhu}}, \bibinfo {author} {\bibfnamefont {R.~J.}\ \bibnamefont
  {Scherrer}}, \ and\ \bibinfo {author} {\bibfnamefont {G.}~\bibnamefont
  {Starkman}},\ }\href {\doibase 10.1016/j.physletb.2020.135300} {\bibfield
  {journal} {\bibinfo  {journal} {Phys. Lett. B}\ }\textbf {\bibinfo {volume}
  {803}},\ \bibinfo {pages} {135300} (\bibinfo {year} {2020})},\ \Eprint
  {http://arxiv.org/abs/1907.06674} {arXiv:1907.06674 [astro-ph.CO]}
  \BibitemShut {NoStop}%
\bibitem [{\citenamefont {Singh~Sidhu}\ \emph
  {et~al.}(2019{\natexlab{b}})\citenamefont {Singh~Sidhu}, \citenamefont
  {Starkman},\ and\ \citenamefont {Harvey}}]{SinghSidhu:2019znk}%
  \BibitemOpen
  \bibfield  {author} {\bibinfo {author} {\bibfnamefont {J.}~\bibnamefont
  {Singh~Sidhu}}, \bibinfo {author} {\bibfnamefont {G.}~\bibnamefont
  {Starkman}}, \ and\ \bibinfo {author} {\bibfnamefont {R.}~\bibnamefont
  {Harvey}},\ }\href {\doibase 10.1103/PhysRevD.100.103015} {\bibfield
  {journal} {\bibinfo  {journal} {Phys. Rev. D}\ }\textbf {\bibinfo {volume}
  {100}},\ \bibinfo {pages} {103015} (\bibinfo {year} {2019}{\natexlab{b}})},\
  \Eprint {http://arxiv.org/abs/1905.10025} {arXiv:1905.10025 [astro-ph.HE]}
  \BibitemShut {NoStop}%
\bibitem [{\citenamefont {Dhakal}\ \emph {et~al.}(2023)\citenamefont {Dhakal},
  \citenamefont {Prohira}, \citenamefont {Cappiello}, \citenamefont {Beacom},
  \citenamefont {Palo},\ and\ \citenamefont {Marino}}]{Dhakal:2022rwn}%
  \BibitemOpen
  \bibfield  {author} {\bibinfo {author} {\bibfnamefont {P.}~\bibnamefont
  {Dhakal}}, \bibinfo {author} {\bibfnamefont {S.}~\bibnamefont {Prohira}},
  \bibinfo {author} {\bibfnamefont {C.~V.}\ \bibnamefont {Cappiello}}, \bibinfo
  {author} {\bibfnamefont {J.~F.}\ \bibnamefont {Beacom}}, \bibinfo {author}
  {\bibfnamefont {S.}~\bibnamefont {Palo}}, \ and\ \bibinfo {author}
  {\bibfnamefont {J.}~\bibnamefont {Marino}},\ }\href {\doibase
  10.1103/PhysRevD.107.043026} {\bibfield  {journal} {\bibinfo  {journal}
  {Phys. Rev. D}\ }\textbf {\bibinfo {volume} {107}},\ \bibinfo {pages}
  {043026} (\bibinfo {year} {2023})},\ \Eprint
  {http://arxiv.org/abs/2209.07690} {arXiv:2209.07690 [hep-ph]} \BibitemShut
  {NoStop}%
\bibitem [{\citenamefont {Davoudiasl}\ \emph {et~al.}(2020)\citenamefont
  {Davoudiasl}, \citenamefont {Denton},\ and\ \citenamefont
  {Gehrlein}}]{Davoudiasl:2020ypv}%
  \BibitemOpen
  \bibfield  {author} {\bibinfo {author} {\bibfnamefont {H.}~\bibnamefont
  {Davoudiasl}}, \bibinfo {author} {\bibfnamefont {P.~B.}\ \bibnamefont
  {Denton}}, \ and\ \bibinfo {author} {\bibfnamefont {J.}~\bibnamefont
  {Gehrlein}},\ }\href {\doibase 10.1103/PhysRevD.102.091701} {\bibfield
  {journal} {\bibinfo  {journal} {Phys. Rev. D}\ }\textbf {\bibinfo {volume}
  {102}},\ \bibinfo {pages} {091701} (\bibinfo {year} {2020})},\ \Eprint
  {http://arxiv.org/abs/2007.04989} {arXiv:2007.04989 [hep-ph]} \BibitemShut
  {NoStop}%
\bibitem [{\citenamefont {Du}\ \emph {et~al.}(2023)\citenamefont {Du},
  \citenamefont {Lee}, \citenamefont {Wang},\ and\ \citenamefont
  {Zurek}}]{GWDuLeeWangZurek:2023dhk}%
  \BibitemOpen
  \bibfield  {author} {\bibinfo {author} {\bibfnamefont {Y.}~\bibnamefont
  {Du}}, \bibinfo {author} {\bibfnamefont {V.~S.~H.}\ \bibnamefont {Lee}},
  \bibinfo {author} {\bibfnamefont {Y.}~\bibnamefont {Wang}}, \ and\ \bibinfo
  {author} {\bibfnamefont {K.~M.}\ \bibnamefont {Zurek}},\ }\href@noop {} {\
  (\bibinfo {year} {2023})},\ \Eprint {http://arxiv.org/abs/2306.13122}
  {arXiv:2306.13122 [astro-ph.CO]} \BibitemShut {NoStop}%
\bibitem [{\citenamefont {Page}\ \emph {et~al.}(2022)\citenamefont {Page},
  \citenamefont {Homan}, \citenamefont {Nava-Callejas}, \citenamefont
  {Cavecchi}, \citenamefont {Beznogov}, \citenamefont {Degenaar}, \citenamefont
  {Wijnands},\ and\ \citenamefont {Parikh}}]{hyperburstPage:2022ikz}%
  \BibitemOpen
  \bibfield  {author} {\bibinfo {author} {\bibfnamefont {D.}~\bibnamefont
  {Page}}, \bibinfo {author} {\bibfnamefont {J.}~\bibnamefont {Homan}},
  \bibinfo {author} {\bibfnamefont {M.}~\bibnamefont {Nava-Callejas}}, \bibinfo
  {author} {\bibfnamefont {Y.}~\bibnamefont {Cavecchi}}, \bibinfo {author}
  {\bibfnamefont {M.~V.}\ \bibnamefont {Beznogov}}, \bibinfo {author}
  {\bibfnamefont {N.}~\bibnamefont {Degenaar}}, \bibinfo {author}
  {\bibfnamefont {R.}~\bibnamefont {Wijnands}}, \ and\ \bibinfo {author}
  {\bibfnamefont {A.~S.}\ \bibnamefont {Parikh}},\ }\href {\doibase
  10.3847/1538-4357/ac72a8} {\bibfield  {journal} {\bibinfo  {journal}
  {Astrophys. J.}\ }\textbf {\bibinfo {volume} {933}},\ \bibinfo {pages} {216}
  (\bibinfo {year} {2022})},\ \Eprint {http://arxiv.org/abs/2202.03962}
  {arXiv:2202.03962 [astro-ph.HE]} \BibitemShut {NoStop}%
\bibitem [{\citenamefont {{Tumino}}(2018)}]{FusionRatesTumino2018}%
  \BibitemOpen
  \bibfield  {author} {\bibinfo {author} {\bibfnamefont {e.~a.}\ \bibnamefont
  {{Tumino}}, \bibfnamefont {A.}},\ }\href {\doibase 10.1038/s41586-018-0149-4}
  {\bibfield  {journal} {\bibinfo  {journal} {Nature}\ }\textbf {\bibinfo
  {volume} {557}},\ \bibinfo {pages} {687} (\bibinfo {year}
  {2018})}\BibitemShut {NoStop}%
\bibitem [{\citenamefont {{Caughlan}}\ and\ \citenamefont
  {{Fowler}}(1988)}]{FusionRatesCaughlanFowler1988}%
  \BibitemOpen
  \bibfield  {author} {\bibinfo {author} {\bibfnamefont {G.~R.}\ \bibnamefont
  {{Caughlan}}}\ and\ \bibinfo {author} {\bibfnamefont {W.~A.}\ \bibnamefont
  {{Fowler}}},\ }\href {\doibase 10.1016/0092-640X(88)90009-5} {\bibfield
  {journal} {\bibinfo  {journal} {Atomic Data and Nuclear Data Tables}\
  }\textbf {\bibinfo {volume} {40}},\ \bibinfo {pages} {283} (\bibinfo {year}
  {1988})}\BibitemShut {NoStop}%
\bibitem [{\citenamefont {Mori}\ \emph {et~al.}(2019)\citenamefont {Mori},
  \citenamefont {Famiano}, \citenamefont {Kajino}, \citenamefont {Kusakabe},\
  and\ \citenamefont {Tang}}]{FusionRatesWD:Mori2018krw}%
  \BibitemOpen
  \bibfield  {author} {\bibinfo {author} {\bibfnamefont {K.}~\bibnamefont
  {Mori}}, \bibinfo {author} {\bibfnamefont {M.~A.}\ \bibnamefont {Famiano}},
  \bibinfo {author} {\bibfnamefont {T.}~\bibnamefont {Kajino}}, \bibinfo
  {author} {\bibfnamefont {M.}~\bibnamefont {Kusakabe}}, \ and\ \bibinfo
  {author} {\bibfnamefont {X.}~\bibnamefont {Tang}},\ }\href {\doibase
  10.1093/mnrasl/sly188} {\bibfield  {journal} {\bibinfo  {journal} {Mon. Not.
  Roy. Astron. Soc.}\ }\textbf {\bibinfo {volume} {482}},\ \bibinfo {pages}
  {L70} (\bibinfo {year} {2019})},\ \Eprint {http://arxiv.org/abs/1810.01025}
  {arXiv:1810.01025 [astro-ph.HE]} \BibitemShut {NoStop}%
\bibitem [{\citenamefont {Dohi}\ \emph {et~al.}(2022)\citenamefont {Dohi},
  \citenamefont {Nishimura}, \citenamefont {Sotani}, \citenamefont {Noda},
  \citenamefont {Liu}, \citenamefont {Liu}, \citenamefont {Nagataki},\ and\
  \citenamefont {Hashimoto}}]{FusionRatesSupburstDohi:2022gmm}%
  \BibitemOpen
  \bibfield  {author} {\bibinfo {author} {\bibfnamefont {A.}~\bibnamefont
  {Dohi}}, \bibinfo {author} {\bibfnamefont {N.}~\bibnamefont {Nishimura}},
  \bibinfo {author} {\bibfnamefont {H.}~\bibnamefont {Sotani}}, \bibinfo
  {author} {\bibfnamefont {T.}~\bibnamefont {Noda}}, \bibinfo {author}
  {\bibfnamefont {H.}~\bibnamefont {Liu}}, \bibinfo {author} {\bibfnamefont
  {H.-L.}\ \bibnamefont {Liu}}, \bibinfo {author} {\bibfnamefont
  {S.}~\bibnamefont {Nagataki}}, \ and\ \bibinfo {author} {\bibfnamefont
  {M.-a.}\ \bibnamefont {Hashimoto}},\ }\href {\doibase
  10.3847/1538-4357/ac8dfe} {\bibfield  {journal} {\bibinfo  {journal}
  {Astrophys. J.}\ }\textbf {\bibinfo {volume} {937}},\ \bibinfo {pages} {124}
  (\bibinfo {year} {2022})},\ \Eprint {http://arxiv.org/abs/2208.14622}
  {arXiv:2208.14622 [astro-ph.HE]} \BibitemShut {NoStop}%
\bibitem [{\citenamefont {Cooper}\ \emph {et~al.}(2009)\citenamefont {Cooper},
  \citenamefont {Steiner},\ and\ \citenamefont
  {Brown}}]{FusionRatesWD:Cooper:2009ps}%
  \BibitemOpen
  \bibfield  {author} {\bibinfo {author} {\bibfnamefont {R.~L.}\ \bibnamefont
  {Cooper}}, \bibinfo {author} {\bibfnamefont {A.~W.}\ \bibnamefont {Steiner}},
  \ and\ \bibinfo {author} {\bibfnamefont {E.~F.}\ \bibnamefont {Brown}},\
  }\href {\doibase 10.1088/0004-637X/702/1/660} {\bibfield  {journal} {\bibinfo
   {journal} {Astrophys. J.}\ }\textbf {\bibinfo {volume} {702}},\ \bibinfo
  {pages} {660} (\bibinfo {year} {2009})},\ \Eprint
  {http://arxiv.org/abs/0903.3994} {arXiv:0903.3994 [astro-ph.HE]} \BibitemShut
  {NoStop}%
\bibitem [{\citenamefont {Jiang}\ \emph {et~al.}(2007)\citenamefont {Jiang},
  \citenamefont {Rehm}, \citenamefont {Back},\ and\ \citenamefont
  {Janssens}}]{FusionRatesWDHindrance:2007}%
  \BibitemOpen
  \bibfield  {author} {\bibinfo {author} {\bibfnamefont {C.~L.}\ \bibnamefont
  {Jiang}}, \bibinfo {author} {\bibfnamefont {K.~E.}\ \bibnamefont {Rehm}},
  \bibinfo {author} {\bibfnamefont {B.~B.}\ \bibnamefont {Back}}, \ and\
  \bibinfo {author} {\bibfnamefont {R.~V.~F.}\ \bibnamefont {Janssens}},\
  }\href {\doibase 10.1103/PhysRevC.75.015803} {\bibfield  {journal} {\bibinfo
  {journal} {Phys. Rev. C}\ }\textbf {\bibinfo {volume} {75}},\ \bibinfo
  {pages} {015803} (\bibinfo {year} {2007})}\BibitemShut {NoStop}%
\bibitem [{\citenamefont {{Taniguchi}}\ and\ \citenamefont
  {{Kimura}}(2021)}]{FusionRatesFullMicroscopicModel:Taniguchi}%
  \BibitemOpen
  \bibfield  {author} {\bibinfo {author} {\bibfnamefont {Y.}~\bibnamefont
  {{Taniguchi}}}\ and\ \bibinfo {author} {\bibfnamefont {M.}~\bibnamefont
  {{Kimura}}},\ }\href {\doibase 10.1016/j.physletb.2021.136790} {\bibfield
  {journal} {\bibinfo  {journal} {Physics Letters B}\ }\textbf {\bibinfo
  {volume} {823}},\ \bibinfo {eid} {136790} (\bibinfo {year} {2021})},\ \Eprint
  {http://arxiv.org/abs/2106.04321} {arXiv:2106.04321 [nucl-th]} \BibitemShut
  {NoStop}%
\bibitem [{\citenamefont {Kasliwal}\ \emph {et~al.}(2012)\citenamefont
  {Kasliwal} \emph {et~al.}}]{KasliwalCaRichTransients:2011se}%
  \BibitemOpen
  \bibfield  {author} {\bibinfo {author} {\bibfnamefont {M.~M.}\ \bibnamefont
  {Kasliwal}} \emph {et~al.},\ }\href {\doibase 10.1088/0004-637X/755/2/161}
  {\bibfield  {journal} {\bibinfo  {journal} {Astrophys. J.}\ }\textbf
  {\bibinfo {volume} {755}},\ \bibinfo {pages} {161} (\bibinfo {year}
  {2012})},\ \Eprint {http://arxiv.org/abs/1111.6109} {arXiv:1111.6109
  [astro-ph.HE]} \BibitemShut {NoStop}%
\bibitem [{\citenamefont {Smirnov}\ \emph {et~al.}(2022)\citenamefont
  {Smirnov}, \citenamefont {Goobar}, \citenamefont {Linden},\ and\
  \citenamefont {M\"ortsell}}]{SmirnovLindenCaRichTransients:2022zip}%
  \BibitemOpen
  \bibfield  {author} {\bibinfo {author} {\bibfnamefont {J.}~\bibnamefont
  {Smirnov}}, \bibinfo {author} {\bibfnamefont {A.}~\bibnamefont {Goobar}},
  \bibinfo {author} {\bibfnamefont {T.}~\bibnamefont {Linden}}, \ and\ \bibinfo
  {author} {\bibfnamefont {E.}~\bibnamefont {M\"ortsell}},\ }\href@noop {} {\
  (\bibinfo {year} {2022})},\ \Eprint {http://arxiv.org/abs/2211.00013}
  {arXiv:2211.00013 [astro-ph.CO]} \BibitemShut {NoStop}%
\bibitem [{\citenamefont {Chakrabarty}\ and\ \citenamefont
  {et~al.}(2018)}]{supburst:NICERearly}%
  \BibitemOpen
  \bibfield  {author} {\bibinfo {author} {\bibfnamefont {D.}~\bibnamefont
  {Chakrabarty}}\ and\ \bibinfo {author} {\bibnamefont {et~al.}},\ }in\
  \href@noop {} {\emph {\bibinfo {booktitle} {American Astronomical Society
  Meeting Abstracts \#231}}},\ \bibinfo {series} {American Astronomical Society
  Meeting Abstracts}, Vol.\ \bibinfo {volume} {231}\ (\bibinfo {year} {2018})\
  p.\ \bibinfo {pages} {157.13}\BibitemShut {NoStop}%
\bibitem [{\citenamefont {{Agrawal}}(2006)}]{supburst:Astrosat}%
  \BibitemOpen
  \bibfield  {author} {\bibinfo {author} {\bibfnamefont {P.~C.}\ \bibnamefont
  {{Agrawal}}},\ }\href {\doibase 10.1016/j.asr.2006.03.038} {\bibfield
  {journal} {\bibinfo  {journal} {Advances in Space Research}\ }\textbf
  {\bibinfo {volume} {38}},\ \bibinfo {pages} {2989} (\bibinfo {year}
  {2006})}\BibitemShut {NoStop}%
\end{thebibliography}%
